\documentclass[reprint,amsmath,amssymb,aps,prx,noeprint,twoside]{revtex4-2}

\usepackage{graphicx}
\usepackage{dcolumn}
\usepackage[caption=false]{subfig}

\usepackage{tikz}
\usetikzlibrary{positioning,calc}
\usepackage{xcolor}

\usepackage{bm}
\usepackage{bbm}

\DeclareFontFamily{U}{matha}{\hyphenchar\font45}
\DeclareFontShape{U}{matha}{m}{n}{<-> matha10}{}
\DeclareSymbolFont{matha}{U}{matha}{m}{n}
\DeclareMathSymbol{\rightharpoonup}{3}{matha}{"E1}
\newcommand{\harpoffset}{-0.35ex}
\newcommand{\overharpoon}[1]{%
  \overset{\smash{\raisebox{\harpoffset}{$\scriptstyle\rightharpoonup$}}}{#1}%
}

\usepackage{setspace}
\usepackage{hyperref}
\hypersetup{
  colorlinks=true,
  linkcolor=blue, filecolor=blue, urlcolor=blue, citecolor=blue,
  pdftitle={Interpretable AI Analysis of Strongly Correlated Electrons (with Supplement)},
  pdfauthor={Changkai Zhang}
}
\usepackage{geometry}
\newgeometry{left=0.6in,right=0.6in,top=0.9in,bottom=0.7in}
\usepackage{fancyhdr}
\renewcommand{\headrulewidth}{0.5pt}

\usepackage{times}
\usepackage[utf8]{inputenc}
\usepackage{CJKutf8}
\newcommand{\zh}[1]{\begin{CJK}{UTF8}{gbsn}#1\end{CJK}}

\newlength\mylen
\setcitestyle{citesep={,\kern-\mylen\kern1pt}}
\ifPDFTeX \DeclareUnicodeCharacter{2212}{-} \fi

\usepackage{docmute}

\makeatletter
\AtBeginDocument{%
  \providecommand{\@author@finish}{}%
  \providecommand{\title@column}{}%
  \providecommand{\titleblock@produce}{}%
  \let\REV@orig@maketitle\maketitle
  \def\maketitle{%
    \@author@finish
    \title@column\titleblock@produce
    \suppressfloats[t]%
  }%
}
\makeatother

\usepackage{balance}   

\begin{document}

\pagestyle{fancy}\fancyhf{}
\fancyhead[LE,RO]{PREPRINT FOR PHYSICAL REVIEW (2025)}
\fancyhead[RE]{ZHANG AND VON DELFT}
\fancyhead[LO]{AI ANALYSIS OF CORRELATED ELECTRONS}
\fancyfoot[C]{MANUSCRIPT - \thepage}
\fancypagestyle{first}{%
  \lhead{}\rhead{}\chead{\large PREPRINT FOR PHYSICAL REVIEW (2025)}
}


\preprint{APS/123-QED}


\title{\vspace*{12pt}Interpretable Artificial Intelligence (AI) Analysis of Strongly Correlated Electrons\vspace{3pt}}%

\author{Changkai Zhang (\zh{张昌凯})}
\author{Jan von Delft\vspace{2pt}}
\affiliation{Arnold Sommerfeld Center for Theoretical Physics,
Center for NanoScience,~and Munich Center for Quantum Science and Technology,\\
Ludwig-Maximilians-Universität München, 80333 Munich, Germany}
\vspace{2pt}


\begin{abstract}
\setstretch{1.08}
Artificial Intelligence (AI) has become an exceptionally powerful tool for analyzing scientific data. In particular, attention-based architectures have demonstrated a remarkable capability to capture complex correlations and to furnish interpretable insights into latent, otherwise inconspicuous patterns. This progress motivates the application of AI techniques to the analysis of strongly correlated electrons, which remain notoriously challenging to study using conventional theoretical approaches. Here, we propose novel AI workflows for analyzing snapshot datasets from tensor-network simulations of the two-dimensional (2D) Hubbard model over a broad range of temperature and doping. The 2D Hubbard model is an archetypal strongly correlated system, hosting diverse intriguing phenomena including Mott insulators, anomalous metals, and high-$T_c$ superconductivity. Our AI techniques yield fresh perspectives on the intricate quantum correlations underpinning these phenomena and facilitate universal omnimetry for ultracold-atom simulations of the corresponding strongly correlated systems.
\end{abstract}

\maketitle

\vspace{-2em}

\thispagestyle{first}


\setstretch{1.08}

\section{Introduction}

The invention of Artificial Intelligence (AI) has revolutionized the way we interrogate and interpret scientific data. The attention scheme \cite{Bahdanau&Bengio2014,Rocktaschel&Blunsom2015,Xu&Bengio2015,Wang&Zhao2016,Kim&Rush2017} has been proven highly effective in transduction tasks in conjunction with recurrent or convolutional networks. Afterwards, the transformer \cite{Vaswani&Polosukhin2017} --- an architecture built solely upon the attention scheme --- was demonstrated to be compelling in capturing global dependencies in sequential data. Over the past decade, transformer-like architectures have dramatically enhanced the capability of the AI models across various domains, including natural language processing (NLP) \cite{Dong&Xu2018,Devlin&Toutanova2019,Ren&Liu2019,Brown&Amodei2020,Gulati&Pang2020,Latif&Qadir2023}, computer vision \cite{Carion&Zagoruyko2020,Dosovitskiy&Houlsby2020,Liu&Guo2021,Khan&Shah2022}, bioinformatics \cite{Jaegle&Carreira2021,Jumper&Hassabis2021,Rives&Fergus2021}, and numerous other areas \cite{Chen&Mordatch2021,Wen&Sun2022,Zhou&Zhang2020,Velickovic&Bengio2018}. Compared with alternative designs, the attention mechanism in the transformer excels particularly at encoding the correlation structure of the input data. This feature motivates the application of the transformer models in studying strongly correlated electrons.

Strongly correlated systems \cite{Dagotto&Dagotto2005,Avella&Mancini2012,Avella&Mancini2013,Avella&Mancini2015} are governed by considerably strong interactions, inducing collective behaviors that defy descriptions hinged on individual (quasi-) particles. Canonical examples include Mott insulators \cite{Mott&Mott1949,Mcwhan&Rice1973,imadaMetalinsulatorTransitions1998}, high-$T_c$ superconductors \cite{Bednorz&Mueller1986-SC-LaBaCuO,Wu&Chu1987,Tokura&Uchida1989,Schilling&Ott1993}, heavy-fermion materials \cite{Kondo&Kondo1964,Andres&Ott1975,Steglich&Schafer1979,Kamihara&Hosono2008}, fractional quantum Hall systems \cite{Tsui&Gossard1982,Laughlin&Laughlin1983,Stormer&Gossard1999}, spin liquids \cite{Anderson&Anderson1973,Haldane&Haldane1983,Shimizu&Saito2003,Szirmai&Nafradi2020}, and quark-gluon plasmas \cite{Gross&Wilczek1973,DavidPolitzer&DavidPolitzer1974,Koch&Rafelski1986,Wilson&Wilson2005}. The consequent high levels of quantum entanglement and correlations render these systems notoriously challenging for conventional theoretical approaches. With advances in computational hardware, a handful of numerical algorithms --- among them Quantum Monte Carlo (QMC) \cite{Blankenbecler&Scalapino1981-AFQMC,Sugiyama1986-AFQMC,Qin&Problem2020,Schollwoeck&White2024-Hubbard-ehdoped}, Dynamical Mean-Field Theory (DMFT) \cite{Capone2006-Hubbard-DMFT,Knizia&Chan2012-DMET,Zheng2016-DMET-Hubbard,Vanhala2018-Hubbard-DMFT}, Density Matrix Renormalization Group (DMRG) \cite{White1992-DMRG,White&Scalapino1998,Schollwock&Schollwock2011,Stoudenmire&White2012,Gleis&VonDelft2023}, and various ground-state \cite{Cirac2004-PEPS,Cirac2008-PEPS,HCJiang2008-SimpleUpdate,Barthel2009-PEPS-Contract,Cirac2010-fermionicPEPS,Corboz2010-PEPS-NN,Corboz2010-PEPS-NNN,Zhang&VonDelft2025} or finite-temperature \cite{White&White2009-METTS,Stoudenmire&White2010-METTS,Li&Su2011-LinearizedTRG,Czarnik2012-PEPS-finiteT-first,Czarnik2015-PEPS-finiteT-variational,Li&Weichselbaum2018-XTRG,Li&vonDelft2019-Heisenberg-XTRG,Li2022-tanTRG,Zhang&VonDelft2025-XTRG} Tensor Network (TN) methods --- have been devised to tackle strongly correlated systems. Moreover, quantum simulation apparatuses based on ultra-cold atoms \cite{Mazurenko&Greiner2017,Chiu&Greiner2018,Koepsell2019-Ultracold-FermiHubbard,Koepsell2020-Ultracold-tech,Chalopin&Hilker2025,Xu&Greiner2025} have achieved substantial progress in emulating strongly interacting lattice systems. Together, these techniques offer valuable many-body data from which the AI models can learn and distill meaningful insights.

Among the plethora of strongly correlated electron systems, the two-dimensional (2D) Hubbard model \cite{Hubbard1967,Anderson1987-Hubbard-RVB} stands out as a paradigmatic arena for a variety of intriguing phenomena, such as Mott physics, anomalous metals, and high-$T_c$ superconductivity. The Hubbard model encapsulates the essential physics of itinerant electrons on a lattice with strong on-site Coulomb repulsion. Over the past few decades, the 2D Hubbard model has been subject to intensive investigations both numerically \cite{2DHubbard-Benchmark,NNHubbard-Conclusive,Robinson&Tsvelik2019,Qin&Gull2022,Qin&Problem2020,Schollwoeck&White2024-Hubbard-ehdoped,HCJiang2019-Hubbard-DMRG,Jiang&Jiang2020,White&Schollwoeck2020-DMRG-Hubbard-PlaquettePairing,Jiang&Devereaux2024-Hubbard-ehdoped,Corboz2019-Hubbard-PEPS,Zhang&VonDelft2025,Wietek&Stoudenmire2021-METTS,Li2022-tanTRG,Bohrdt&Knap2019,Zhang&VonDelft2025-XTRG} and experimentally \cite{Mazurenko&Greiner2017,Chiu&Greiner2018,Grusdt2019-Hubbard-StringPattern,Koepsell2019-Ultracold-FermiHubbard,Salomon&Gross2019,Koepsell2020-Ultracold-tech,Chen&vonDelft2021-Hubbard-XTRG,Koepsell2021-Ultracold-Polaron,Sompet&Bloch2022,Hirthe&Hilker2023,Xu&Greiner2023,Chalopin&Bloch2024,Pasqualetti&Folling2024,Bourgund&Hilker2025,Chalopin&Hilker2025,Xu&Greiner2025}. Robust anti-ferromagnetic (AFM) orders have been confirmed near half-filling \cite{Jiang&Jiang2020,Wietek&Stoudenmire2021-METTS,Dong&Millis2022,Xu&Zhang2022,Xiao&Zhang2023,Sousa-junior&DosSantos2024,Zhang&VonDelft2025}, while in the doped regime --- especially with carrier hopping beyond neighboring sites --- diverse charge and spin orders, often coexisting with or competing against pairing tendencies, have been identified \cite{Jiang&Devereaux2024-Hubbard-ehdoped,White&Schollwoeck2020-DMRG-Hubbard-PlaquettePairing,Jiang&Jiang2020,Jiang&White2023,Li2022-tanTRG,Zhang&VonDelft2025}. These properties broadly echo the observations in the cuprate superconductors.

Despite this decent progress, the majority of the existing researches focused on local and low-order spin and/or charge correlations, especially two-point correlators. However, mounting evidence indicates that high-order \cite{Grusdt2021-Ultracold-Correlator,Miles&Kim2021,Bourgund&Hilker2025}, non-local \cite{Cheuk&Zwierlein2016,Hilker&Gross2017}, polaronic \cite{Grusdt&Demler2018,Koepsell2019-Ultracold-FermiHubbard,Salomon&Gross2019,Grusdt&Pollet2020} or otherwise string-like \cite{Weng&Ting1997,Grusdt2019-Hubbard-StringPattern,Ho&Ho2020} correlations play a pivotal role in deciphering the complicated phase diagram of the 2D Hubbard model. This recognition highlights the promise of the AI techniques for the global vision of the underlying quantum correlations.

In this Article, we study the 2D Hubbard model using specifically designed AI models. We start with assembling a dedicated dataset of snapshots across categories of temperatures and doping levels by sampling the thermal density matrix via TN simulations. Then, we propose two AI architectures classifying snapshots into the respective categories: the \emph{pro architecture}, an analog of the encoder-only transformer, and the \emph{core architecture}, a streamlined variant that attains comparable performance, better support for parallelism and improved interpretability. 

Next, we perform multiple analyses on the trained core model. We use a confusion analysis to measure the quality of the classification tasks and obtain insights into the aggregate strength of quantum correlations in each category. Exploiting the \emph{semi-linear} structure of the attention stack in the core architecture, we propose an interpretation in terms of an effective Markovian dynamics, demonstrating the alignment of the attention design with intrinsic features of the physical system. Further examinations on the orthogonality relationships of the embedding and the attention maps are provided in the supplemental material \cite{supplemental}.

Finally, we demonstrate an application of our core AI models as a universal omnimeter for ultracold-atom quantum simulations. The AI classifiers produce probablistic scores (logits) for each category, which serve as posterior likelihoods conditioned on a snapshot acquired in the experiment. Averaging these outputs over a snapshot ensemble from repeated observations thus provides an empirical probability distribution over the categories. Once the categories are calibrated with pre-determined physical quantities, the expectation values weighted by the probability distribution yield an accurate estimate of the corresponding quantities for the ensemble.

\section{Lattice Model \& Dataset}

Lattice models serve as common platforms for the physics of crystalline materials, wherein charge carriers reside on and hop between discrete lattice sites. In many materials of interest, itinerant electrons predominantly occupy the outer-most $s$ orbital for transport. Consequently, the local Hilbert space at each lattice site is spanned by four basis: empty $\mid\mkern-4mu\varnothing\rangle$, spin-up $\mid\uparrow\rangle$, spin-down $\mid\downarrow\rangle$, and doubly occupied $\mid\uparrow\downarrow\rangle$ state.

In our study, we focus on the quintessential 2D Hubbard model on an 8×8 square lattice with open boundary conditions, defined via the following Hamiltonian
\begin{equation}
  \label{Hamiltonian}
    \mathcal{H} = -\sum_{i,j,\sigma} t_{ij} \left[\, c^\dagger_{i\sigma} c_{j\sigma} + \text{h.c.} \,\right] + U\sum_i n_{i\uparrow} n_{i\downarrow}.
    \vspace{-4pt}
\end{equation}
Here, $c^\dagger_{i\sigma}$ ($c_{i\sigma}$) creates (annihilates) an electron with spin $\sigma$ on site $i$, and $n_{i\sigma} = c^\dagger_{i\sigma} c_{i\sigma}$ denotes the corresponding number operator. The first term in Eq.~\eqref{Hamiltonian} describes the kinetic energy associated with electron hopping between sites $i$ and $j$ with amplitude $t_{ij}$, while the second term accounts for the on-site Coulomb repulsion with strength $U$. Throughout this work, we consider the minimal Hubbard model where $t_{ij} = 1$ for nearest-neighbor pairs and $t_{ij} = 0$ otherwise. Also, we set $U=10$ as established to be realistic for cuprate materials \cite{Hirayama2018-Hubbard-parameter,Hirayama2018-Hubbard-electronic}.

\begin{figure}[htp!]
    \vspace{0.5em}
    \centering
    \includegraphics[scale=0.39]{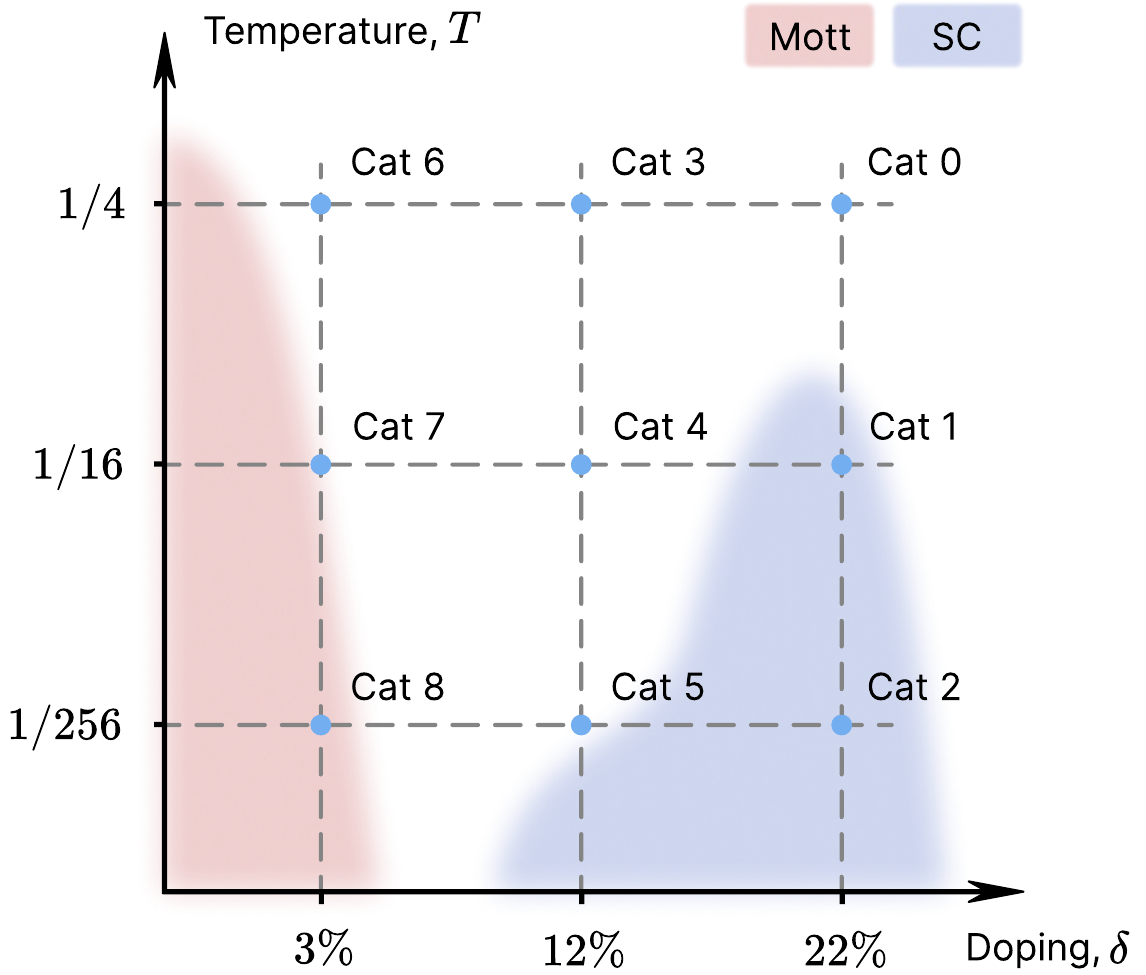}
    \begin{minipage}{0.45\textwidth}
    \caption{A schematic depiction of the locations in phase space for the nine categories (Cat), created by combining three choices of temperatures (high, medium, and low) with three doping regimes (over-doped, medium-doped, and under-doped). The red and blue freehand-shaded areas mark the AFM Mott insulating phase and the high-$T_c$ superconducting phase, respectively, as expected for the Hubbard model. The charge doping varies with temperature (see also Fig.~\ref{fig:Omnimetry}); precise values are provided in the supplemental material \cite{supplemental}.}
    \label{fig:Dataset}
    \end{minipage}
    \vspace{-0.5em}
\end{figure}

In many elemental metals, electron interactions are effectively weak ($U\!\approx\!0$) due to electric-field screening by the surrounding lattice ions, yielding conventional metallic behavior at half-filling (one electron per site). By contrast, in materials like high-$T_c$ cuprates, the on-site Coulomb repulsion becomes abnormally strong for electrons, opening a large energy gap that penalizes double occupancy. Hence, electron motion is substantially hindered by a large potential barrier and the system exhibits Mott insulating behavior at the macroscopic level.

Extra charge carriers can be introduced by adding (electron doping) or removing (hole doping) electrons relative to half-filling. Upon sufficient doping, charge transport in the material sets in and superconductivity may emerge, signified by enhanced pairing correlations at low temperatures. This evolution underlies the schematic phase diagram as shown in Fig.~\ref{fig:Dataset}, where red and blue shaded areas mark the AFM Mott insulating phase and the high-$T_c$ superconducting dome, respectively.

The (unnormalized) thermal density matrix $\rho = e^{-\beta \mathcal{H}}$ characterizes the statistical state of the lattice system, with inverse temperature $\beta = 1/T$. Note that $\rho$ admits a Taylor expansion at high temperature (small $\beta$) for a given Hamiltonian, and that
\begin{equation}
    \rho(2\beta) = \rho(\beta) \cdot \rho(\beta).
\end{equation}
Accordingly, one may cool the system down by repeatedly squaring the thermal density matrix starting from a high-temperature construct. This idea underlies the eXponential Tensor Renormalization Group (XTRG) method \cite{Li&Weichselbaum2018-XTRG,Li&vonDelft2019-Heisenberg-XTRG,Chen&vonDelft2021-Hubbard-XTRG,Lin&Shi2022-XTRG,Zhang&VonDelft2025-XTRG}, which offers a comprehensive thermal description of the lattice system over a broad temperature range. We thus employ XTRG to produce thermal density matrices at high, medium, and low temperatures at over-doped, medium-doped, and under-doped regions (hole-doped), yielding nine categories as indicated in Fig.~\ref{fig:Dataset}.

\begin{figure*}[htp!]
    \vspace{-0.1em}
    \centering
    \hspace{0.03\textwidth}
    \subfloat{\includegraphics[scale=0.32]{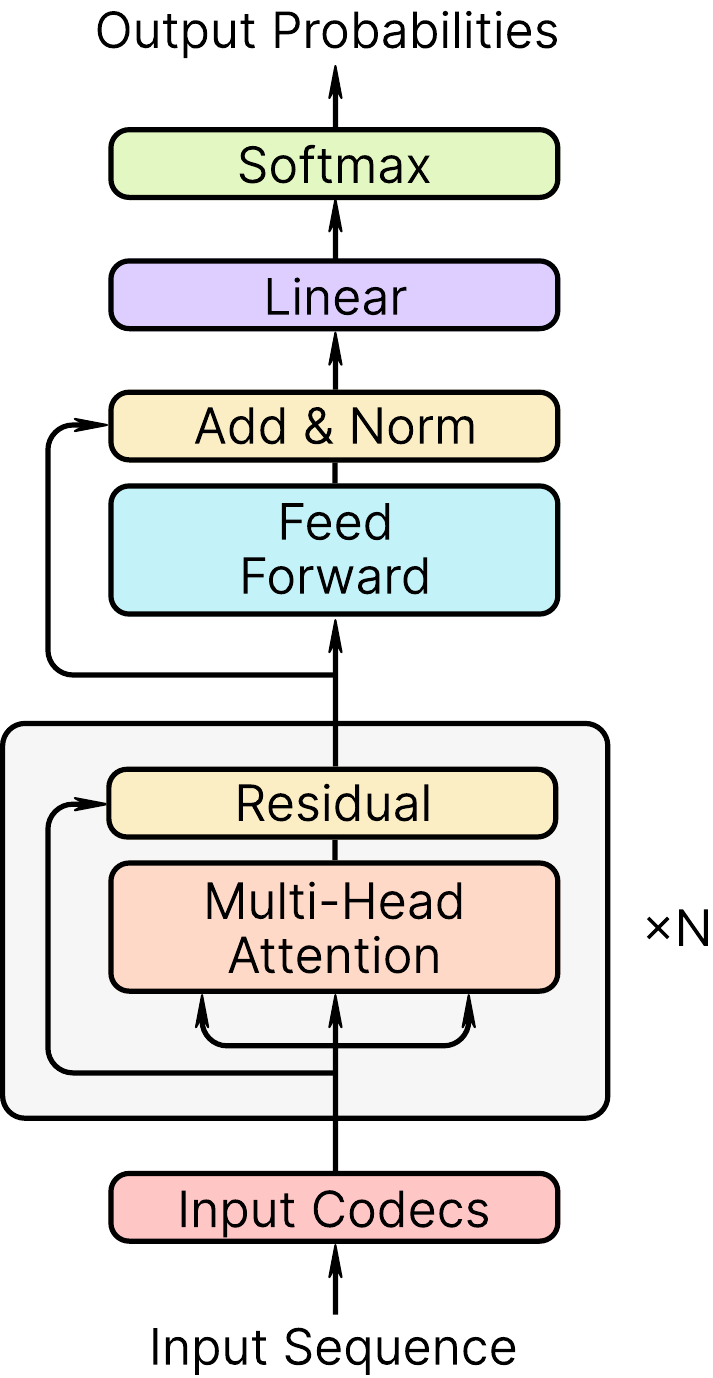}}
    \hspace{0.08\textwidth}
    \subfloat{\includegraphics[scale=0.32]{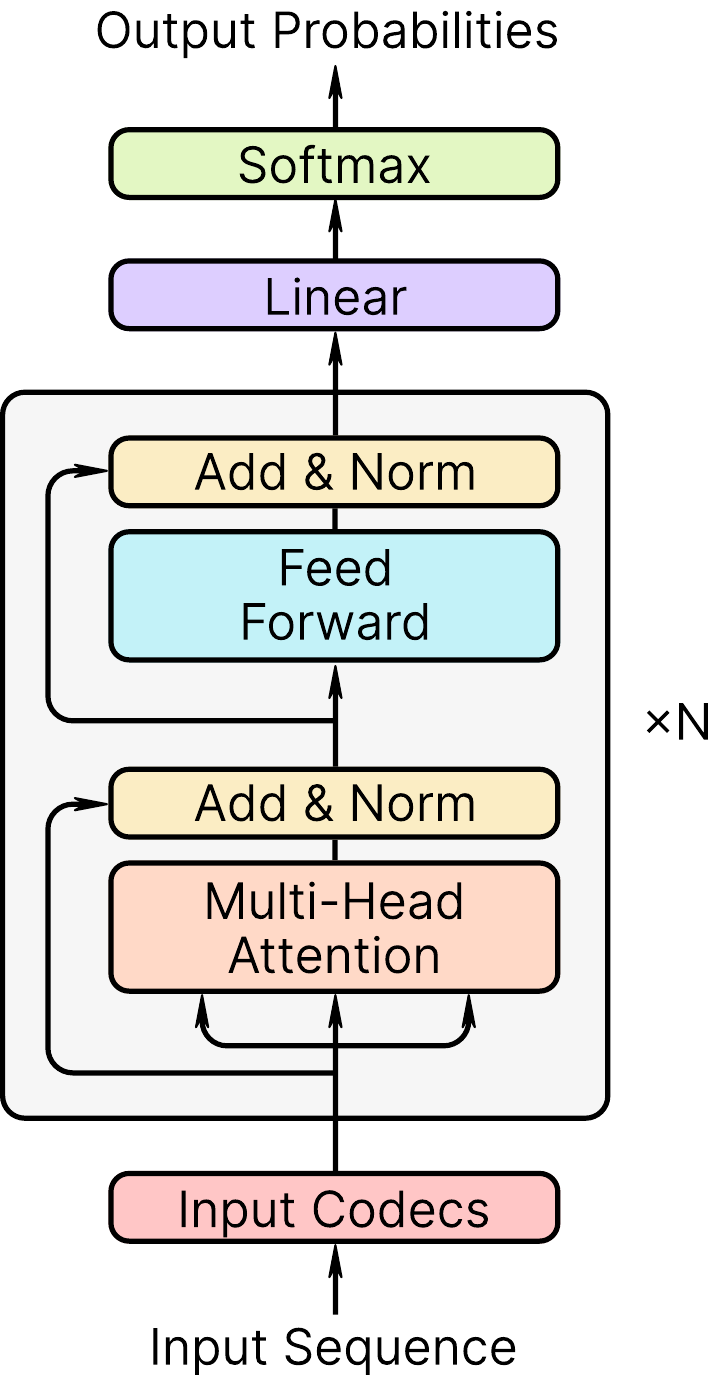}}\\[0.5em]
    \begin{minipage}{0.9\textwidth}
    \caption{Schematic illustrations of the core (left) and the pro (right) architecture for classification of sequential inputs. Both architectures comprise input codecs, multi-head attention blocks, feed-forward networks and a final linear classification head. The pro architecture is an analog of the encoder-only transformer, while the core architecture leaves out the feed-forward networks between attention blocks which enhances parallelism and improves interpretability.}
    \label{fig:Architecture}
    \end{minipage}
    \vspace{-0.3em}
\end{figure*}

We then perform standard site-wise sampling on each thermal density matrix \cite{Zhang&VonDelft2025-XTRG} to obtain snapshots (Fock bases of the many-body Hilbert space) of the lattice system. The snapshots, each consisting of 8×8 cells containing either $|\varnothing\rangle$, $\mid\uparrow\rangle$, $\mid\downarrow\rangle$, or $\mid\uparrow\downarrow\rangle$, are next flattened according to row-major order into a sequence of 64 elements. For each location in the phase space, we generate 1000 snapshots, yielding a dataset of 9 categories and 9000 snapshots in total. This dataset is randomly partitioned into training (90\%) and test (10\%) subsets for the subsequent AI workflows.

\vspace*{-1.5em}
\section{Architectures}

Our AI architectures originate from the transformer paradigm while being tailored for categorical classification. Given an input snapshot, the objective is to infer a probability distribution over categories. Rather than appending a dedicated \texttt{CLS} token \cite{Devlin&Toutanova2019,Khan&Shah2022} as a label, we adopt a streamlined design that endows the involved attention mechanism with a distinctive and physically meaningful interpretation.

Figure~\ref{fig:Architecture} depicts schematic layouts of the models deployed in this study. Both architectures share the same foundational components --- input codecs, a stack of multi-head attention, feed-forward networks, and a terminal linear classification head. The pro architecture (right) is inspired by an encoder-only transformer, whereas the core architecture (left) leaves out the feed-forward networks between attention blocks, thereby giving enhanced parallelism and improved interpretability.

Our exposition below assumes familiarity with standard components and techniques in the practices of the transformer architecture \cite{Vaswani&Polosukhin2017}, and hence will focus on our distinctive designs. Full technical details can be found in the supplemental material \cite{supplemental}.

\subsection{Tokenization \& Input Codecs}

The tokenization is straightforward in our setting, as the \emph{vocabulary} (local Hilbert space) comprises only four distinct \emph{words} (local states). We therefore assign \texttt{0}, \texttt{1}, \texttt{2}, and \texttt{3} to the empty, spin-up, spin-down, and doubly occupied states, respectively. Under this encoding, each snapshot in the dataset now becomes a sequence of integers (tokens). Formally, let $\mathcal{S} \!=\! \{\texttt{0}, \texttt{1}, \texttt{2}, \texttt{3}\}$ denote the tokenized local state space. The input sequence becomes $\overharpoon{\sigma}\in\mathcal{S}^L$, where $L$ is the flattened sequence length.

The input codecs accept and map each tokenized sequence into the model's latent parameter space in two stages. Each token $\sigma$ is first transformed into a $d_\text{model}$-dimensional embedding vector $\bm{e}(\sigma)\!\in\!\mathbb{R}^{d_\text{model}}$ via a learnable embedding module; this embedding depends solely on the token (local state) and is agnostic to its location in the sequence. To preserve positional information, we then add a positional vector $\bm{\varrho}_i\!\in\!\mathbb{R}^{d_\text{model}}$ for site $i$ to each embedding vector. For this purpose, we employ the sinusoidal positional encoding \cite{Vaswani&Polosukhin2017}, which has proved effective across a wide range of applications.

The input codecs thus assemble a feature matrix $\varSigma$ with elements $\varSigma_i^\mu \mkern-2mu=\mkern-2mu e^\mu(\sigma_i) \mkern-2mu+\mkern-2mu \varrho^\mu_i$ for each snapshot, where $i$ indexes lattice sites and $\mu$ indexes latent dimensions. Contingent on the dataset under consideration, $d_\text{model}$ should be adjusted for a balance of expressivity against overfitting. Note that in \cite{Vaswani&Polosukhin2017} the input embeddings are multiplied by a factor of $\sqrt{d_\text{model}}$ to scale up the weights; in contrast, this operation is empirically detrimental in our application domain, plausibly due to the exceedingly small vocabulary size relative to the sequence length.

\subsection{Locality-Biased Attention}

The (multi-head) attention mechanism plays a central role in harnessing global correlation awareness for both architectural designs. We adopt the prevalent scaled dot-product attention scheme \cite{Bahdanau&Bengio2014,Vaswani&Polosukhin2017} to acquire raw attention scores, and subsequently impose a locality bias for an improved training profile. Even though positional information has been encoded amid the input codecs, we find that the prototypical attention setup, which is primarily designed for 1D sequences, struggles in perceiving 2D spatial relationships. Hence, an explicit locality bias assists in this regard.

We start with linear projections of the input embeddings into query, key, and value vectors
\begin{equation}
    \bm{\mathcal{Q}}_i = \bm{\varSigma}_i W_{\!Q},\quad \bm{\mathcal{K}}_i = \bm{\varSigma}_i W_{\!K},\quad \bm{\mathcal{V}}_{\!i} = \bm{\varSigma}_i W_{\!V}
\end{equation}
with $W_{\!Q}$, $W_{\!K}$, and $W_{\!V}$ being learnable weight matrices. Here, we suppress the latent-space index $\mu$ and take matrix multiplications implicit. The attention between site $i$ and $j$ thus reads
\begin{equation}
    A_{ij} = \text{softmax}_j (\bm{\mathcal{Q}}_i \bm{\mathcal{K}}_j / \mathfrak{T}),
\end{equation}
where $\mathfrak{T}$ denotes the \emph{model temperature} (conceptually distinct from the physical temperature) which controls the sharpness of the attention distribution. In our exercises, $\mathfrak{T} = \sqrt{d_k}$ (see below for the definition of $d_k$) works reasonably well.

The multi-head attention is realized by partitioning the latent space into $h$ subspaces, each with dimension $d_k = d_\text{model}/h$. Attention is computed independently within each head, after which the head outputs are concatenated and linearly transformed to yield the final raw attention. In our actual practice, the multi-head configurations fail to outperform their single-head counterpart, likely attributable to the limited size of the training set.

Many realistic physical systems exhibit locality: objects only significantly influence their immediate neighbors, leading to a decay of the interactions and correlations with spatial separation. Accordingly, we apply a locality bias to the raw attention scores \vspace*{-1em}
\begin{equation}
    \mathcal{A}_{ij} = \text{softmax}_j(A_{ij} \circ G_{ij}),
\end{equation}
where $G_{ij}$ is a hand-crafted bias function that decays with the physical distance $d_{ij}$ between sites $i$ and $j$, and the circle~$\circ$~denotes element-wise (Hadamard) multiplication. This locality bias encourages the mechanism to focus on nearby sites, effectively accelerating the convergence during the training process. In our implementation, we choose a Gaussian kernel
\begin{equation}
    G_{ij} = \exp\left\{-d_{ij}^2 / 2\varsigma^2\right\},
\end{equation}
with standard deviation $\varsigma = \lambda/2$ and $\lambda$ a characteristic length scale of the system (e.g., $\lambda\!=\!8$ for the 8×8 square lattice considered here). The eventual performance is not highly affected by the specific choice of the bias function $G_{ij}$. For instance, a power-law decay kernel works almost equally well.

Finally, as a standard technique to stabilize the gradient propagation, we apply a residual connection \cite{He&Sun2016} by adding a $\bm{\varSigma}_i$ to the output of the attention block as
\begin{equation}
    \mathrm{attn}(\bm{\varSigma}_i) = \bm{\varSigma}_i + \textstyle{\sum_j} \mathcal{A}_{ij} \bm{\mathcal{V}}_{\!j}.
\end{equation}
Afterwards, layer normalization \cite{Ba&Hinton2016} is employed in the pro architecture, whereas the core architecture omits this step for reasons that will become clear in the ensuing interpretation.

\subsection{Feed-Forward \& Classification}

The feed-forward networks (FFNs) constitute one of the principal sources of non-linearity in the model. Each FFN is a site-wise fully-connected three-layer perceptron comprising an input layer, a hidden layer, and an output layer. The input and output layers have width $d_\text{model}$, while the hidden layer has width $d_\text{hidden}$. A ReLU activation is applied between the two affine maps to introduce non-linearity. Concretely, the FFN reads
\begin{equation}
    \mathrm{FFN}(\bm{\varSigma}_i) = \mathrm{ReLU}(\bm{\varSigma}_i W_{\!1} + \bm{b}_1)W_{\!2} + \bm{b}_2,
\end{equation}
where $W_{\!1}$, $W_{\!2}$, $\bm{b}_1$, and $\bm{b}_2$ are learnable weights and biases. The same linear maps are shared across all sites $i$, whereas different FFN blocks carry independent parameters. A residual connection and layer normalization follow each FFN.

The arrangement of FFNs constitutes the pivotal difference between the pro and core architectures. The pro variant inserts an FFN of dimension $d_\text{hidden}\!=\!d_\text{ff}$ after each attention block, while the core variant defers non-linearity to a single FFN of dimension $d_\text{hidden}\!= \!N\!\times\!d_\text{ff}$ after the entire stack of $N$ attention blocks. We call the latter design \emph{semi-linear} attention stack (for the reason that will be clear in impending interpretations).

Under this parametrization, the core and pro architectures contain an almost equivalent amount of learnable model parameters. However, the postponed feed-forward network markedly reduces the depth of non-linearity within the core model, which benefits the upcoming interpretation since physical objects commonly propagate linearly. Also, centralizing the FFN boosts parallelism during both training and inference.

The classification head ingests the abstract embedding $\bm{\varSigma}^{\mkern2mu\raisebox{1pt}{\scriptsize\text{ff}}}_i$ processed through the preceding attention and feed-forward networks, and returns a categorical distribution over the target labels. Concretely, the logit $y_{i,c}$ and the corresponding probability $p_{i,c}$ are computed as
\begin{equation}
    y_{i,c} = \bm{\varSigma}^{\mkern2mu\raisebox{1pt}{\scriptsize\text{ff}}}_i W_{\!c} + \bm{b}_c,\quad p_{i,c} = \text{softmax}_c(y_{i,c})
\end{equation}
where $W_c$ and $\bm{b}_c$ denote the learnable weight and bias associated with category $c$. Each site $i$ produces its own distribution; this is warranted by the attention mechanism, which injects into each site contextual information aggregated from all other sites. The overall prediction is obtained by the argmax of the averaged per-site distributions $p_c = \text{avg}_i\, p_{i,c}$ over all lattice sites.

\begin{figure*}[htp!]
    \vspace{0.8em}
    \centering
    \includegraphics[width=0.99\textwidth]{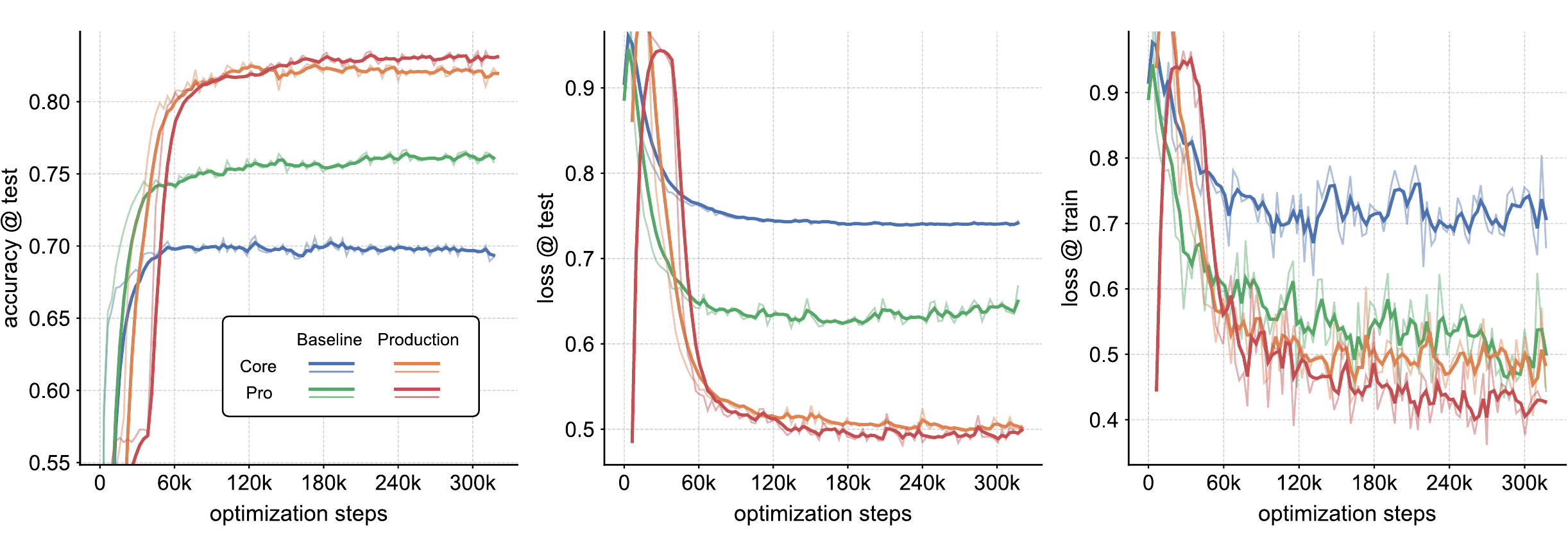}
    \begin{minipage}{0.9\textwidth}
    \caption{Training profiles and benchmarks of the production and baseline models following the core and pro architectures. Metrics are displayed as raw data (thin lines with muted color) and with an exponential smoothing factor $\alpha\!=\!0.4$ (thick lines with deep color). For baseline models (with trivialized attention), the pro architecture consistently outperforms the core variant across all metrics, congruous with the anticipated benefits of elevated non-linearity in the pro model. By contrast, for production models (with full-functional attention), the core architecture achieves merely negligible gaps in performance, indicating an alignment of the semi-linear attention with the intrinsic properties of the dataset.}
    \label{fig:Benchmarks}
    \end{minipage}
\end{figure*}

\subsection{Hyperparameters}

For the XTRG-generated snapshot dataset of the Hubbard model, we adopt the following hyperparameters: embedding dimension $d_\text{model}\!=\!128$; single-head attention $h\!=\!1$; feed-forward dimension $d_\text{ff}\!=\!1024$; and $N\!=\!2$ attention blocks. Increasing either the number of heads or the number of attention blocks empirically induces severe overfitting and should be considered only after expanding the dataset.

Furthermore, we implement an ablation toggle that trivializes all attention blocks by hard-setting $A_{ij}\!=\!1$ when enabled. Under this switch, the effective attention reduces to the fixed locality kernel, thereby delegating the classification task entirely to the FFNs. This toggle is useful both as a baseline and during the warm-up phase of training.

\section{Training \& Benchmarks}

\textbf{Hardware.} --- All the production and baseline models were trained on an NVIDIA 3090 GPU. The project utilizes PyTorch version 2.2.2 and CUDA 11.8.

\textbf{Initialization.} --- All learnable parameters are initialized with Xavier (Glorot) initialization \cite{Glorot&Bengio2010}. For production runs, the first 200 epochs serve as a warm-up phase during which the ablation toggle is enabled, effectively pre-training the FFNs with trivialized attention.

\textbf{Batch \& Epochs.} --- We use a batch size of 256 and train for a total of 10,000 epochs. Checkpoints are saved every 100 epochs, and the one with the lowest validation loss is selected as the final deliverable.

\textbf{Objective.} --- We optimize the ubiquitous cross-entropy loss as our optimization objective. The loss is computed as the negative log-likelihood averaged over all sites
\begin{equation}
    \text{loss} = \text{avg}_i\left[\, -\textstyle{\sum_c} \;\xi_{c}\log p_{i,c}\:\right],
\end{equation}
where $\xi_{c}$ denotes the ground-truth one-hot label for category $c$ (broadcasted across all sites), or specifically
\begin{equation}
    \xi_{c} = 
    \begin{cases}
        \ 1\quad \text{for correct category}\ \:\! c,\\
        \ 0\quad \text{otherwise}.
    \end{cases}
\end{equation}

\textbf{Optimizer.} --- We use the standard Adam optimizer \cite{Kingma&Ba2014} with $\beta_1 = 0.9$, $\beta_2 = 0.999$, and $\epsilon = 10^{-9}$. The learning rate is fixed to $5\times10^{-6}$. We find that both architectural designs are sensitive to this setting: materially larger or smaller values tend to induce premature plateaus at elevated loss.

\textbf{Regularization.} --- We apply dropouts \cite{Srivastava&Salakhutdinov2014} with a rate of 0.01 to input codecs, attention blocks, FFNs and all residual connections. Contrary to the common practice in the NLP applications, we disable the label smoothing \cite{Szegedy&Wojna2016} as over-confidence is not a primary concern for a physically-generated dataset.

\textbf{Training Yield.} --- Owing to stochastic initialization, training outcomes exhibit variability. Empirically, roughly one in seven attempts attains a top-performing model.

\textbf{Benchmarks.} --- Figure~\ref{fig:Benchmarks} summarizes the training profiles 
of four models we trained: the production and baseline models under the core and pro architectures. The baselines are trained with the aforementioned ablation toggle switched on, such that their attention mechanisms are effectively disabled.

Benchmarks are reported as test-set accuracy and loss, along with training loss. Test accuracy is defined as the fraction of correctly classified snapshots in the held-out test set. All four models achieve stable convergence given sufficient training, and generalization is satisfactory as indicated by the proximity of training and test losses.

Moreover, the production models consistently surpass their baseline counterparts, attesting to the efficacy of the attention mechanism. For the baselines, the pro variant outperforms the core across all metrics, consistent with the anticipated benefits of enhanced non-linearity. By contrast, in production runs, the core model closes the gap to within negligible differences, indicating that FFNs interleaved between attention blocks are largely redundant and that the semi-linear attention stack aligns well with the intrinsic structure of the dataset.

\section{Interpretation}

For decades, achieving a principled interpretation of the internal mechanism of AI models has been one of the highest endeavors in the field \cite{Bibal&Frenay2016}. Unlike opaque black-box approaches \cite{Guidotti&Pedreschi2018}, the attention mechanism offers a natural lens on a model's focus and decision-making. The calculated attention scores $\mathcal{A}_{ij}$ are typically construed as a measure of \emph{importance} of token $j$ to token $i$ (or equivalently, the \emph{attention} paid by token $i$ to token $j$) \cite{Rocktaschel&Blunsom2015,Xu&Bengio2015}. This heuristic has been widely utilized in a variety of application domains \cite{Choi&Sun2016,Martins&Astudillo2016,Wang&Zhao2016,Lee&Kim2017,Xie&Hovy2017,Dehghani&Kaiser2018,Brunner&Wattenhofer2019,Chen&Ji2019,Clark&Manning2019,Coenen&Wattenberg2019} for analysis, diagnosis and debugging.

However, a comprehensive study \cite{Jain&Wallace2019} showed that attention weights often fail to provide consistent or exclusive explanations of model predictions; in particular, alternative attention patterns can yield essentially identical performance. These observations have ignited a prolonged debate \cite{Jain&Wallace2019,Vashishth&Faruqui2019,Vig&Vig2019,Wiegreffe&Pinter2019,Zhang&Odena2019,Pruthi&Lipton2020,Bai&Wang2021,Galassi&Torroni2021,Bibal&Watrin2022} on whether --- or to what extent --- the attention meaningfully reveals a model's reasoning process.

Despite the ongoing controversy, consensus remains that attention maps furnish at least \emph{an} (if not \emph{the}) explanation for the inner workings of the model \cite{Wiegreffe&Pinter2019}. Thereafter, further aggregation methods, such as attention rollout \cite{Abnar&Zuidema2020,Yuan&Dou2021,Xu&Liu2023} and attention flow \cite{Abnar&Zuidema2020,Metzger&Finkbeiner2023,Azarkhalili&Libbrecht2025}, have been proposed to propagate attention scores across multiple layers. The attention rollout, in particular, essentially performs a layer-wise matrix multiplication of the (residual-augmented) attention matrices. Considering the fact that these matrices are all row-stochastic, it becomes natural to interpret them as Markovian transition kernels \cite{Yuan&Dou2021}. This probabilistic viewpoint forms the basis of our interpretation.

Our interpretation focuses exclusively on the core architecture, as it depends critically on the \emph{semi-linear} nature of the attention stack (the precise meaning of which will be clarified in the upcoming subsections).

\subsection{Classical \& Quantum Markov Process}

Before heading to the interpretation, we first formalize the relevant constructs for both classical and quantum Markov processes on the lattice system. For the classical scenario, consider a discrete-time Markov dynamics in which, at each update, the state $s_i$ at site $i$ may overwrite the state $s_j$ at site $j$; the corresponding transition probability is $\mathcal{A}_{ij}$. All lattice sites update synchronously in one time step.

Suppose that a collection of observables is associated with each local state, and write $\varSigma^\mu_i$ for the $\mu$-th observable evaluated on the state at site $i$. Under the Markov evolution described above, the observables update after one step as
\begin{equation}
  \label{eq:markov}
    \varSigma^\mu_i \leftarrow {\textstyle\sum_{j}} \mathcal{A}_{ij} \varSigma^\mu_j.
\end{equation}

For the quantum scenario, the local state can be associated with a (pure) local density matrix $\rho_i\! = \: \mid\mkern-5mu\sigma_i\rangle\langle\sigma_i\mkern0.5mu|\equiv|i\rangle\langle i|$. The quantum Markov process is specified by a completely-positive trace-preserving (CPTP) map $\mathcal{E}$ (also known as a quantum channel) comprising a set of Kraus operators
\begin{equation}
    K_j = {\textstyle\sum_\imath} \sqrt{\mathcal{A}_{\imath j}}\ |j\rangle\langle \imath|,
\end{equation}
whose action on the local state is
\begin{equation}
    \mathcal{E}(\rho_i) = {\textstyle\sum_j} K_j \rho_i K_j^\dagger = {\textstyle\sum_j}\mathcal{A}_{ij} \rho_j.
\end{equation}
Assign to each $\rho_i$ the same family of observables $\varSigma^\mu_i$. Under the channel $\mathcal{E}$, these observables also evolve according to Eq.~\eqref{eq:markov}. Therefore, in both classical and quantum constructs, Eq.~\eqref{eq:markov} captures the one-step evolution of observables under the Markov process with the transition kernel $\mathcal{A}_{ij}$.

\begin{figure*}[htp!]
    \vspace{-0.6em}
    \centering
    \hspace{-0.6em}
    \subfloat{\includegraphics[scale=0.6]{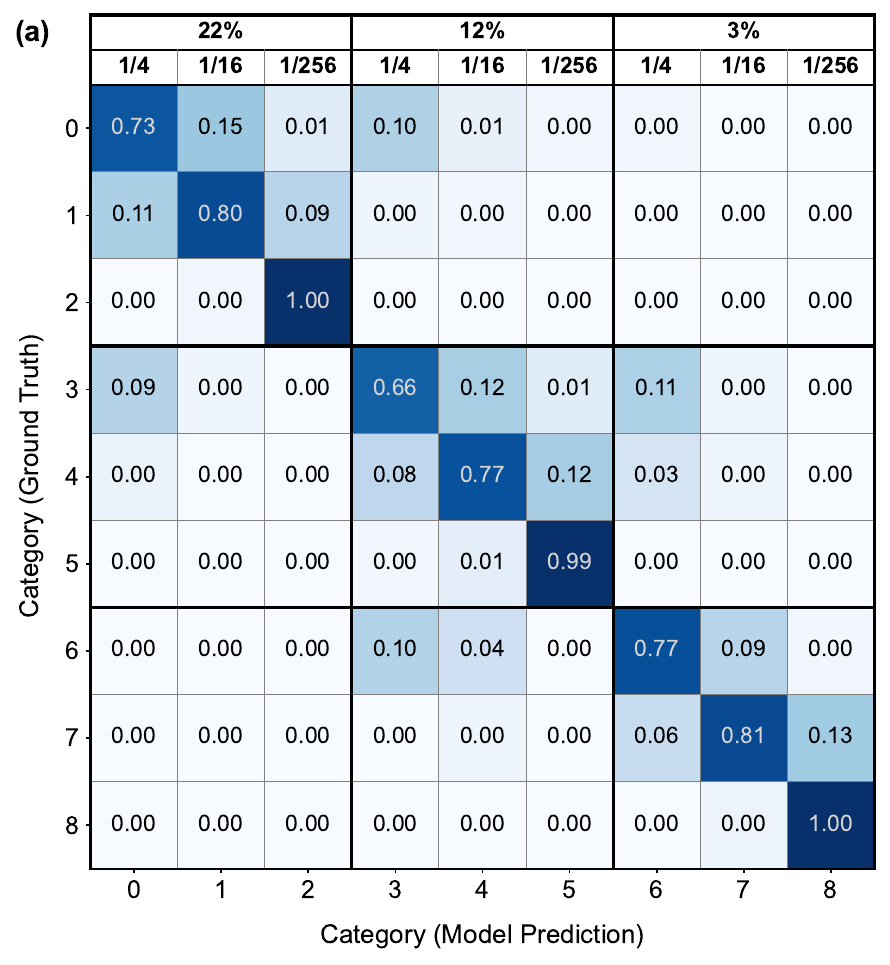}}
    \hspace{0.02\textwidth}
    \subfloat{\includegraphics[scale=0.6]{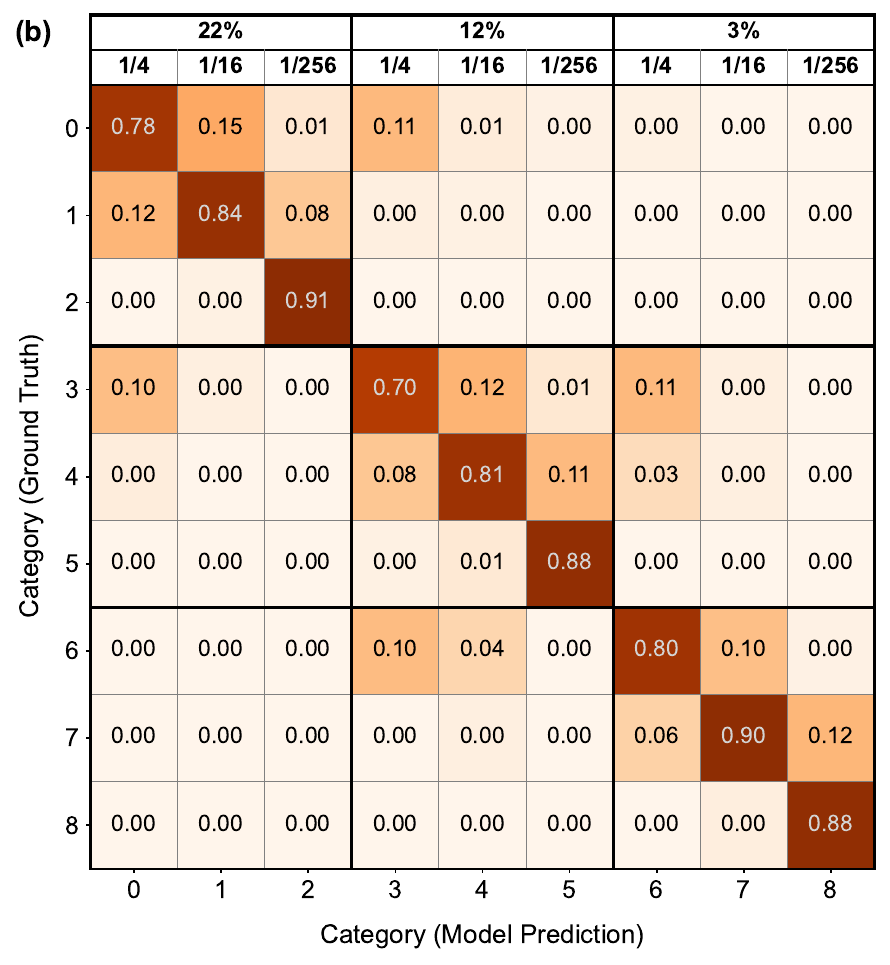}}\\[-0.1em]
    \hspace{0.4em}
    \begin{minipage}{0.9\textwidth}
    \caption{(a) Sensitivity matrix (row-normalized confusion matrix) and (b) precision matrix (column-normalized confusion matrix) for the core model. Color intensity indicates the degree of sensitivity (a) and precision (b), with exact values written within each cell. Top rows indicate the charge doping and temperature of the corresponding categories. Diagonal entries show (a) the probability of correct classification for each category and (b) the probability of a predicted category being correct.}
    \label{fig:Confusion}
    \end{minipage}
\end{figure*}

\subsection{Interpretation of the Attention Stack}

We now make explicit the link between the attention stack in the core architecture and the Markovian description above. Viewing the embedded features as observables, the right-hand side of Eq.~\eqref{eq:markov} coincides with the application of the attention matrix $\mathcal{A}_{ij}$ to the input embedding $\varSigma^\mu_i$. In conjunction with the embedding projector $W_{\!V}$ and the residual connection, the output of a stack of $N\!=\mkern-2mu2$ attention blocks can be expressed as
\begin{equation}
  \label{eq:attn_stack}
  \begin{aligned}
    \bm{\varSigma}_i^{\mkern2mu\raisebox{1pt}{\scriptsize\text{attn}}} = \bm{\varSigma}_i & + {\textstyle\sum_{k}}\mathcal{A}^{(1)}_{ik}\bm{\varSigma}_k W_{\!V}^{(1)}\\
    & + {\textstyle\sum_{j}} \mathcal{A}^{(2)}_{ij} \left[\bm{\varSigma}_j + {\textstyle\sum_{k}}\mathcal{A}^{(1)}_{jk}\bm{\varSigma}_k W_{\!V}^{(1)}\right] W_{\!V}^{(2)}\\
    = \bm{\varSigma}_i & + {\textstyle\sum_{j}} \mathcal{A}^{(1)}_{ij} \bm{\varSigma}_j W_{\!V}^{(1)} + {\textstyle\sum_{j}} \mathcal{A}^{(2)}_{ij} \bm{\varSigma}_j W_{\!V}^{(2)}\\
    & + {\textstyle\sum_{j,k}} \mathcal{A}^{(2)}_{ij} \mathcal{A}^{(1)}_{jk} \bm{\varSigma}_k W_{\!V}^{(1)} W_{\!V}^{(2)},
  \end{aligned}
\end{equation}
where $\mathcal{A}_{ij}^{(\ell)}$ and $W_{\!V}^{(\ell)}$ denote, respectively, the attention matrix and the embedding projector from the $\ell$-th attention block. On the right-hand side of the final equality, four terms appear: the first term carries the original embedding; the second and third terms propagate features according to the first and second attention kernels; and the final term captures the cascaded propagation through both blocks. These terms can be interpreted as four distinct \emph{Markovian propagation modes} acting on the input $\bm{\varSigma}_i$. The same expansion generalizes analogously to deeper stacks.

For trivial embedding projectors $W_{\!V}^{(\ell)}\mkern-5mu = \mkern-5mu\mathbbm{1}$ (where $\mathbbm{1}$ is the identity matrix), Eq.~\eqref{eq:attn_stack} reduces to
\begin{equation}
  \label{eq:attn_stack_linear}
    \bm{\varSigma}_i^{\mkern2mu\raisebox{1pt}{\scriptsize\text{attn}}} = {\textstyle\sum_{j}} \mathcal{R}_{ij} \bm{\varSigma}_j = {\textstyle\sum_{j}} \left[{\textstyle\prod_\ell} (\mathbbm{1} + \mathcal{A}^{(\ell)})_{ij}\right] \bm{\varSigma}_j,
\end{equation}
where $\mathcal{R}_{ij}$ is precisely the standard attention rollout \cite{Abnar&Zuidema2020} (without normalization). We refer to this limiting case as a \emph{linear} attention stack. However, in practice, the embedding projectors are generally non-trivial, yielding the \emph{semi-linear} attention stack. This is the unique feature of the core architecture; by contrast, in the pro variant, the layer normalization and the FFN introduce non-linearity between attention blocks.

This perspective furnishes an interesting interpretation of the core architecture. First, the input codecs learn a feature embedding whose components can be construed as physically relevant observables attached to each local state. The attention stack then effects a superposition of Markovian propagation modes that evolve these observables across the lattice, while residual connections preserve the original features. Finally, the classification head operates on the resultant evolved features (observables). In essence, the model learns an \emph{effective} Markov dynamics --- encoded in the attention kernels and embedding projections --- that best aligns the propagated observables with the downstream classification objective.

\begin{figure*}[htp!]
    \centering
    \subfloat{\includegraphics[scale=0.565]{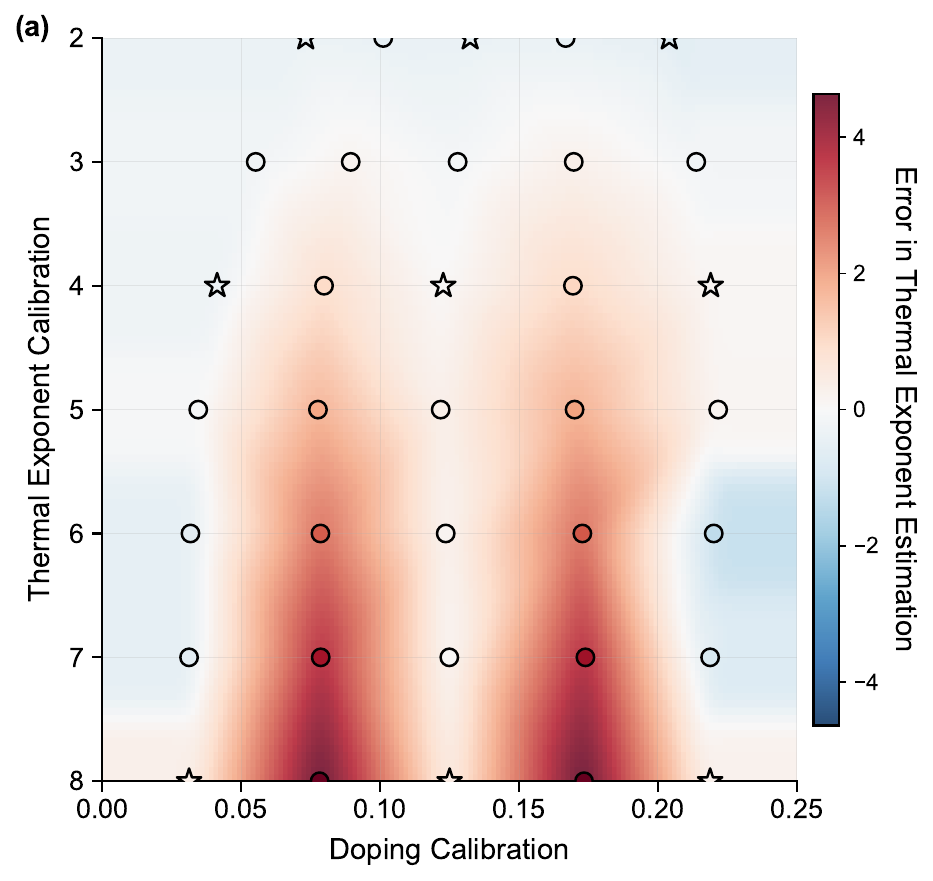}}
    \hspace{0.018\textwidth}
    \subfloat{\includegraphics[scale=0.565]{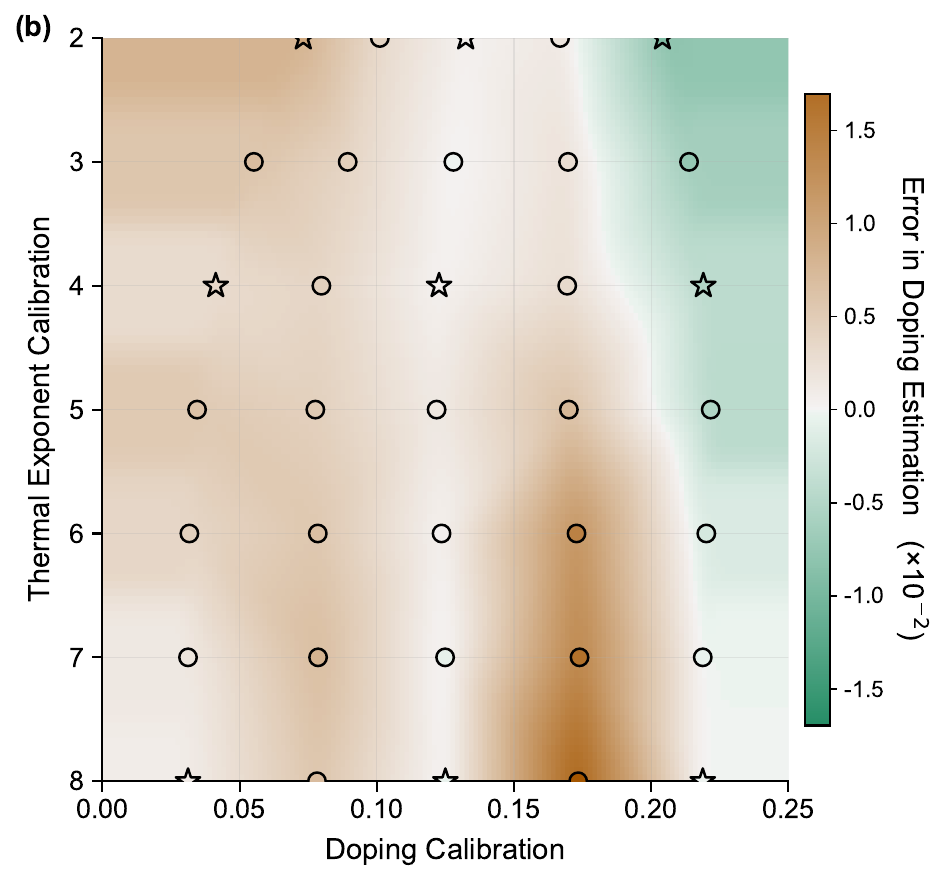}}\\[-0.1em]
    \hspace{0.4em}
    \begin{minipage}{0.9\textwidth}
    \caption{Error in the omnimeter estimation of (a) the thermal exponent and (b) charge doping. Open circles/pentagrams mark the locations in phase space of the snapshot ensembles under evaluation, with pentagrams (circles) indicating data included (not included) in the training set. Color scales are obtained via interpolation. Overall performance is good, except for the bands at the unseen doping levels (around 7\% and 17\% in (a) and around 17\% in (b)).}
    \label{fig:Omnimetry}
    \end{minipage}
    \vspace{0.4em}
\end{figure*}

\section{Confusion Analysis}

The core (production) model attains an overall 83\% accuracy on the test subset, as delineated in Fig.~\ref{fig:Benchmarks}. However, this aggregate metric masks substantial variation across categories; a nuanced assessment requires a full-scale confusion analysis of per-category sensitivity (true positive rate) and precision (positive predictive value).

We construct the confusion matrix $\varXi$, where each entry $\varXi_{cc'}$ counts snapshots whose ground-truth category is $c$ but are classified as $c'$. Row-normalizing $\varXi$ yields the sensitivity matrix, which estimates the probability $p(c'|c)$ that a snapshot from category $c$ is predicted as $c'$. Column-normalizing $\varXi$ produces the precision matrix, which estimates the probability $p(c|c')$ that a snapshot predicted as $c'$ actually originates from category $c$.

Figures~\ref{fig:Confusion}(a) and \ref{fig:Confusion}(b) display the sensitivity and precision matrices, respectively. The 9×9 matrices are partitioned into a 3×3 (doping) block of 3×3 (temperature) cells and exhibit a pronounced block-diagonal structure, indicating that misclassifications occur predominantly within the same doping level. Moreover, sensitivity increases systematically as temperature decreases (reaching almost 100\% at the lowest temperature), whereas precision does not exhibit an equally monotonic trend.

To rationalize these tendencies, we identify two principal sources of randomness in the dataset: thermal and quantum fluctuations. Thermal fluctuations are essentially structureless and uncorrelated, while quantum fluctuations can admit non-trivial quantum correlations. The latter effectively enhance the system's entanglement entropy, thereby providing additional information that aids discrimination across categories.

Under this perspective, sensitivity may be interpreted as the \emph{prominence} of a category's correlation pattern --- greater correlation strength raises the likelihood of correct classification; whereas precision reflects the \emph{uniqueness} of that pattern --- greater distinctiveness reduces the chance that snapshots from other categories are misattributed to it.

Consequently, the observed rise in sensitivity at lower temperatures suggests increasingly prominent correlation structures, consistent with the suppressed thermal noise (and relatively accentuated quantum correlations). By contrast, the more modest gains in precision at the lowest temperatures imply that low-temperature patterns also occur at a higher temperature, in line with the known persistence of Mott physics and superconducting correlations into a moderate-temperature regime.

Lastly, we remark that misclassification is not catastrophic in our context, since the dataset is itself intrinsically stochastic and exhibits substantial randomness. Therefore, an argmax-based decision rule may mislabel even under a Bayes-optimal classifier. For instance, let's suppose a perfect model assigns a snapshot 50\% probability to category $c$ and 40\% to category $c'$; the argmax strategy will thereby deterministically predict category $c$, although this snapshot may quite plausibly originate from category $c'$. Hence, one should regard the probability distribution $p_c$ as the faithful output of the model.

\section{Universal Omnimetry}

One of the straightforward applications of our AI classifier is to perform measurements on an arbitrary ensemble of snapshots. Each category in the dataset is affiliated with a set of known observables, and the classifier outputs a distribution over categories, thereby inducing an estimate of the corresponding observables. We refer to this procedure as \emph{omnimetry}, since all affiliated observables are inferred simultaneously.

For a demonstration of this new technique, we augment the dataset with additional snapshots drawn from regions of phase space that were not included in the training set (i.e., \emph{unseen} by the model). A sufficiently generalizable model should then produce a distribution over categories that reflects the resemblance of the correlation patterns in the input ensemble against those learned during training. Thus, the performance of the omnimeter serves as a probe of the underlying correlation structure across distinct partitions of phase space.

The workflow starts with a \emph{calibration} of observables for all categories in the training set. Let $\omega_c^{(\alpha)}$ denote the $\alpha$-th observable affiliated with category $c$. In our study of the finite-temperature Hubbard model, the affiliated observables are temperature and charge doping. As XTRG algorithm produces thermal density matrices at temperatures $T = 1/2^{n_T}$, we utilize the \emph{thermal exponent} $n_T$ as a representative observable in place of temperature. Additional observables can, of course, be accommodated, provided the dataset supports and a corresponding calibration is available.

Next, given an ensemble of snapshots $\{x\}$ generated under fixed conditions, the core classifier returns, for each snapshot, a categorical distribution $p_c(x)$ over $c$. Averaging these distributions across the ensemble yields a collective distribution measuring the probability that this ensemble corresponds to category $c$. The target observable for the ensemble is then estimated by the distribution-weighted average of the calibration
\begin{equation}
    \langle \omega^{(\alpha)} \rangle = {\textstyle\sum_c}\, \omega_c^{(\alpha)}\, \text{avg}_x p_c(x).
\end{equation}

Figure~\ref{fig:Omnimetry}(a) shows the error of the omnimeter's estimates of the thermal exponent $n_T$ for ensembles drawn from various locations in phase space. Overall performance is satisfactory --- particularly at doping levels partially covered in the training set. Notably, two red bands appear at unseen doping levels (around 7\% and 17\%), with inferred temperatures systematically higher than the ground truth. This bias is plausibly attributed to correlation patterns in those regions that were absent during training; the model interprets these as signatures of elevated thermal fluctuations and hence predicts higher temperature. This behavior, in turn, corroborates that the classifier has genuinely learned to associate correlation patterns with thermodynamic conditions.

Figure~\ref{fig:Omnimetry}(b) reports the error in estimating charge doping. Performance is again strong, aside from a distinct band at the unseen doping of around 17\%. A similar mechanism applies: the model finds that the unseen patterns resemble those at around 22\% doping, consistent with both doping levels lying within the superconducting regime.

These artifacts can be eliminated by augmenting the training data in the relevant portions of phase space. As a trailer, we can announce that a 25-category classifier which includes the problematic doping levels in its training set substantially mitigates these issues, reducing the relative error in both temperature and doping to below 10\%. Further details will be available in the supplemental material \cite{supplemental} as well as an upcoming dedicated technical report.

Thermometry remains a central challenge in ultracold-atom experiments \cite{Chalopin&Bloch2024,Xu&Greiner2025}, and our AI omnimeter offers a competitive upgrade. Modern quantum gas microscopy \cite{Koepsell2019-Ultracold-FermiHubbard,Koepsell2020-Ultracold-tech} produces ensembles of site-resolved snapshots of the analog cold-atom simulator, which can be compared directly with our numerical snapshot dataset. Whereas current thermometers often rely on matching hand-selected metrics and correlators \cite{Hartke&Hartke2020,Chalopin&Bloch2024,Pasqualetti&Folling2024,Xu&Greiner2025}, our approach automatically discovers discriminative patterns and aggregates them into robust temperature estimates. We therefore anticipate that this approach can materially enhance the reliability of thermometry in ultracold-atom platforms.

\section{Summary \& Outlook}

In this Article, we establish an end-to-end technological stack for AI-assisted analysis of strongly correlated electron systems on a lattice. The workflow starts with tensor-network simulations that generate thermal density matrices and, in turn, an extensive snapshot dataset. This dataset is then processed by our tailored AI architectures featuring locality-biased, semi-linear attention with principled interpretability grounded in effective Markovian dynamics and strong capacity to capture latent correlation patterns. The trained model is subsequently subjected to a comprehensive confusion analysis, revealing the prominence and uniqueness of correlation structures across thermodynamic conditions. Finally, the model is deployed as an omnimeter to infer multiple observables from arbitrary ensembles of snapshots.

Our research demonstrates the viability of bespoke AI technologies for interrogating challenging strongly correlated systems. Moreover, the approach is versatile and readily extends to lattice models for diverse physical scenarios. The observation that the core model attains performance comparable to the pro variant suggests further opportunities to optimize the transformer architecture. Also, additional dynamical information may be extracted from the attention stack itself. Besides, the universal omnimetry furnishes a generic measurement methodology for quantum many-body experiments equipped with local-state quantum microscopy \cite{Koepsell2019-Ultracold-FermiHubbard,Koepsell2020-Ultracold-tech}. Beyond classifiers, alternative AI paradigms --- e.g., generative models --- merit exploration for deeper analysis and new applications. We therefore anticipate that this work opens a promising new avenue for the study of strongly correlated systems and will motivate further researches along these lines.

\vspace*{-1em}

\section*{Acknowledgements}

We thank Annabelle Bohrdt, Fabian Döschl, Hannah Lange, Roger Melko and Henning Schlömer for helpful discussions and feedback. This research was funded in part by the Deutsche Forschungsgemeinschaft under Germany's Excellence Strategy EXC-2111 (Project No.~390814868), and is part of the Munich Quantum Valley, supported by the Bavarian state government through the Hightech Agenda Bayern Plus.

\makeatletter
\def\bibsection{%
  \par
  \baselineskip26\p@
  \bib@device{\linewidth}{82\p@}%
  \nobreak\@nobreaktrue
  \addvspace{19\p@}%
  \par
}
\makeatother

\bibliography{library, special}

\balance          
\clearpage        

\clearpage
\pdfbookmark[1]{Supplemental Material}{bm:supp}
\hypertarget{supp-start}{}

\providecommand{\externaldocument}[2][]{}

\setcounter{page}{1}
\setcounter{section}{0}
\setcounter{equation}{0}
\setcounter{figure}{0}
\setcounter{table}{0}

\renewcommand{\thefigure}{S\arabic{figure}}
\renewcommand{\thesection}{S-\Roman{section}}
\renewcommand{\theequation}{S\arabic{equation}}

\newcommand{\BatchStart}{\refstepcounter{figure}\edef\BatchMainNumber{\arabic{figure}}}
\newcommand{\BatchCaption}[2]{%
  \begingroup\renewcommand{\thefigure}{\BatchMainNumber-#1}\caption{#2}\endgroup
  \addtocounter{figure}{-1}}
\newcommand{\BatchEnd}{}

\pagestyle{fancy}\fancyhf{}
\renewcommand{\headrulewidth}{0.5pt}
\fancyfoot[C]{SUPPLEMENTAL MATERIAL - \thepage}
\fancypagestyle{first}{\renewcommand{\headrulewidth}{0pt}}


\makeatletter
\begingroup

  \@ifundefined{titleblock@produce}{}{%
    \@ifundefined{maketitle}{%
      \newcommand{\maketitle}{%
        \begingroup
          \onecolumngrid
          \titleblock@produce
          \twocolumngrid
        \endgroup
        \thispagestyle{first}%
      }%
    }{%
      \renewcommand{\maketitle}{%
        \begingroup
          \onecolumngrid
          \titleblock@produce
          \twocolumngrid
        \endgroup
        \thispagestyle{first}%
      }%
    }%
  }%

  \setstretch{1}

\preprint{APS/123-QED}


\title{\vspace*{12pt}Supplemental Material - Interpretable Artificial Intelligence (AI)\\ Analysis of Strongly Correlated Electrons\vspace{3pt}}%

\author{Changkai Zhang (\zh{张昌凯})}
\author{Jan von Delft\vspace{2pt}}
\affiliation{Arnold Sommerfeld Center for Theoretical Physics,
Center for NanoScience,~and Munich Center for Quantum Science and Technology,\\
Ludwig-Maximilians-Universität München, 80333 Munich, Germany}


\maketitle

\thispagestyle{first}


\setstretch{1.08}

In the supplemental material, we provide (\ref{sec:transformer}) an introduction for physicists to the transformer architecture; (\ref{sec:dataset}) detailed specifications of the snapshot dataset for the Hubbard model; (\ref{sec:derangements}) a performance benchmark of both architectures on an artificial derangement dataset; (\ref{sec:attention_maps}) an analysis of orthogonality and attention maps; and (\ref{sec:omnimeter}) a performance preview of a 25-category omnimeter.

\section{Introduction to the Transformer}
\label{sec:transformer}

In this section, we present an elementary introduction for physicists to the transformer architecture, grounded in the seminal work \emph{Attention is All You Need} \cite{Vaswani&Polosukhin2017}. The transformer was originally developed for sequence transduction tasks in natural language processing (NLP). Subsequently, an encoder-only variant \cite{Devlin&Toutanova2019} has been proposed for generative or classification tasks, which we further develop into the \emph{pro} architecture in the main text. Here, we focus on this particular instantiation of the transformer as applied to physical lattice models, wherein snapshots can be regarded as sequences in the \emph{language} of the physical system. And classifying a given snapshot into one of the nine categories in phase space is akin to e.g. classifying a sentence into one of several sentiment classes in NLP.

\textbf{Tokenization.} --- Tokens are the pre-defined elementary units of the input sequence. In NLP, tokens are typically words, whitespaces, punctuations, etc. For snapshots of a lattice system, tokens are the local states $\sigma$ on each lattice site, e.g., empty, spin-up, spin-down, and double-occupied states for the Fermi-Hubbard model. The input sequence is then a one-dimensional array of tokens obtained by flattening the two-dimensional (2D) lattice snapshot in a row-major order. We assign \texttt{0}, \texttt{1}, \texttt{2}, and \texttt{3} to the four local states, respectively. Therefore, a snapshot of a lattice system with $L$ sites is now \emph{tokenized} into an input sequence $\overharpoon{\sigma}\mkern-2mu\in\mkern-2mu\{\texttt{0}, \texttt{1}, \texttt{2}, \texttt{3}\}^L$.

\textbf{Input Embedding.} --- The tokenizer described above assigns a unique integer to each token (local state). However, these integers are purely nominal and do not encode any semantic information about the corresponding local state. A more informative strategy is to represent each token by a vector of \emph{features}. Specifically, one could use an array $(\pi_c, n_c, s_z, S, \ldots)$ comprising, e.g., parity $\pi_c$, number of particles $n_c$, spin-$z$ $s_z$, total spin $S$, etc., to represent each local state. In this example, the spin-up state would be encoded as $(-1, 1, +\frac{1}{2}, \frac{1}{2}, \ldots)$, and the other local states follow analogously. We refer to this hand-crafted representation as an \emph{input encoding} of the sequence.

However, an input encoding requires manual identification of the relevant features for each local state and may thus be constrained by prior knowledge about the system. A more flexible approach is to allow the model to learn a suitable representation of each token directly from data. This is achieved via an \emph{input embedding}, in which all features are learnable parameters. In practice, the embedding layer is essentially a lookup table that stores an embedding vector $\bm{e}(\sigma)\!\in\!\mathbb{R}^{d_\text{model}}$ for each local state $\sigma$. The dimension $d_\text{model}$ (number of features) of the embedding vectors is a hyperparameter to be chosen when constructing the model. The components of $\bm{e}(\sigma)$ are denoted $e^\mu(\sigma)$, where $\mu=1,2,\ldots,d_\text{model}$ is the feature index.

\textbf{Positional Encoding.} --- The embedding vector of a token depends only on the local state it represents and carries no information about its position in the sequence. We therefore need a separate mechanism to inject positional information. The transformer architecture contains neither recurrent nor convolutional structures --- common devices in other architectures for capturing sequential order --- and instead relies on a \emph{positional encoding}. Analogous to the input encoding, the positional encoding is a fixed (non-learnable) map that converts each position $i$ in the input sequence into a positional vector $\bm{\varrho}_i\!\in\!\mathbb{R}^{d_\text{model}}$. A common choice for the positional encoding is to use sine and cosine functions of different frequencies:
\begin{equation}
    \varrho_i^\mu = \begin{cases}
        \sin(i / 10000^{2\mu/d_\text{model}}), & \text{if $\mu$ is even}, \\
        \cos(i / 10000^{2\mu/d_\text{model}}), & \text{if $\mu$ is odd},
    \end{cases}
\end{equation}
%
The motivation for this sinusoidal form is to enable the model to infer relative positions between tokens, since any $\bm{\varrho}_{i+k}$ can be expressed as a linear function of $\bm{\varrho}_i$. Similar to the input embedding, it is also possible to employ a learnable \emph{positional embedding}. However, in practice, we do not observe a benefit from this upgrade, consistent with the findings in \cite{Vaswani&Polosukhin2017}.

\textbf{Input Codecs.} --- The input codecs consolidate the input embedding and positional encoding. The embedding vector $\bm{e}(\sigma_i)\equiv\bm{e}_i$ of the token at position $i$ is scaled by $w_e$ and added to the positional encoding $\bm{\varrho}_i$ to yield the final input representation $\varSigma^\mu_i = w_e e^\mu_i + \varrho^\mu_i$, i.e. $\bm{\varSigma}_i = w_e \bm{e}_i + \bm{\varrho}_i\mkern-1mu\in\mkern-1mu\mathbb{R}^{d_\text{model}}$. Consequently, for every snapshot, the tokenized input sequence $\overharpoon{\sigma}$ of $L$ tokens is transformed into a feature matrix
\begin{center}
\vspace*{-2.8em}
\begin{equation}
    \begin{bmatrix}
        \sigma_1 \\
        \sigma_2 \\
        \vdots \\
        \sigma_L
    \end{bmatrix}
    \rightarrow
    \begin{bmatrix}
        \varSigma_1^1 & \varSigma_1^2 & \cdots & \varSigma_1^{d_\text{model}} \\
        \varSigma_2^1 & \varSigma_2^2 & \cdots & \varSigma_2^{d_\text{model}} \\
        \vdots & \vdots & \ddots & \mkern-20mu\vdots \\
        \varSigma_L^1 & \varSigma_L^2 & \cdots & \varSigma_L^{d_\text{model}}
    \end{bmatrix}
\end{equation}
\end{center}
\vspace{0.1em}
whose rows enumerate positions and columns enumerate features. In \cite{Vaswani&Polosukhin2017}, the weight is set to $w_e=\sqrt{d_\text{model}}$. In their NLP tasks, the vocabulary size (number of unique tokens) is approximately 37{,}000, while the sequence length is about 25{,}000. It is therefore natural to emphasize the input embedding relative to the positional encoding. In our physical applications, however, the vocabulary size is usually small (e.g., 4 for the Fermi-Hubbard model) whereas the sequence length can be comparatively large (e.g., 64 for an 8×8 lattice). Hence, we instead use $w_e=1$ to place the input embedding and positional encoding on an equal footing.

\textbf{Modular Design.} --- Contemporary AI systems commonly adopt a modular design, wherein the overall architecture comprises a sequence of \emph{modules} (depicted as rectangular blocks in Fig.~\ref{fig:Architecture}) with standardized inputs and/or outputs, enabling algorithms to be assembled in a building-block fashion. In the transformer architecture, all the constituent modules consume and/or emit data in the same format of feature matrix $\varSigma\mkern-2mu\in\mkern-2mu\mathbb{R}^{L\times d_\text{model}}$. This uniform interface greatly simplifies the construction of deep models via a straight-forward stacking of modules. Accordingly, it is natural to regard $\bm{\varSigma}_i$ as a \emph{register} memory or a module \emph{argument}, rather than a specific mathematical entity with fixed values, and one should understand its significance and the contents stored according to the context.

\textbf{Dot-Product Attention.} --- The principal workhorse of the transformer is the attention mechanism, which enables the model to capture long-range (global) correlations across the input sequence. The attention module receives and converts $\bm{\varSigma}_i$ into three sets of vectors: the \emph{queries} $\bm{\mathcal{Q}}_i\in\mathbb{R}^{d_k}$, the \emph{keys} $\bm{\mathcal{K}}_i\in\mathbb{R}^{d_k}$, and the \emph{values} $\bm{\mathcal{V}}_i\in\mathbb{R}^{d_v}$. For self-attention, one commonly takes $d_v = d_k$, and obtains the queries, keys, and values via linear projections of the input codecs:
\begin{equation}
  \begin{aligned}
    Q_i^\nu &= \sum_{\mu=1}^{d_\text{model}} \varSigma_i^\mu W_{\!Q}^{\mu\nu},\quad\text{or}\quad\bm{\mathcal{Q}}_i = \bm{\varSigma}_i W_{\!Q}\in\mathbb{R}^{d_k},\\
    K_i^\nu &= \sum_{\mu=1}^{d_\text{model}} \varSigma_i^\mu W_{\!K}^{\mu\nu},\quad\text{or}\quad\bm{\mathcal{K}}_i = \bm{\varSigma}_i W_{\!K}\in\mathbb{R}^{d_k},\\
    V_i^\nu &= \sum_{\mu=1}^{d_\text{model}} \varSigma_i^\mu W_{\!V}^{\mu\nu},\quad\text{or}\quad\bm{\mathcal{V}}_i = \bm{\varSigma}_i W_{\!V}\in\mathbb{R}^{d_v},
  \end{aligned}
\end{equation}
where $W_{\!Q}$, $W_{\!K}\!\in\!\mathbb{R}^{d_\text{model}\times d_k}$, and $W_{\!V}\!\in\!\mathbb{R}^{d_\text{model}\times d_v}$ are learnable linear projection matrices. The same set of projection matrices are shared across all positions $i$.

The objective of the attention module is to compute \emph{similarities} between queries and keys and to reweight the values accordingly. In the \emph{scaled dot-product attention} \cite{Vaswani&Polosukhin2017}, similarities are measured according to the dot products of queries with keys, so the \emph{attention score} $A_{ij}$ between the $i$-th query and the $j$-th key is given by
\begin{equation}
    \label{eq:attention_scores}
    A_{ij} = \text{softmax}_j\left(\bm{\mathcal{Q}}_i\bm{\mathcal{K}}_j / \mathfrak{T}\right) = {\frac{1}{Z_i}} \exp(\bm{\mathcal{Q}}_i\bm{\mathcal{K}}_j / \mathfrak{T}),
\end{equation}
where \vspace{-1em}
\begin{equation}
    Z_i = \sum_{j=1}^L \exp(\bm{\mathcal{Q}}_i\bm{\mathcal{K}}_j / \mathfrak{T})
\end{equation}
is the normalization factor, and $\mathfrak{T}$ the \emph{model temperature} parameter (conceptually distinct from the actual physical temperature) that controls the distribution of attention scores. Following \cite{Vaswani&Polosukhin2017}, we set $\mathfrak{T}=\sqrt{d_k}$. An inner product should be inferred in the expression \vspace{-0.8em}
\begin{equation}
    \bm{\mathcal{Q}}_i\bm{\mathcal{K}}_j = \sum_{\nu=1}^{d_k} \mathcal{Q}_i^\nu \mathcal{K}_j^\nu,
\end{equation}
and in Eq.~\ref{eq:attention_scores}, the subscript $j$ in $\text{softmax}_j$ indicates that the softmax operation is taken along the $j$ index. The dot-product attention thus outputs \vspace{-0.5em}
\begin{equation}
    \text{attn}(\bm{\varSigma}_i\mid W_{\!Q},W_{\!K},W_{\!V}) = \sum_{j=1}^L A_{ij} \bm{\mathcal{V}}_j.
    \vspace{-0.5em}
\end{equation}

Intuitively, the attention mechanism can be recognized as a \emph{fuzzy} dictionary lookup. Rather than executing a \emph{hard} retrieval that selects the value associated with a single, exactly matching key, the mechanism instead computes attention scores that quantify the \emph{degree of similarity} or correspondence between the query and all available keys. These scores are then used to form a weighted combination of the associated values, thereby producing a context-dependent output representation.

\textbf{Multi-Head Attention.} --- The above single attention can be extended to \emph{multi-head attention} by partitioning the query, key, and value vectors into $h$ parallel subspaces (heads) with per-head width $d_k$ such that $d_\text{model}=h\cdot d_k$. The dot-product attention is then applied independently within each head. Concretely, the $\eta$-th head ($\eta=1,2,\ldots,h$) computes
\begin{equation}
    \bm{\varSigma}_i^{(\eta)} = \text{attn}(\bm{\varSigma}_i\mid W_{\!Q}^\eta, W_{\!K}^\eta, W_{\!V}^\eta)\,\in\mathbb{R}^{d_v}.
\end{equation}
Note that each head has its own set of projection matrices $W_{\!Q}^\eta$, $W_{\!K}^\eta\!\in\!\mathbb{R}^{d_\text{model}\times d_k}$, and $W_{\!V}^\eta\!\in\!\mathbb{R}^{d_\text{model}\times d_v}$. The outputs of all heads are subsequently concatenated along the feature dimension to form a vector in $\mathbb{R}^{h\cdot d_v}$, before passing through a linear projection with a learnable matrix $W_{\!O}\in\mathbb{R}^{h\cdot d_v\times d_\text{model}}$. The final output of the multi-head attention layer is thus
\begin{equation}
    \bm{\varSigma}_i^{\mkern2mu\raisebox{1pt}{\scriptsize\text{attn}}} = \text{concat}(\bm{\varSigma}_i^{(1)},\ldots,\bm{\varSigma}_i^{(h)}) W_{\!O}\,\in\mathbb{R}^{d_\text{model}}.
\end{equation}
For a single-head attention, the output projection $W_{\!O}$ is redundant, since it can be absorbed into $W_{\!V}$. We therefore omit this projection in the main text.

In practice, a convenient and computationally efficient implementation maintains three \emph{shared} projection matrices $W_{\!Q}$, $W_{\!K}$, and $W_{\!V}$ of shape $d_\text{model}\times d_\text{model}$. The overall queries, keys, and values are computed, and then partitioned into $h$ heads:
\begin{equation}
  \definecolor{pastelred}{RGB}{250, 211, 208}
  \definecolor{pastelorange}{RGB}{247, 220, 204}
  \definecolor{pastelgreen}{RGB}{212, 247, 204}
  \begin{aligned}
    \mathcal{Q} = \left[
    \begin{array}{ccccc}
    \colorbox{pastelred}{\makebox[2.1em]{\raisebox{-1.35em}{\rule{0pt}{3.5em}}$\mathcal{Q}^{(1)}$}} & 
    \colorbox{pastelred}{\makebox[2.1em]{\raisebox{-1.35em}{\rule{0pt}{3.5em}}$\mathcal{Q}^{(2)}$}} & 
    \colorbox{pastelred}{\makebox[2.1em]{\raisebox{-1.35em}{\rule{0pt}{3.5em}}$\mathcal{Q}^{(3)}$}} & 
    \cdots & 
    \colorbox{pastelred}{\makebox[2.1em]{\raisebox{-1.35em}{\rule{0pt}{3.5em}}$\mathcal{Q}^{(h)}$}} \\
    \end{array}
    \right],\\
    \mathcal{K} = \left[
    \begin{array}{ccccc}
    \colorbox{pastelorange}{\makebox[2.1em]{\raisebox{-1.35em}{\rule{0pt}{3.5em}}$\mathcal{K}^{(1)}$}} & 
    \colorbox{pastelorange}{\makebox[2.1em]{\raisebox{-1.35em}{\rule{0pt}{3.5em}}$\mathcal{K}^{(2)}$}} & 
    \colorbox{pastelorange}{\makebox[2.1em]{\raisebox{-1.35em}{\rule{0pt}{3.5em}}$\mathcal{K}^{(3)}$}} & 
    \cdots & 
    \colorbox{pastelorange}{\makebox[2.1em]{\raisebox{-1.35em}{\rule{0pt}{3.5em}}$\mathcal{K}^{(h)}$}} \\
    \end{array}
    \right],\\
    \mathcal{V} = \left[
    \begin{array}{ccccc}
    \colorbox{pastelgreen}{\makebox[2.1em]{\raisebox{-1.35em}{\rule{0pt}{3.5em}}$\mathcal{V}^{(1)}$}} & 
    \colorbox{pastelgreen}{\makebox[2.1em]{\raisebox{-1.35em}{\rule{0pt}{3.5em}}$\mathcal{V}^{(2)}$}} & 
    \colorbox{pastelgreen}{\makebox[2.1em]{\raisebox{-1.35em}{\rule{0pt}{3.5em}}$\mathcal{V}^{(3)}$}} & 
    \cdots &
    \colorbox{pastelgreen}{\makebox[2.1em]{\raisebox{-1.35em}{\rule{0pt}{3.5em}}$\mathcal{V}^{(h)}$}} \\
    \end{array}
    \right].
  \end{aligned}
\end{equation}
Here, the $i$-th row of the matrices $\mathcal{Q}$, $\mathcal{K}$, and $\mathcal{V}\in\mathbb{R}^{L\times d_\text{model}}$ corresponds to the vectors $\bm{\mathcal{Q}}_i$, $\bm{\mathcal{K}}_i$, and $\bm{\mathcal{V}}_i$, respectively. Consequently, the output of the $\eta$-th head becomes
\begin{equation}
  \definecolor{pastelred}{RGB}{250, 211, 208}
  \definecolor{pastelorange}{RGB}{247, 220, 204}
  \definecolor{pastelgreen}{RGB}{212, 247, 204}
  \definecolor{pastelblue}{RGB}{204, 214, 247}
  \colorbox{pastelblue}{\makebox[2.1em]{\raisebox{-1.35em}{\rule{0pt}{3.5em}}$\varSigma^{(\eta)}$}} = \text{softmax}\left[\mkern5mu
    \colorbox{pastelred}{\makebox[2.1em]{\raisebox{-1.35em}{\rule{0pt}{3.5em}}$\mathcal{Q}^{(\eta)}$}}\mkern5mu
    \colorbox{pastelorange}{\makebox[3.5em]{\raisebox{-0.65em}{\rule{0pt}{2.1em}}$[\mathcal{K}^\top]^{(\eta)}$}} / \mathfrak{T}\mkern5mu
  \right]
    \colorbox{pastelgreen}{\makebox[2.1em]{\raisebox{-1.35em}{\rule{0pt}{3.5em}}$\mathcal{V}^{(\eta)}$}},
\end{equation}
where the $\text{softmax}$ is applied row-wise. This vectorized implementation is more efficient in practice, as it leverages highly optimized parallel linear algebra routines. The final output of the multi-head attention block is thus
\begin{equation}
  \definecolor{pastelblue}{RGB}{204, 214, 247}
    \varSigma^{\mkern2mu\raisebox{1pt}{\scriptsize\text{attn}}} = \left[
    \begin{array}{ccccc}
    \colorbox{pastelblue}{\makebox[2.1em]{\raisebox{-1.35em}{\rule{0pt}{3.5em}}$\varSigma^{(1)}$}} & 
    \colorbox{pastelblue}{\makebox[2.1em]{\raisebox{-1.35em}{\rule{0pt}{3.5em}}$\varSigma^{(2)}$}} & 
    \colorbox{pastelblue}{\makebox[2.1em]{\raisebox{-1.35em}{\rule{0pt}{3.5em}}$\varSigma^{(3)}$}} & 
    \cdots &
    \colorbox{pastelblue}{\makebox[2.1em]{\raisebox{-1.35em}{\rule{0pt}{3.5em}}$\varSigma^{(h)}$}} \\
    \end{array}
    \right] W_{\!O}.
\end{equation}

\textbf{Feed-Forward Network.} --- Following the multi-head attention block, a position-wise feed-forward network (FFN) is applied independently to each received $\bm{\varSigma}_i$. Again, the same FFN (i.e., the same parameters) is shared across all positions $i$. Conceptually, the FFN is a three-layer fully-connected perceptron comprising an input layer, a widened hidden layer, and an output layer. The input and output layers have dimension $d_\text{model}$, matching the output of multi-head attention, while the hidden layer has a larger width $d_\text{hidden}$ to enhance the model's representational capacity. A schematic illustration is provided in Fig.~\ref{fig:ffn}.

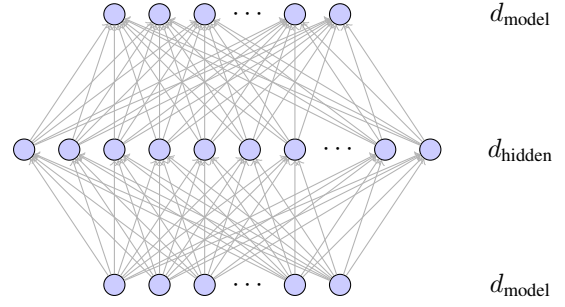
\begin{figure}[h]
\centering
\begin{tikzpicture}[
    neuron/.style={circle, draw, fill=blue!20, minimum size=8pt, inner sep=0pt},
    layer/.style={align=center},
    connection/.style={->, thin, gray!60}
]

\def\layersep{1.8cm}
\def\neuronsep{0.6cm}

\foreach \x in {1,2,3,5,6} {
    \node[neuron] (I-\x) at (\x*\neuronsep, 2*\layersep) {};
}
\node at (4*\neuronsep, 2*\layersep) {$\cdots$};

\foreach \x in {1,2,3,4,5,6,7,9,10} {
    \node[neuron] (H-\x) at (\x*\neuronsep - 1.2cm, \layersep) {};
}
\node at (8*\neuronsep - 1.2cm, \layersep) {$\cdots$};

\foreach \x in {1,2,3,5,6} {
    \node[neuron] (O-\x) at (\x*\neuronsep, 0) {};
}
\node at (4*\neuronsep, 0) {$\cdots$};

\foreach \i in {1,2,3,5,6} {
    \foreach \j in {1,2,3,4,5,6,7,9,10} {
        \draw[connection] (H-\j) -- (I-\i);
    }
}

\foreach \i in {1,2,3,4,5,6,7,9,10} {
    \foreach \j in {1,2,3,5,6} {
        \draw[connection] (O-\j) -- (H-\i);
    }
}

\node[align=left] at (10*\neuronsep, 2*\layersep) {$d_{\text{model}}$};
\node[align=left] at (10*\neuronsep, \layersep) {$d_{\text{hidden}}$};
\node[align=left] at (10*\neuronsep, 0) {$d_{\text{model}}$};

\end{tikzpicture}
\caption{Schematic diagram of a three-layer feed-forward network (FFN) used in the transformer architecture. The input and output layers have dimension $d_{\text{model}}$, while the hidden layer has dimension $d_{\text{hidden}}$. Each neuron in a given layer is connected to all neurons in the adjacent layers.}
\label{fig:ffn}
\end{figure}

Specifically, the FFN applies the following transformation to each position $i$ in the sequence:
\begin{equation}
    \text{FFN}(\bm{\varSigma}_i) = \text{ReLU}(\bm{\varSigma}_i W_{\mkern-2mu1} + \bm{b}_1) W_{\mkern-2mu2} + \bm{b}_2,
\end{equation}
where $W_{\mkern-2mu1} \in \mathbb{R}^{d_{\text{model}} \times d_{\text{hidden}}}$ and $W_{\mkern-2mu2} \in \mathbb{R}^{d_{\text{hidden}} \times d_{\text{model}}}$ are learnable projection matrices, and $\bm{b}_1 \in \mathbb{R}^{d_\text{hidden}}$, $\bm{b}_2 \in \mathbb{R}^{d_\text{model}}$ (learnable) bias vectors. The ReLU (\emph{Rectified Linear Unit}) activation function, defined as $\text{ReLU}(x) = \max(0, x)$, supplies the crucial non-linearity in the FFN. Figure~\ref{fig:relu} depicts the overall shape of the ReLU function.

\begin{figure}[h!]
\centering
\begin{tikzpicture}[scale=1.2]
    \coordinate (origin) at (0,0);
    
    \definecolor{seabornred}{RGB}{196, 34, 34}
    
    \draw[thick, ->] (-3,0) -- (3,0) node[right] {$x$};
    \draw[thick, ->] (0,-0.5) -- (0,3) node[above] {$\text{ReLU}(x)$};
    
    \draw[gray!30, thin] (-3,-0.5) grid[step=0.5] (3,3);
    
    \draw[very thick, seabornred] (-3,0) -- (0,0);  
    \draw[very thick, seabornred] (0,0) -- (2.5,2.5);  
    
    \fill[seabornred] (0,0) circle (2pt);
    
    \foreach \x in {-2,-1,1,2} {
        \draw (\x,0.05) -- (\x,-0.05) node[below] {$\x$};
    }
    \foreach \y in {1,2} {
        \draw (0.05,\y) -- (-0.05,\y) node[left] {$\y$};
    }
    
    \node[seabornred, above right] at (0.7,0.3) {$y = x$ for $x > 0$};
    \node[seabornred, above] at (-1.5,0.3) {$y = 0$ for $x \leqslant 0$};
    
    
\end{tikzpicture}
\caption{Schematic diagram of the ReLU (Rectified Linear Unit) activation function. The function zeros out the negative inputs and increases linearly for positive inputs, providing essential non-linearity while maintaining computational simplicity.}
\label{fig:relu}
\end{figure}
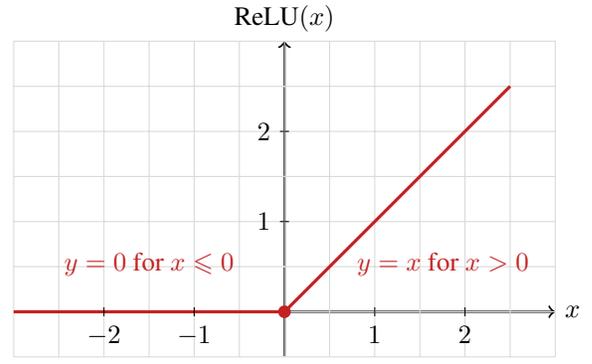

\begin{figure*}[htp!]
\centering
\vspace{2em}
\includegraphics[scale=0.81]{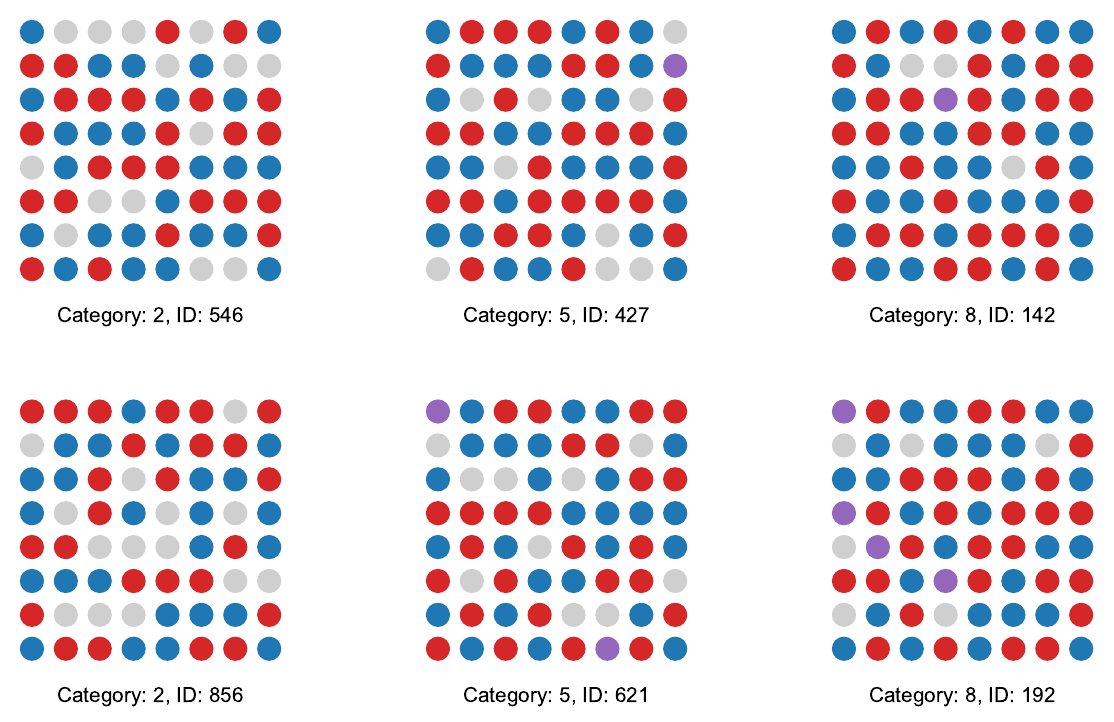}
\begin{minipage}{0.9\linewidth}
    \caption{Example snapshots from the 9-category XTRG dataset at the lowest temperature for the minimal Hubbard model on an 8×8 lattice. Each snapshot is represented as a 2D grid, where each site is color-coded according to its local state: empty (grey), spin-up (red), spin-down (blue), and doubly occupied (purple).}
    \label{fig:hubbard_snapshots}
\end{minipage}
\end{figure*}

\textbf{Output Projection.} --- Following the final FFN of the transformer, vectors $\bm{\varSigma}_i$ are passed to an output projection module responsible for producing the model's logits. This output projection is usually implemented as a learnable linear transformation into the output space $\mathcal{G}$ of the downstream task (e.g. categories for classification or vocabulary in text generation). For each position $i$, the output logits $y_i\in\mathbb{R}^{|\mathcal{G}|}$ are computed as
\begin{equation}
    \bm{y}_i = \bm{\varSigma}_i W_{\!y} + \bm{b}_{y},
\end{equation}
where $W_{\!y}\in\mathbb{R}^{d_\text{model}\times |\mathcal{G}|}$ denotes the weight matrix and $\bm{b}_{y}\in\mathbb{R}^{|\mathcal{G}|}$ the bias. The logits are subsequently normalized via a softmax function to yield probabilities $\bm{p}_i \!=\! \text{softmax}(\bm{y}_i) \!\in\!\mathbb{R}^{|\mathcal{G}|}$. In generative scenarios, $\bm{p}_i$ guides the sampling of the output token at position $i$. For classification or regression tasks, one may aggregate the outputs across all positions (e.g., via averaging $\bm{p} = \text{avg}_i(\bm{p}_i)$) or employ a dedicated $\texttt{CLS}$ token \cite{Devlin&Toutanova2019,Khan&Shah2022} to derive a single, sequence-level prediction.

\textbf{Residual Connection.} --- Deep neural networks are prone to vanishing gradients during training; the \emph{residual connection} \cite{He&Sun2016} is a standard remedy that markedly stabilizes the optimization process. The key idea is to introduce a shortcut path that bypasses the block and adds the input directly to the block's output. Concretely, for an input vector $\bm{\varSigma}_i$, the output of a block with transformation function $\mathcal{F}(\bm{\varSigma}_i)$ is modified to
\begin{equation}
    \bm{\varSigma}_{i} \mkern2mu\leftarrow\mkern1mu \bm{\varSigma}_i + \mathcal{F}(\bm{\varSigma}_i).
\end{equation}
In this formulation, even if the gradient through $\mathcal{F}$ becomes vanishingly small, the identity pathway preserves well-conditioned gradient flow, thereby facilitating effective backpropagation. In our architectural depiction (namely Fig.~\ref{fig:Architecture}) in the main text, the \texttt{Residual} block and the \texttt{Add} in the \texttt{Add \& Norm} block both represent this residual connection.

\textbf{Layer Normalization.} --- \emph{Layer normalization} (LayerNorm) \cite{Ba&Hinton2016} further stabilizes and accelerates training by normalizing the magnitude across the feature dimension per position. For an input vector $\bm{\varSigma}_i\in\mathbb{R}^{d_\text{model}}$, LayerNorm computes
\begin{equation}
    \text{LayerNorm}(\bm{\varSigma}_i) = a\cdot \frac{\bm{\varSigma}_i - \text{mean}(\bm{\varSigma}_i)}{\text{std}(\bm{\varSigma}_i) + \epsilon} + b,
\end{equation}
where $\text{mean}(\bm{\varSigma}_i)$ and $\text{std}(\bm{\varSigma}_i)$ denote the mean and standard deviation of the elements of $\bm{\varSigma}_i$, respectively; $a$ and $b$ are learnable scale and shift parameters; and $\epsilon=10^{-6}$ is a small constant that prevents division by zero. In our architectural diagram in the main text, the \texttt{Norm} in the \texttt{Add \& Norm} block corresponds to this LayerNorm operation.

\section{Specifications of the Snapshot Dataset}
\label{sec:dataset}
\vspace{-0.1em}

Here, we report the temperature $T$, charge doping $\delta$, and double occupancy $n_{\uparrow\downarrow}$ for the nine categories in the XTRG snapshot dataset of the minimal Hubbard model on an 8×8 lattice, as summarized in Table~\ref{tab:dataset}. Several representative snapshots from each doping level at the lowest temperature are shown in Fig.~\ref{fig:hubbard_snapshots}.

\begin{table}[h!]
\centering
\renewcommand{\arraystretch}{1.2}
\begin{tabular}{c|ccc}
\noalign{\hrule height 1pt}
\rule{0pt}{1.2\normalbaselineskip}
~\textbf{Categories}~~ & \hspace{1.5em}$T$\hspace{1.2em} & \hspace{1.5em}$\delta$\hspace{1.6em} & \hspace{1.2em}$n_{\uparrow\downarrow}$\hspace{1em}
\\[\dimexpr0.2\normalbaselineskip]
\hline
\rule{0pt}{1.2\normalbaselineskip}\hspace{-3pt}
Cat 0 & 1/4   & 0.2041 & 0.0123 \\
Cat 1 & 1/16  & 0.2190 & 0.0151 \\
Cat 2 & 1/256 & 0.2188 & 0.0156 \\
Cat 3 & 1/4   & 0.1324 & 0.0149 \\
Cat 4 & 1/16  & 0.1227 & 0.0186 \\
Cat 5 & 1/256 & 0.1250 & 0.0182 \\
Cat 6 & 1/4   & 0.0732 & 0.0179 \\
Cat 7 & 1/16  & 0.0413 & 0.0228 \\
Cat 8 & 1/256 & 0.0312 & 0.0220 \\[\dimexpr0.2\normalbaselineskip]
\noalign{\hrule height 1pt}
\end{tabular}
\begin{minipage}{0.9\linewidth}
\caption{Temperature $T$, charge doping $\delta$, and double occupancy $n_{\uparrow\downarrow}$ of the nine categories in the XTRG snapshot dataset for the minimal Hubbard model.}
\label{tab:dataset}
\end{minipage}
\end{table}

\begin{figure*}[htp!]
    \vspace{0.8em}
    \centering
    \includegraphics[width=0.99\textwidth]{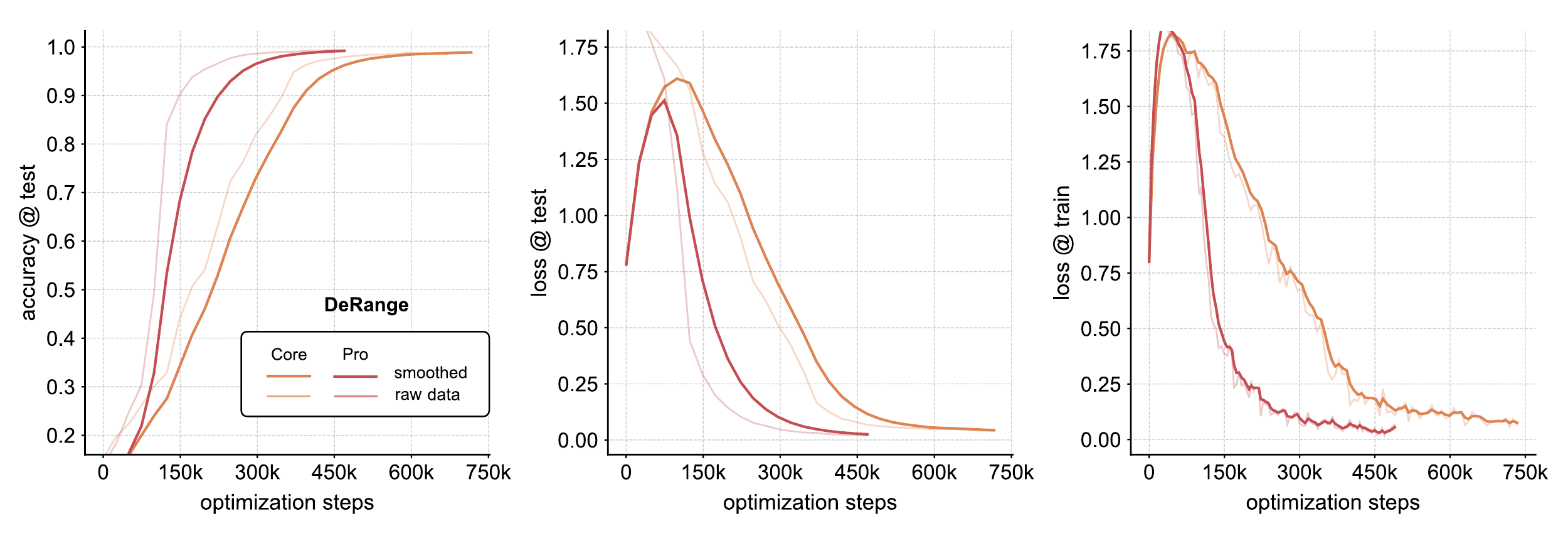}
    \begin{minipage}{0.9\textwidth}
    \caption{Training profiles of the core and pro models for the derangement dataset. Metrics are displayed as raw data (thin lines with muted color) and with an exponential smoothing factor $\alpha\!=\!0.4$ (thick lines with deep color). Both models achieve near-perfect training accuracy, demonstrating their capability to capture the imposed correlation structures, although the core model requires more optimization steps to converge. This indicates the efficacy of the semi-linear attention mechanism for modeling correlations in snapshot-type datasets.}
    \label{fig:performance}
    \end{minipage}
    \vspace{0.3em}
\end{figure*}

\section{Benchmark on Derangements}
\label{sec:derangements}

In this section, we benchmark our pro and core architectures on a synthetic dataset of derangements to demonstrate their ability to capture latent correlation structure. A derangement is a permutation in which no element remains in its original position. For example, given \texttt{[1, 2, 3]}, the derangements are \texttt{[2, 3, 1]} and \texttt{[3, 1, 2]}. We construct artificial 8×8 snapshots in which the left half (left four columns) is generated uniformly at random, while the right half (right four columns) is obtained by applying a (column-wise) derangement to the left half.

\begin{table}[hp!]
\centering
\renewcommand{\arraystretch}{1.2}
\begin{tabular}{c|ccc}
\noalign{\hrule height 1pt}
\rule{0pt}{1.2\normalbaselineskip}
~\textbf{Categories}~~ & \hspace{1.5em} \textbf{Derangements} \hspace{1.5em}
\\[\dimexpr0.2\normalbaselineskip]
\hline
\rule{0pt}{1.2\normalbaselineskip}\hspace{-3pt}
Cat 0 & Random \\
Cat 1 & \texttt{[1,0,3,2]} \\
Cat 2 & \texttt{[1,3,0,2]} \\
Cat 3 & \texttt{[2,0,3,1]} \\
Cat 4 & \texttt{[2,3,0,1]} \\
Cat 5 & \texttt{[2,3,1,0]} \\
Cat 6 & \texttt{[3,2,0,1]} \\[\dimexpr0.2\normalbaselineskip]
\noalign{\hrule height 1pt}
\end{tabular}
\begin{minipage}{0.9\linewidth}
\vspace{0.4em}
\caption{Derangements for the seven categories in the synthetic dataset. Numbers in the derangements denote the columns (not to be mistaken with tokens or local states). Category 0 contains snapshots generated completely randomly as a comparison baseline.}
\label{tab:derange_dataset}
\end{minipage}
\end{table}

Table~\ref{tab:derange_dataset} enumerates the six derangements used in our dataset, together with category~0 comprising fully random snapshots as a comparative baseline. Each category contains 10{,}000 snapshots of size 8×8. Representative examples are shown in the upper panel of Fig.~\ref{fig:derange_cat1}-(1-6). The derangement defining each category is indicated above the panel, and corresponding columns --- i.e., columns that are identical by construction --- are highlighted with matching background colors. In these synthetic snapshots, there is 100\% correlation between corresponding columns across the left and right halves, and no correlations otherwise. The objective is to assess whether a trained model can assign snapshots to the correct category purely from these correlation patterns, i.e. whether it can correctly identify the permutation used to generate the right four columns from the left four random ones.

We train both the core and pro architectures on the derangement dataset without a locality bias; training profiles are summarized in Fig.~\ref{fig:performance}. Both models attain near-perfect training accuracy, indicating successful identification of the imposed correlations, although the core model requires more optimization steps to converge. This observation further supports the efficacy of the semi-linear attention mechanism for modeling correlations in snapshot-type datasets.

Next, we examine the attention maps produced by the core model. Figs.~\ref{fig:derange_cat1}-(1-6) display, for each derangement category, a snapshot together with the attention scores from the first attention layer. These visualizations reveal where the model \emph{looks} when processing each position of the input. Each panel comprises an 8×8 array of subplots, each showing an 8×8 grid of cells. Within each subplot, the attention scores $\mathcal{A}_{ij}$ are encoded by a color scale; the query position $i$ coincides with the subplot's location and is indicated by a red circle.

\begin{figure*}[htp!]
\vspace{-0.5em}
\centering
\subfloat{\includegraphics[scale=0.5]{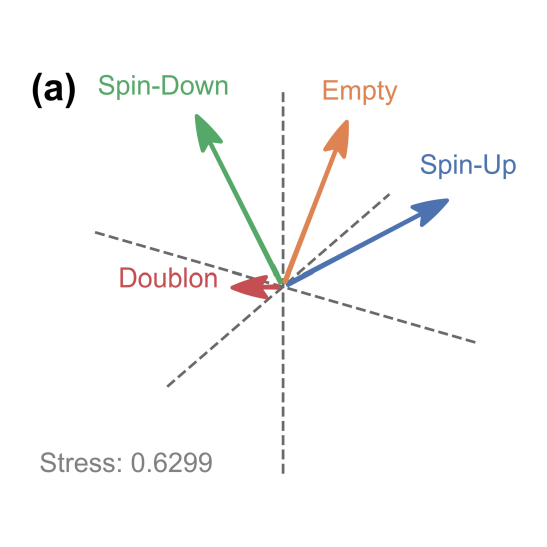}}
\subfloat{\includegraphics[scale=0.5]{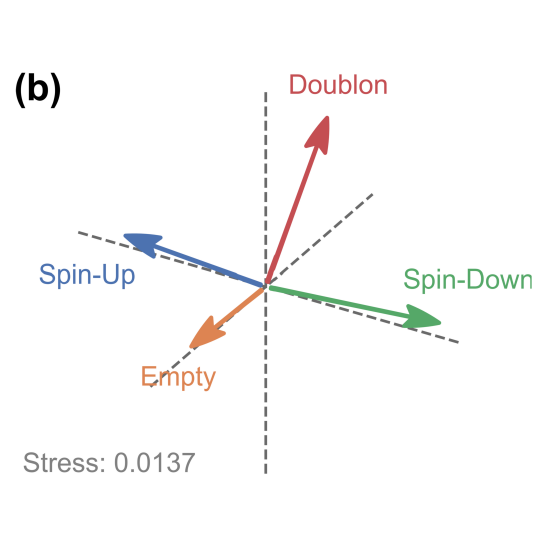}}
\subfloat{\includegraphics[scale=0.5]{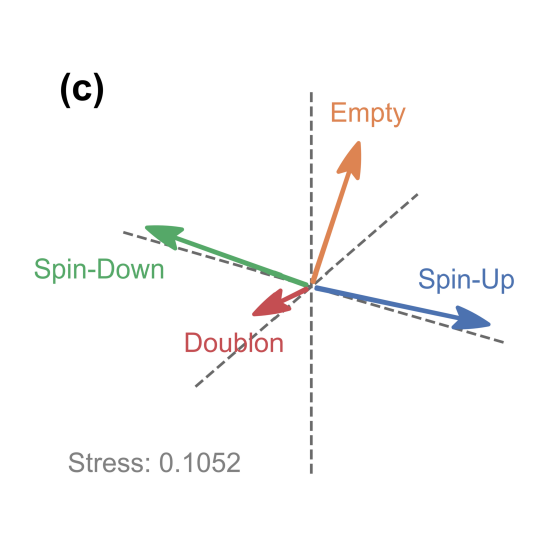}}
\subfloat{\includegraphics[scale=0.5]{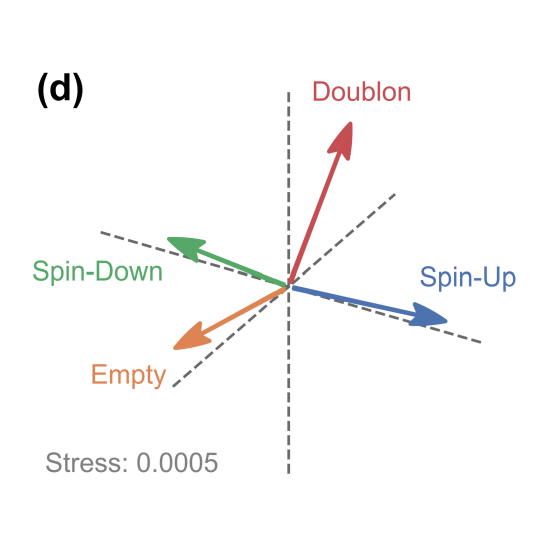}}\\
\begin{minipage}{0.9\linewidth}
    \caption{Orthogonality relationships between the average embeddings under (a) mode 0 (no projection), (b) mode 1, (c) mode 2, and (d) mode 3 projection of the four local states (empty, spin-up, spin-down, and doubly occupied) from the core model for the XTRG Hubbard dataset. The angles between the original embeddings are close to 90°, a clear indication of the distinct local states; after the projections, the spin-up and spin-down states become nearly opposite in direction, reflecting the $\mathrm{SU}(2)$ rotational spin symmetry underlying the snapshots.}
    \label{fig:orthogonality}
\end{minipage}
\end{figure*}

In general, we observe elevated attention at the same row of the query, and in most maps the dominant attention is devoted to the corresponding site (with an identical state by construction) in the opposite half of the snapshot, e.g. the attention maps at the 5th and 6th row, 1st column of Fig.~\ref{fig:derange_cat1}-1 highlight the sites at the same row, 6th column. This behavior indicates that the model has internalized the per-row permutation structure and largely identified the strong cross-half correlations. The maps are not perfectly pristine --- some spurious attention persists, e.g. the 2nd and 3rd rows in Fig.~\ref{fig:derange_cat1}-1 --- likely attributable to the correlations being sufficiently strong that high accuracy is achievable without completely disentangling all dependencies. Nevertheless, the attention visualizations collectively corroborate the model's capacity to recover the underlying correlation structure.

\section{Orthogonality and Attention Maps}
\label{sec:attention_maps}

Orthogonality relations (angles) among the vectors $\bm{\varSigma}_i$ that represent tokens (local states) elucidate how the model internally encodes and discriminates between these states. Since the vectors $\bm{\varSigma}_i$ also depend on the position $i$, we define the \emph{average embedding} $\bm{\varSigma}[\sigma]$ of token $\sigma$ as
\begin{equation}
    \bm{\varSigma}[\sigma] = \frac{1}{N_\sigma} \sum_{i \in \mathcal{I}_\sigma,\ \text{all}\, x} \bm{\varSigma}_i(x),
\end{equation}
where $x$ denotes an input snapshot, $\bm{\varSigma}_i(x)$ the corresponding input embeddings, $\mathcal{I}_\sigma = \{i\mid \sigma_i = \sigma\}$ the set of positions at which token $\sigma$ occurs, and $N_\sigma = |\mathcal{I}_\sigma|$ its multiplicity in the entire ensemble. The average embedding $\bm{\varSigma}[\sigma]$ is therefore the mean embedding vector of token $\sigma$ aggregated over all of its occurrences in the dataset.

Also, for different propagation modes $m$ in the attention stack, the input embeddings $\bm{\varSigma}_i$ are further projected according to mode-specific linear transformations $W_{\!\mathcal{V}}$ (see main text). For our core model with two attention layers (blocks), we identify four projectors $W^{(m)}$: $W^{(m=0)} = I$ (identity, no projection), $W^{(m=1)} = W^{(\ell=1)}_{\!\mathcal{V}}$ (layer 1 projection), $W^{(m=2)} = W^{(\ell=2)}_{\!\mathcal{V}}$ (layer 2 projection), and $W^{(m=3)} = W^{(\ell=1)}_{\!\mathcal{V}} W^{(\ell=2)}_{\!\mathcal{V}}$ (layer 1 \& 2 projection). We then define the projected average embedding of token $\sigma$ under mode $m$ as
\begin{equation}
    \bm{\varSigma}_m[\sigma] = \bm{\varSigma}[\sigma] W^{(m)}.
\end{equation}
In this regard, we can calculate the \emph{overlap} (i.e the normalized inner product, and thus the cosine of their angle $\theta_{\sigma\sigma'}^{(m)}$) between the average embeddings of any pair of tokens $\sigma$ and $\sigma'$ for propagation mode $m$ as
\begin{equation}
    \cos\langle \bm{\varSigma}_m[\sigma], \bm{\varSigma}_m[\sigma'] \rangle = \frac{\bm{\varSigma}_m[\sigma] \cdot \bm{\varSigma}_m[\sigma']}{\|\bm{\varSigma}_m[\sigma]\| \cdot \|\bm{\varSigma}_m[\sigma']\|}.
\end{equation}
Detailed numerics are summarized in Table~\ref{tab:orthogonality}.

To visualize these orthogonality relations for each mode $m$, we embed the four vectors $\bm{\varSigma}_m[\sigma]$, $\sigma=\texttt{0},\texttt{1},\texttt{2},\texttt{3}$, into three-dimensional space via principal component analysis. Concretely, we construct the Gram matrix
\begin{equation}
    G^{(m)}_{\sigma\sigma} = 1, \quad G^{(m)}_{\sigma\sigma'} = \cos\theta_{\sigma\sigma'}^{(m)} \quad\text{for}\quad \sigma \neq \sigma'
\end{equation}
and perform an eigen-decomposition $G^{(m)} = U \Lambda U^\top = X X^\top$ with $X = U \sqrt{\Lambda}$. The rows $\chi_r$ of $X$ provide the coordinates of the visualization vectors. When $\mathrm{rank}\,G^{(m)}$ exceeds three, we retain only the three largest eigenvalues and corresponding eigenvectors to obtain a three-dimensional approximation. The quality of this approximation is quantified by the \emph{stress} metric
\begin{equation}
    \text{stress} = \sqrt{\frac{\sum_{r<s} (G^{(m)}_{rs} - \chi_r^\top \chi_s)^2}{\sum_{r<s} (G^{(m)}_{rs})^2}}.
\end{equation}

Figure~\ref{fig:orthogonality}(a,b,c,d) visualizes the orthogonality relations for modes $m=0,1,2,3$, respectively. The angles between the original embeddings (mode~0) cluster near 90°, indicating that the model has learned to represent the four local states as nearly orthogonal vectors. The associated stress is large, consistent with the impossibility of embedding four almost mutually orthogonal vectors exactly in three dimensions.

\begin{table}[h!]
\centering
\renewcommand{\arraystretch}{1.2}
\vspace{2em}

\begin{tabular}{@{\hspace{0.5em}}c@{\hspace{1em}}|@{\hspace{1em}}c@{\hspace{1em}}c@{\hspace{1em}}}
\noalign{\hrule height 1pt}
\rule{0pt}{1.2\normalbaselineskip}
\textbf{Token Pairs} & \textbf{Overlap} & \textbf{Angle (°)}
\\[\dimexpr0.2\normalbaselineskip]
\hline
\rule{0pt}{1.2\normalbaselineskip}\hspace{-3pt}
(\texttt{0}, \texttt{1}) & 0.3121 & 71.82° \\
(\texttt{0}, \texttt{2}) & 0.2937 & 72.92° \\
(\texttt{0}, \texttt{3}) & 0.1858 & 79.29° \\
(\texttt{1}, \texttt{2}) & 0.1460 & 81.61° \\
(\texttt{1}, \texttt{3}) & 0.06737 & 86.14° \\
(\texttt{2}, \texttt{3}) & 0.1946 & 78.78° \\[\dimexpr0.2\normalbaselineskip]
\noalign{\hrule height 1pt}
\end{tabular}

\vspace{0.5em}
(a) mode 0: original average embeddings

\vspace{2em}

\begin{tabular}{@{\hspace{0.5em}}c@{\hspace{1em}}|@{\hspace{1em}}c@{\hspace{1em}}c@{\hspace{1em}}}
\noalign{\hrule height 1pt}
\rule{0pt}{1.2\normalbaselineskip}
\textbf{Token Pairs} & \textbf{Overlap} & \textbf{Angle (°)}
\\[\dimexpr0.2\normalbaselineskip]
\hline
\rule{0pt}{1.2\normalbaselineskip}\hspace{-3pt}
(\texttt{0}, \texttt{1}) & -0.2111 & 102.2° \\
(\texttt{0}, \texttt{2}) & -0.1068 & 96.13° \\
(\texttt{0}, \texttt{3}) & -0.5090 & 120.6° \\
(\texttt{1}, \texttt{2}) & -0.9195 & 156.9° \\
(\texttt{1}, \texttt{3}) & 0.04032 & 87.69° \\
(\texttt{2}, \texttt{3}) & 0.1173 & 83.26° \\[\dimexpr0.2\normalbaselineskip]
\noalign{\hrule height 1pt}
\end{tabular}

\vspace{0.5em}
(b) mode 1 projected average embeddings

\vspace{2em}

\begin{tabular}{@{\hspace{0.5em}}c@{\hspace{1em}}|@{\hspace{1em}}c@{\hspace{1em}}c@{\hspace{1em}}}
\noalign{\hrule height 1pt}
\rule{0pt}{1.2\normalbaselineskip}
\textbf{Token Pairs} & \textbf{Overlap} & \textbf{Angle (°)}
\\[\dimexpr0.2\normalbaselineskip]
\hline
\rule{0pt}{1.2\normalbaselineskip}\hspace{-3pt}
(\texttt{0}, \texttt{1}) & 0.1127 & 83.53° \\
(\texttt{0}, \texttt{2}) & 0.1275 & 82.67° \\
(\texttt{0}, \texttt{3}) & 0.2651 & 74.63° \\
(\texttt{1}, \texttt{2}) & -0.7540 & 138.9° \\
(\texttt{1}, \texttt{3}) & -0.2313 & 103.4° \\
(\texttt{2}, \texttt{3}) & 0.4466 & 63.48° \\[\dimexpr0.2\normalbaselineskip]
\noalign{\hrule height 1pt}
\end{tabular}

\vspace{0.5em}
(c) mode 2 projected average embeddings

\vspace{2em}

\begin{tabular}{@{\hspace{0.5em}}c@{\hspace{1em}}|@{\hspace{1em}}c@{\hspace{1em}}c@{\hspace{1em}}}
\noalign{\hrule height 1pt}
\rule{0pt}{1.2\normalbaselineskip}
\textbf{Token Pairs} & \textbf{Overlap} & \textbf{Angle (°)}
\\[\dimexpr0.2\normalbaselineskip]
\hline
\rule{0pt}{1.2\normalbaselineskip}\hspace{-3pt}
(\texttt{0}, \texttt{1}) & -0.1238 & 97.11° \\
(\texttt{0}, \texttt{2}) & -0.1439 & 98.27° \\
(\texttt{0}, \texttt{3}) & -0.6658 & 131.7° \\
(\texttt{1}, \texttt{2}) & -0.9630 & 164.4° \\
(\texttt{1}, \texttt{3}) & 0.02786 & 88.40° \\
(\texttt{2}, \texttt{3}) & 0.1491 & 81.43° \\[\dimexpr0.2\normalbaselineskip]
\noalign{\hrule height 1pt}
\end{tabular}

\vspace{0.5em}
(d) mode 3 projected average embeddings

\vspace{1em}
\caption{Average inner product for each token-pair and corresponding angles (in degrees) for (a) mode 0: original embeddings (no projection), (b) mode 1, (c) mode 2, and (d) mode 3 projected average embeddings. All values are rounded to 4 significant digits.}
\label{tab:orthogonality}
\end{table}

After projection (i.e. modes~1, 2, and 3), the spin-up and spin-down embeddings become nearly antipodal, reflecting the underlying $\mathrm{SU}(2)$ spin-rotational symmetry of the snapshots. The low stress corroborates this symmetry-induced constraint, which effectively removes one independent basis state from the local Hilbert space. Moreover, the spinful states are broadly orthogonal to the plane spanned by the empty and doubly occupied states, capturing the distinction between sectors of different total spin.

These orthogonality relations collectively substantiate that the model faithfully captures the physical significance of the basis states in the local Hilbert space.

Beyond orthogonality, attention maps offer complementary insight into the model’s processing of snapshots. Figures~\ref{fig:map_2856}-(1-3) present, for three representative snapshots from the XTRG Hubbard dataset, the attention rollout \cite{Abnar&Zuidema2020} of the core model. As before, each panel comprises an 8×8 array of subplots, each showing an 8×8 grid. Within each subplot, the rolled-out attention (with the identity component subtracted) $\mathcal{R}_{ij}-I_{ij}$ is encoded by a color scale; the query position $i$ coincides with the subplot location and is marked with a red circle.

In contrast to the derangement benchmark, the attention maps for the Hubbard snapshots are substantially more challenging to comprehend. This is expected: correlations in the Hubbard data are far more intricate and less deterministic. A salient feature is produced by the locality bias, whereby attention concentrates on nearby sites. Another notable characteristic is that --- unlike the derangement benchmark where attention typically condenses onto a few positions --- the attention scores for the Hubbard snapshots are markedly more diffuse. This suggests that the Hubbard correlations can be high-order and spatially extended. Additional structures likely exist and remain to be elucidated through more refined analytical methods.

\begin{table*}[h!]
\centering
\renewcommand{\arraystretch}{1.2}
\vspace{2em}

\begin{minipage}{0.3\textwidth}
(a) $\mu = 1.2$,~~$\delta\approx22\%$
\vspace{0.3em}

\begin{tabular}{@{\hspace{1em}}c@{\hspace{1em}}|@{\hspace{1em}}c@{\hspace{1em}}c@{\hspace{1em}}}
\noalign{\hrule height 1pt}
\rule{0pt}{1.2\normalbaselineskip}
$n_T$ & $\varGamma^{zz}$ & $\text{std}(\varGamma^{zz})$
\\[\dimexpr0.2\normalbaselineskip]
\hline
\rule{0pt}{1.2\normalbaselineskip}\hspace{-3pt}
0.0 & -0.01081 & 0.02386 \\
2.0 & -0.03751 & 0.02319 \\
4.0 & -0.06626 & 0.02012 \\
6.0 & -0.07412 & 0.01853 \\
8.0 & -0.07559 & 0.01867 \\[\dimexpr0.2\normalbaselineskip]
\noalign{\hrule height 1pt}
\end{tabular}
\end{minipage}
\hspace{0.7em}
\begin{minipage}{0.3\textwidth}
(b) $\mu = 1.4$,~~$\delta\approx17\%$
\vspace{0.3em}

\begin{tabular}{@{\hspace{1em}}c@{\hspace{1em}}|@{\hspace{1em}}c@{\hspace{1em}}c@{\hspace{1em}}}
\noalign{\hrule height 1pt}
\rule{0pt}{1.2\normalbaselineskip}
$n_T$ & $\varGamma^{zz}$ & $\text{std}(\varGamma^{zz})$
\\[\dimexpr0.2\normalbaselineskip]
\hline
\rule{0pt}{1.2\normalbaselineskip}\hspace{-3pt}
0.0 & -0.01126 & 0.02494 \\
2.0 & -0.04096 & 0.02541 \\
4.0 & -0.07556 & 0.02136 \\
6.0 & -0.08089 & 0.02037 \\
8.0 & -0.08156 & 0.02026 \\[\dimexpr0.2\normalbaselineskip]
\noalign{\hrule height 1pt}
\end{tabular}
\end{minipage}
\hspace*{0.7em}
\begin{minipage}{0.3\textwidth}
(c) $\mu = 1.6$,~~$\delta\approx12\%$
\vspace{0.3em}

\begin{tabular}{@{\hspace{1em}}c@{\hspace{1em}}|@{\hspace{1em}}c@{\hspace{1em}}c@{\hspace{1em}}}
\noalign{\hrule height 1pt}
\rule{0pt}{1.2\normalbaselineskip}
$n_T$ & $\varGamma^{zz}$ & $\text{std}(\varGamma^{zz})$
\\[\dimexpr0.2\normalbaselineskip]
\hline
\rule{0pt}{1.2\normalbaselineskip}\hspace{-3pt}
0.0 & -0.01281 & 0.02607 \\
2.0 & -0.04639 & 0.02623 \\
4.0 & -0.08709 & 0.02200 \\
6.0 & -0.09164 & 0.02141 \\
8.0 & -0.09061 & 0.02056 \\[\dimexpr0.2\normalbaselineskip]
\noalign{\hrule height 1pt}
\end{tabular}
\end{minipage}

\vspace{2em}

\begin{minipage}{0.3\textwidth}
(d) $\mu = 1.8$,~~$\delta\approx7\%$
\vspace{0.3em}

\begin{tabular}{@{\hspace{1em}}c@{\hspace{1em}}|@{\hspace{1em}}c@{\hspace{1em}}c@{\hspace{1em}}}
\noalign{\hrule height 1pt}
\rule{0pt}{1.2\normalbaselineskip}
$n_T$ & $\varGamma^{zz}$ & $\text{std}(\varGamma^{zz})$
\\[\dimexpr0.2\normalbaselineskip]
\hline
\rule{0pt}{1.2\normalbaselineskip}\hspace{-3pt}
0.0 & -0.01377 & 0.02611 \\
2.0 & -0.05229 & 0.02630 \\
4.0 & -0.09902 & 0.02453 \\
6.0 & -0.1053 & 0.02413 \\
8.0 & -0.1052 & 0.02321 \\[\dimexpr0.2\normalbaselineskip]
\noalign{\hrule height 1pt}
\end{tabular}
\end{minipage}
\hspace{0.7em}
\begin{minipage}{0.3\textwidth}
(e) $\mu = 2.0$,~~$\delta\approx3\%$
\vspace{0.3em}

\begin{tabular}{@{\hspace{1em}}c@{\hspace{1em}}|@{\hspace{1em}}c@{\hspace{1em}}c@{\hspace{1em}}}
\noalign{\hrule height 1pt}
\rule{0pt}{1.2\normalbaselineskip}
$n_T$ & $\varGamma^{zz}$ & $\text{std}(\varGamma^{zz})$
\\[\dimexpr0.2\normalbaselineskip]
\hline
\rule{0pt}{1.2\normalbaselineskip}\hspace{-3pt}
0.0 & -0.01381 & 0.02711 \\
2.0 & -0.05450 & 0.02621 \\
4.0 & -0.1078 & 0.02696 \\
6.0 & -0.1169 & 0.02705 \\
8.0 & -0.1192 & 0.02641 \\[\dimexpr0.2\normalbaselineskip]
\noalign{\hrule height 1pt}
\end{tabular}
\end{minipage}

\vspace{1.2em}

\begin{minipage}{0.9\textwidth}
\caption{The reference spin-$z$ correlations $\varGamma^{zz}$ and the corresponding standard deviations, $\text{std}(\varGamma^{zz})$, for different chemical potentials $\mu$ (and thus doping $\delta$) across thermal exponents $n_T$. All correlation values and standard deviations are computed over all snapshots in each category, and rounded to 4 significant digits.}
\label{tab:spin_correlations}
\end{minipage}
\end{table*}

\section{Preview of a 25-category Omnimeter}
\label{sec:omnimeter}

\begin{figure*}[htp!]
    \centering
    \subfloat{\includegraphics[scale=0.52]{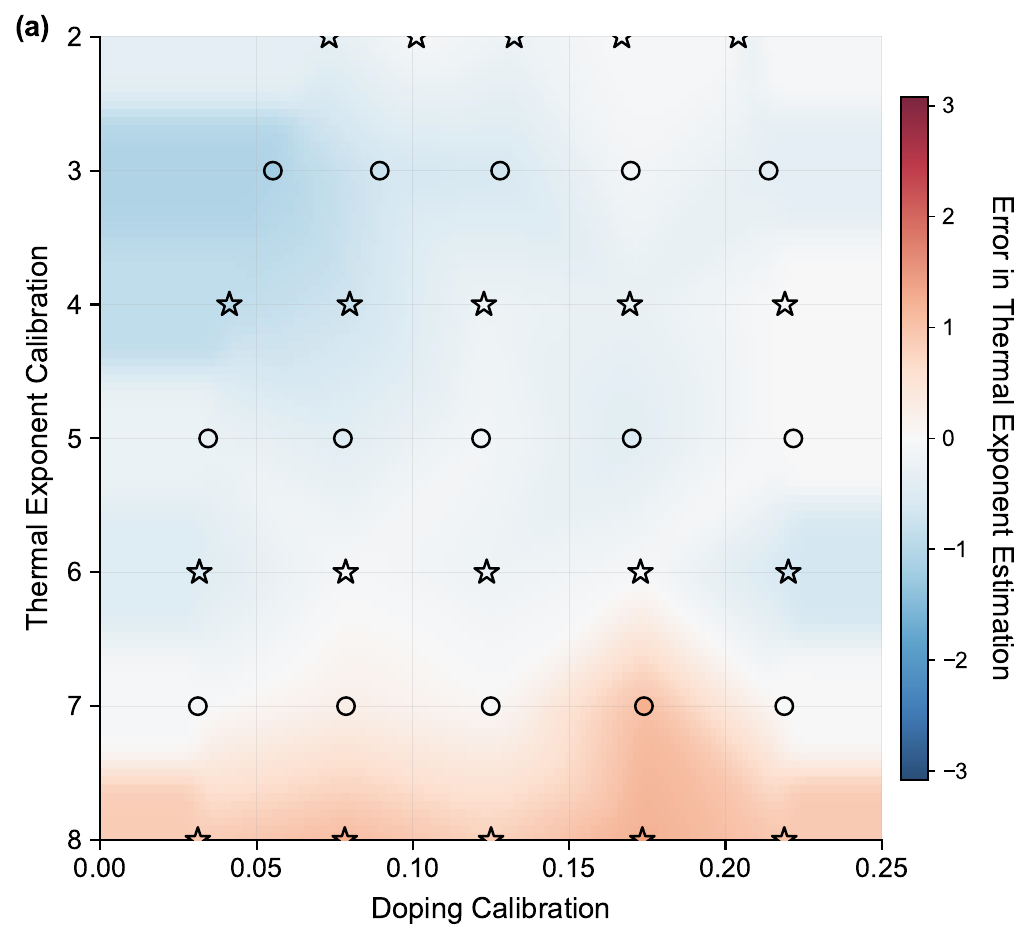}}
    \hspace{0.012\textwidth}
    \subfloat{\includegraphics[scale=0.52]{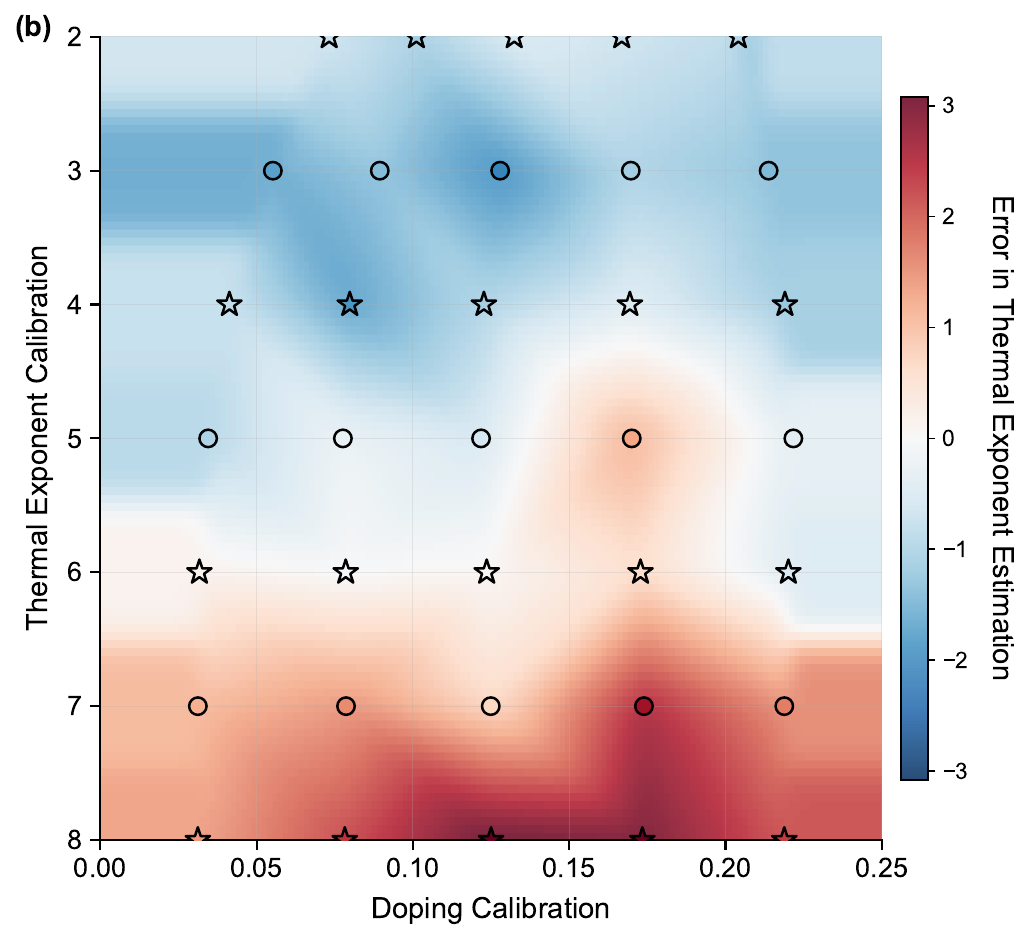}}\\[-0.1em]
    \hspace{0.4em}
    \begin{minipage}{0.9\textwidth}
    \caption{Error of the thermal exponent estimate $n_T = -\log_2 T$ for (a) the 25-category omnimeter and (b) the spin-correlation-based thermometer, evaluated on a random 30-snapshot ensemble. Open circles/pentagrams mark the locations in phase space of the snapshot ensembles under evaluation, with pentagrams (circles) indicating data included (not included) in the training set. Color scales are obtained via interpolation and made equal in both panels for a direct comparison. The omnimeter consistently outperforms the spin-correlation thermometer in most regions of phase space, especially at lower temperatures where correlations saturate.}
    \label{fig:eval_thermometry}
    \end{minipage}
    \vspace{0.4em}
\end{figure*}

In this section, we present a technical preview of a 25-category \emph{omnimeter} for the Hubbard model and demonstrate its advantage over the contemporary spin-correlation-based thermometer \cite{Chalopin&Bloch2024,Xu&Greiner2025}. A comprehensive evaluation of this omnimeter and its applications will be detailed in a forthcoming technical report.

In the main text, we observed that the 9-category omnimeter can fail at doping levels absent from the training set. To address this limitation, we broaden the training coverage to five doping levels, $\delta \simeq 3\%, 7\%, 12\%, 17\%, 22\%$; and for each doping level, we consider five thermal exponents $n_T = -\log_2 T = 0,2,4,6,8$, yielding a total of 25 categories.

Current state-of-the-art thermometry for cold-atom Hubbard experiments relies on a direct comparison between measured spin correlations and calibrated values from numerical simulations \cite{Chalopin&Bloch2024,Xu&Greiner2025}. Concretely, one compares the spin correlations measured from a snapshot obtained from the quantum gas microscope and assigns a temperature based on the closest match to the calibrated correlations. However, this straight-forward approach becomes numerically unstable especially when the correlations saturate at low temperatures (see Table~\ref{tab:spin_correlations}), leading to large fluctuations in the temperature estimates.

Therefore, instead of directly assigning the calibrated value, we adopt a probability-based formulation that delivers more stable estimates. Specifically, we postulate the probability (weight) of a snapshot $x$ at thermal exponent $n_T$ as
\begin{equation}
    \vspace{0.1em}
    p(n_T|x) \propto 1/\|\varGamma^{zz}(x) - \varGamma^{zz}(n_T)\|,
    \vspace{0.1em}
\end{equation}
where $\varGamma^{zz}(x)$ and $\varGamma^{zz}(n_T)$ denote the nearest-neighbor spin-$z$ correlations along the $y$ (vertical) direction measured from snapshot $x$, and the calibrated values (via XTRG) at thermal exponent $n_T$, respectively. This empirical formula ensures that the the thermal exponent with the closer correlation value attains the higher probability, while still allocating non-zero weights to other categories to enhance stability. Other functional forms, e.g. exponential decay, can also be considered; however, alternative choices do not significantly affect the performance.

For an ensemble ${x}$ of snapshots, the posterior $p(n_T|x)$ is obtained by averaging $p(n_T|x)$ over all $x$. The restriction to the $y$ direction is required to match the 2D geometry to the tensor network structure (only neighboring sites along the $y$ direction are guaranteed a bond directly connecting them) \cite{Zhang&VonDelft2025-XTRG}. While one can construct composite estimators that fuse multiple correlation messengers, in practice these do not surpass the stability or accuracy of the single-messenger formulation.

We evaluate the performance of the 25-category omnimeter and the spin-correlation-based thermometer on random ensembles of 30 snapshots for each location in phase space; results are summarized in Fig.~\ref{fig:eval_thermometry}. As before, open circles/pentagrams mark the locations in phase space of the snapshot ensembles under evaluation, with pentagrams (circles) indicating data included (not included) in the training set. Color scales are obtained via interpolation and made equal for both panels to enable a direct, like-for-like comparison.

The AI omnimeter consistently outperforms the thermometer based on spin correlations across most of phase space, with a pronounced advantage at lower temperatures where spin correlations begin to saturate. To rationalize this behavior, Table~\ref{tab:spin_correlations} reports the reference spin-$z$ correlations $\varGamma^{zz}$ together with their standard deviations $\text{std}(\varGamma^{zz})$ at the calibrated locations. At low temperatures (large $n_T$), the standard deviations exceed the separation between adjacent temperature categories, implying that nearest neighbor spin correlations alone cannot reliably discriminate fine temperature increments (e.g. in Table~\ref{tab:spin_correlations}(a), the standard deviations $\text{std}(\varGamma^{zz})$ for $n_T\!=\!6$ and $8$ are around $0.018$, whereas the difference between $\varGamma^{zz}$ is only $\approx0.001$). By contrast, the omnimeter automatically exploits a broader spectrum of correlation features beyond $\varGamma^{zz}$, enabling substantially more accurate temperature estimation.

\BatchStart

\begin{figure*}[hp]
\centering
\vspace{1.2em}
\subfloat{\includegraphics[scale=0.54]{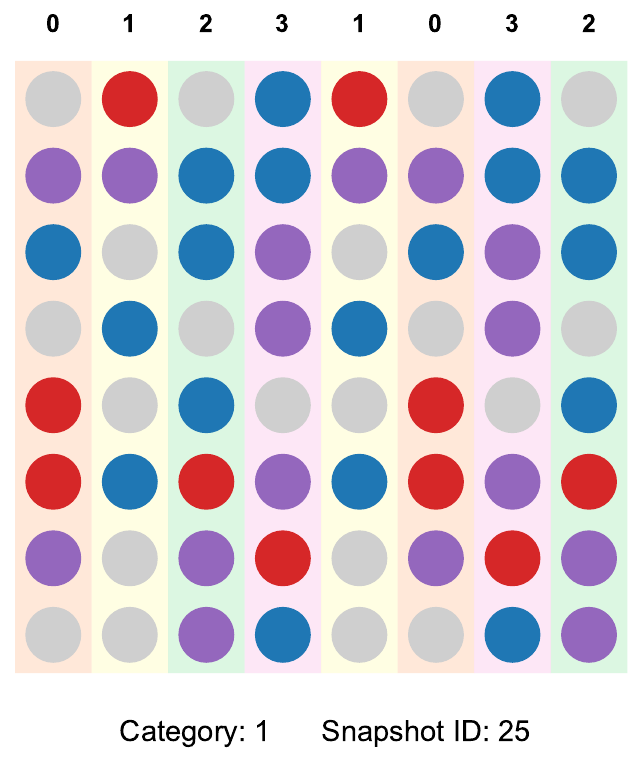}}\\
\subfloat{\includegraphics[scale=0.46]{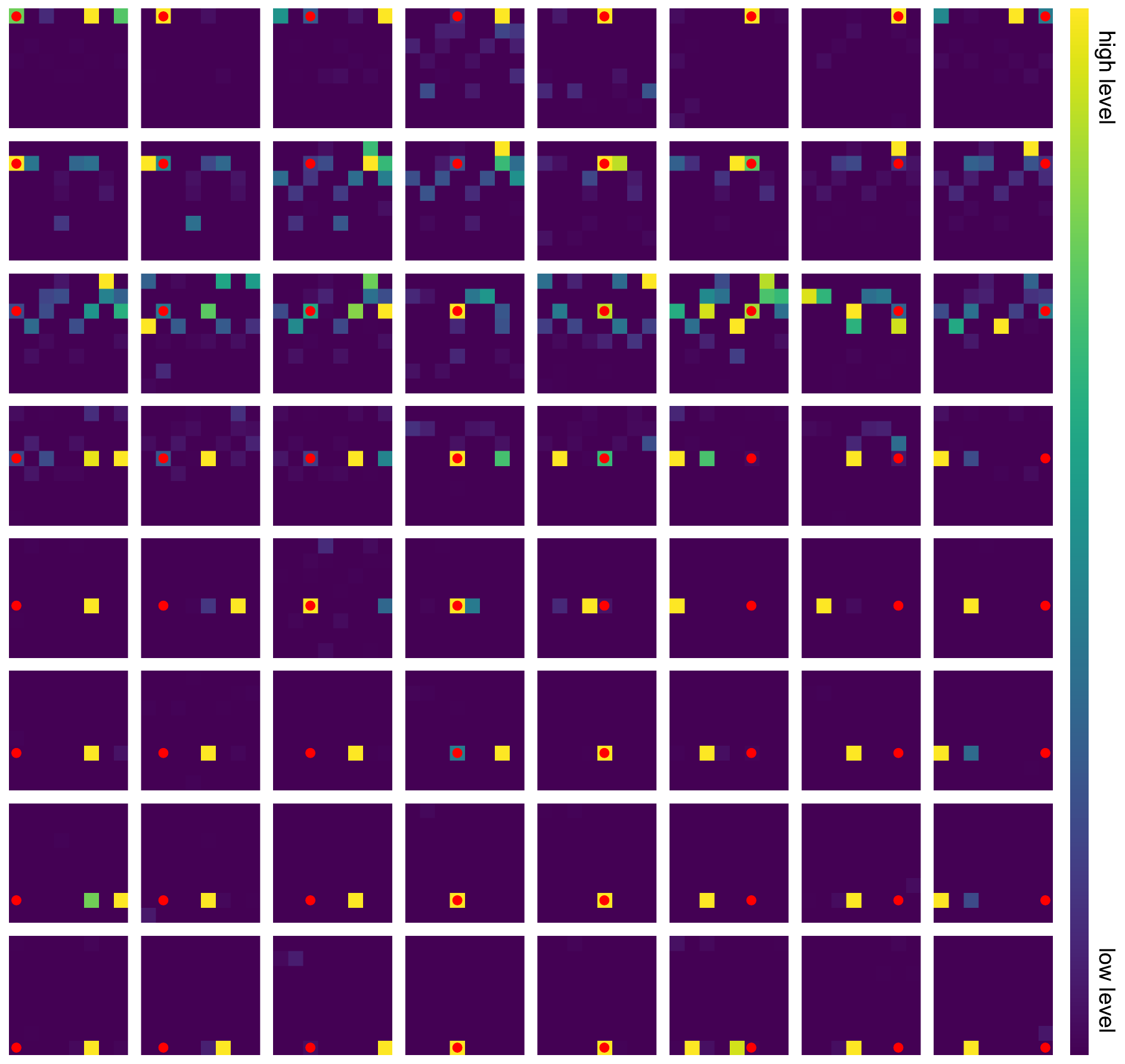}}\\[0.7em]
\begin{minipage}{0.95\textwidth}
    \BatchCaption{1}{Attention map (bottom) of the core model evaluated on a snapshot (top) from category~1 (\texttt{[1,0,3,2]}). The visualization consists of an 8×8 array of subplots, each displaying an 8×8 grid of cells. In each subplot, attention scores $\mathcal{A}_{ij}$ are encoded by a color scale; the query position $i$ coincides with the subplot's position and is marked by a red circle. The prominent highlighting of the attention map indicates that the model has successfully captured the underlying correlations between corresponding columns in the left and right halves.}
    \label{fig:derange_cat1}
\end{minipage}
\end{figure*}

\begin{figure*}[hp]
\centering
\vspace{1.2em}
\subfloat{\includegraphics[scale=0.54]{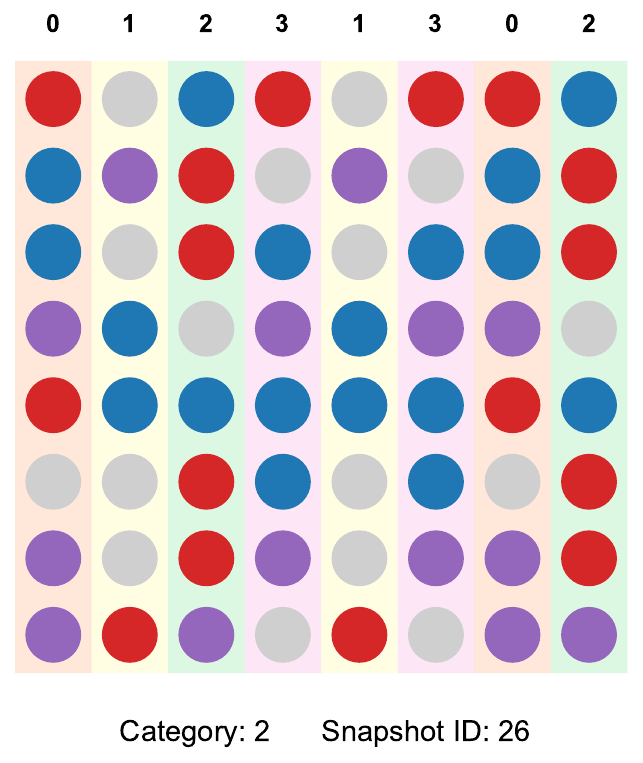}}\\
\subfloat{\includegraphics[scale=0.46]{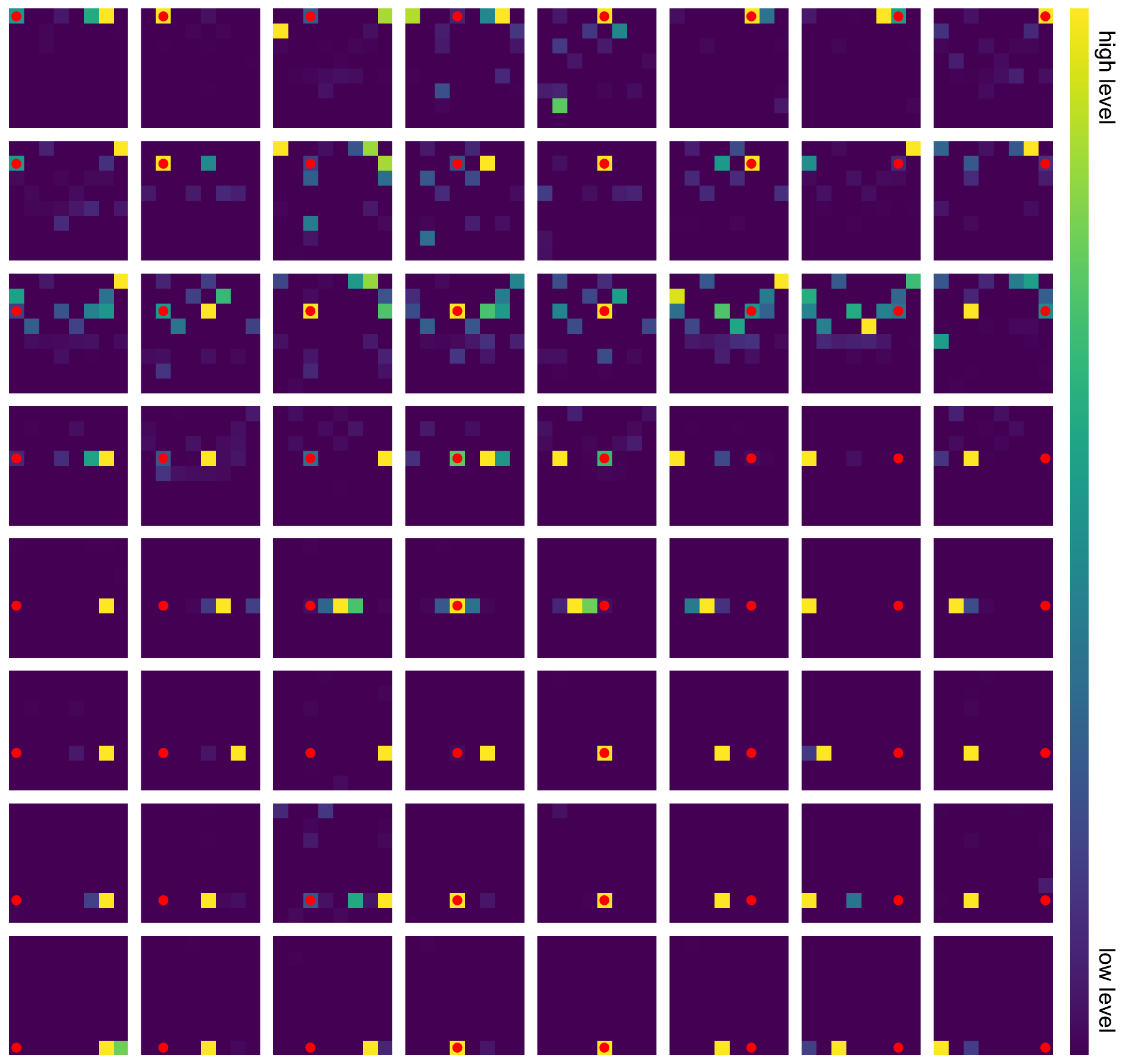}}\\[0.7em]
\begin{minipage}{0.95\textwidth}
    \BatchCaption{2}{Same as Fig.~\ref{fig:derange_cat1}-1, now for a snapshot from category~2 (\texttt{[1,3,0,2]}). Same as Fig.~\ref{fig:derange_cat1}-1, the visualization consists of an 8×8 array of subplots, each displaying an 8×8 grid of cells. In each subplot, attention scores $\mathcal{A}_{ij}$ are encoded by a color scale; the query position $i$ coincides with the subplot's position and is marked by a red circle. The prominent highlighting of the attention map indicates that the model has successfully captured the underlying correlations between corresponding columns in the left and right halves.}
    \label{fig:derange_cat2}
\end{minipage}
\end{figure*}

\begin{figure*}[hp]
\centering
\vspace{1.2em}
\subfloat{\includegraphics[scale=0.54]{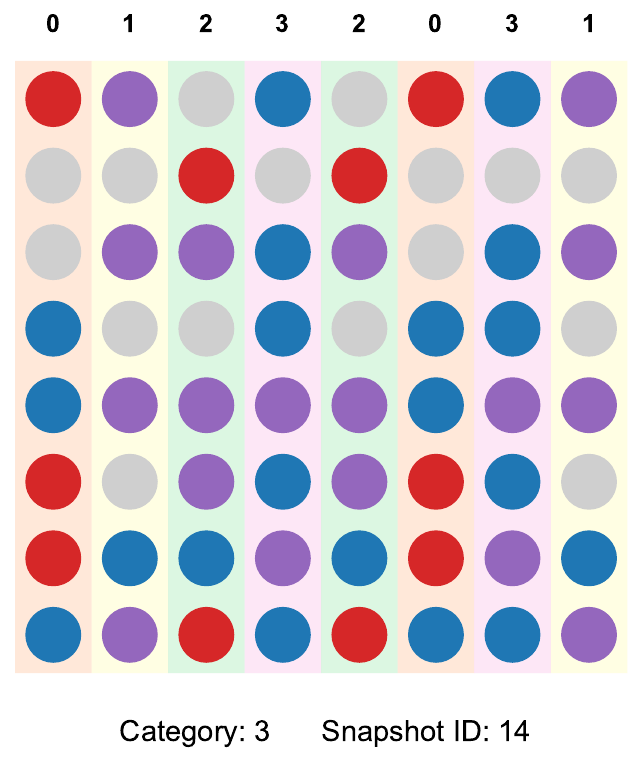}}\\
\subfloat{\includegraphics[scale=0.46]{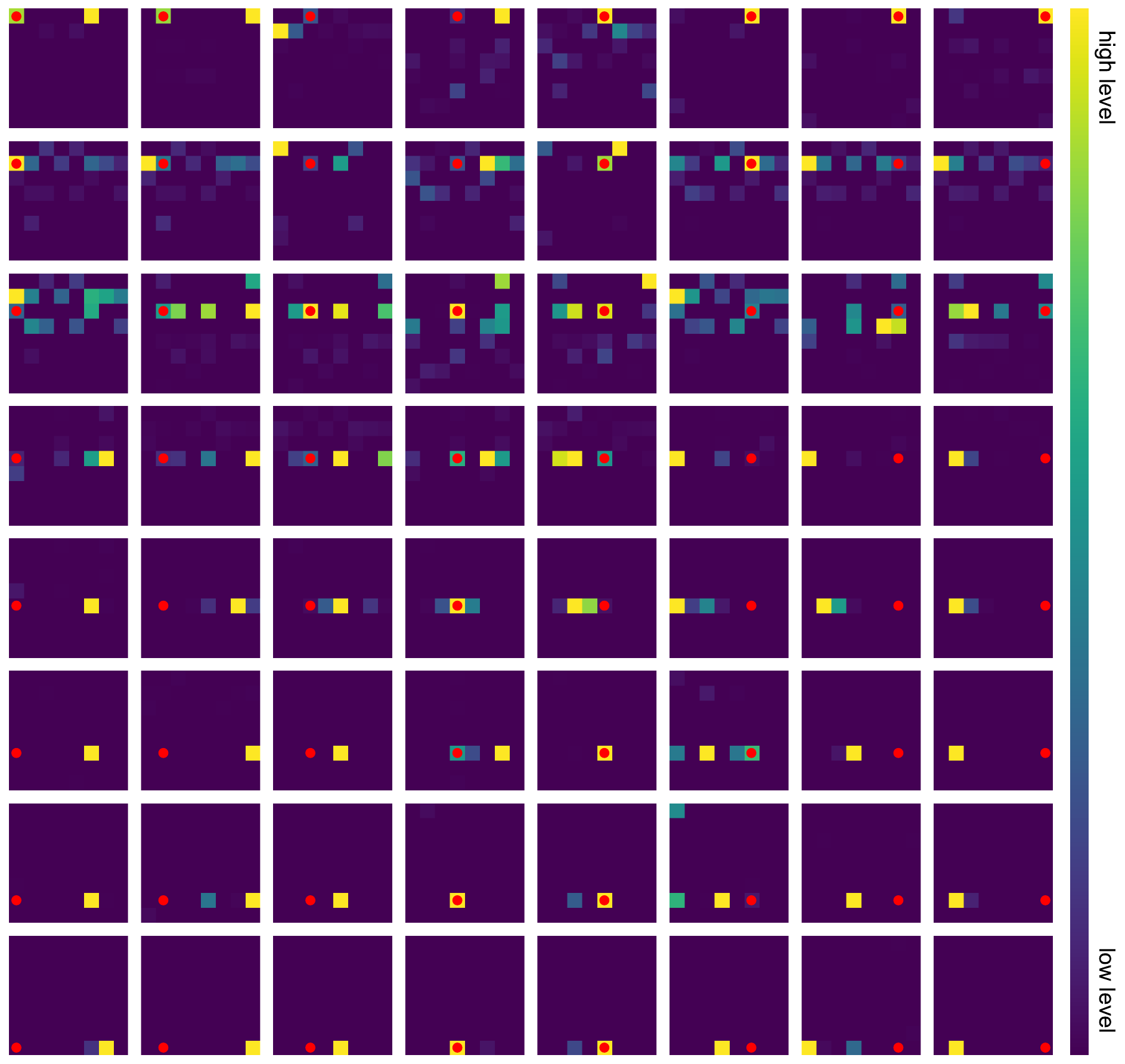}}\\[0.7em]
\begin{minipage}{0.95\textwidth}
    \BatchCaption{3}{Same as Fig.~\ref{fig:derange_cat1}-1, now for a snapshot from category~3 (\texttt{[2,0,3,1]}). Same as Fig.~\ref{fig:derange_cat1}-1, the visualization consists of an 8×8 array of subplots, each displaying an 8×8 grid of cells. In each subplot, attention scores $\mathcal{A}_{ij}$ are encoded by a color scale; the query position $i$ coincides with the subplot's position and is marked by a red circle. The prominent highlighting of the attention map indicates that the model has successfully captured the underlying correlations between corresponding columns in the left and right halves.}
    \label{fig:derange_cat3}
\end{minipage}
\end{figure*}

\begin{figure*}
\centering
\vspace{1.2em}
\subfloat{\includegraphics[scale=0.54]{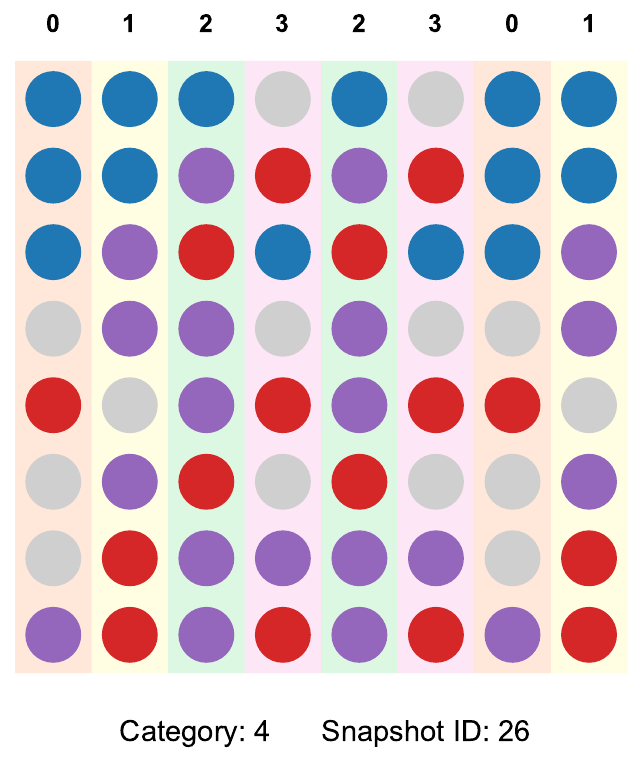}}
\hspace{2em}
\subfloat{\includegraphics[scale=0.46]{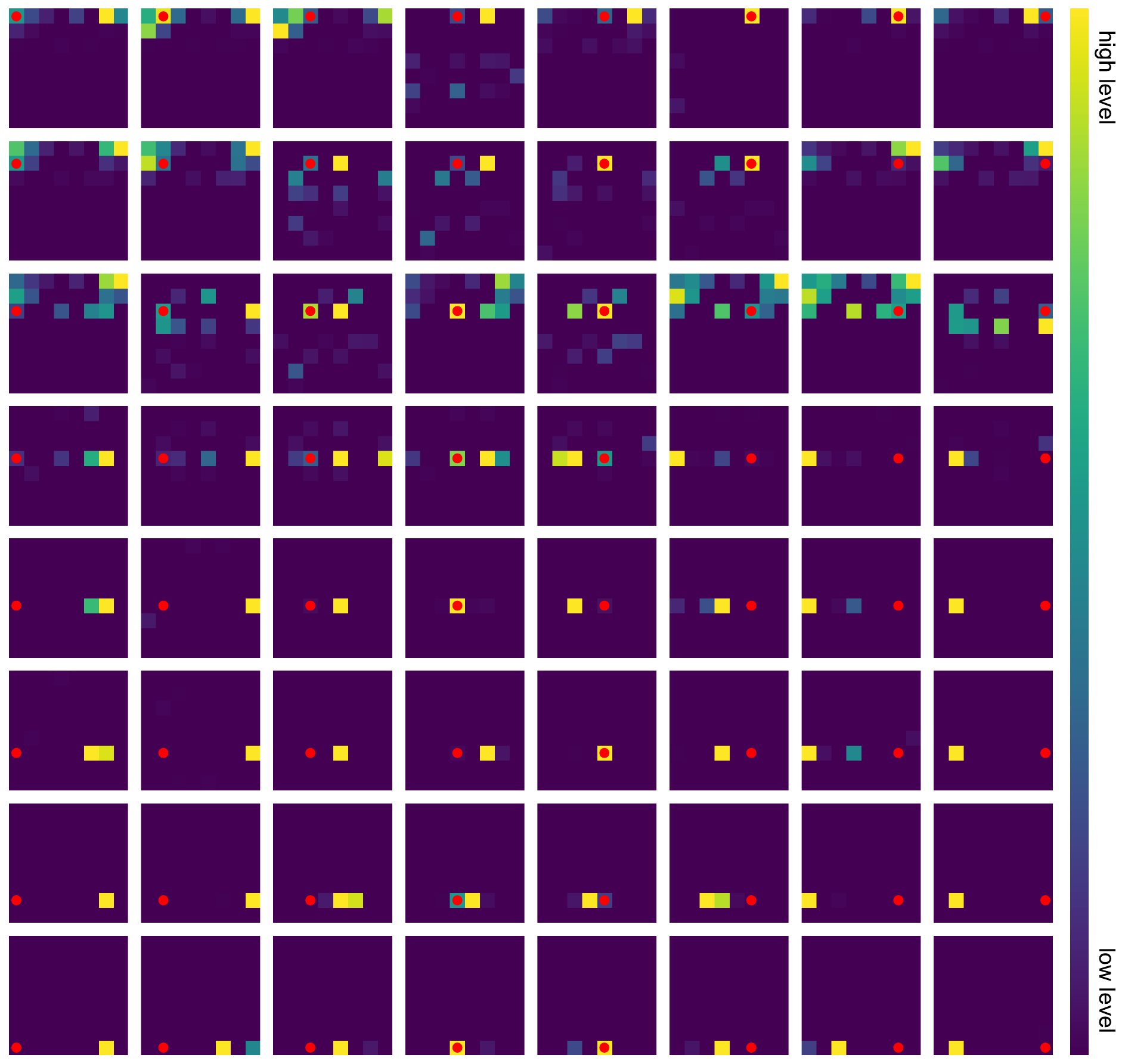}}\\[0.7em]
\begin{minipage}{0.95\textwidth}
    \BatchCaption{4}{Same as Fig.~\ref{fig:derange_cat1}-1, now for a snapshot from category~4 (\texttt{[2,3,0,1]}). Same as Fig.~\ref{fig:derange_cat1}-1, the visualization consists of an 8×8 array of subplots, each displaying an 8×8 grid of cells. In each subplot, attention scores $\mathcal{A}_{ij}$ are encoded by a color scale; the query position $i$ coincides with the subplot's position and is marked by a red circle. The prominent highlighting of the attention map indicates that the model has successfully captured the underlying correlations between corresponding columns in the left and right halves.}
    \label{fig:derange_cat4}
\end{minipage}
\end{figure*}

\begin{figure*}[hp]
\centering
\vspace{1.2em}
\subfloat{\includegraphics[scale=0.54]{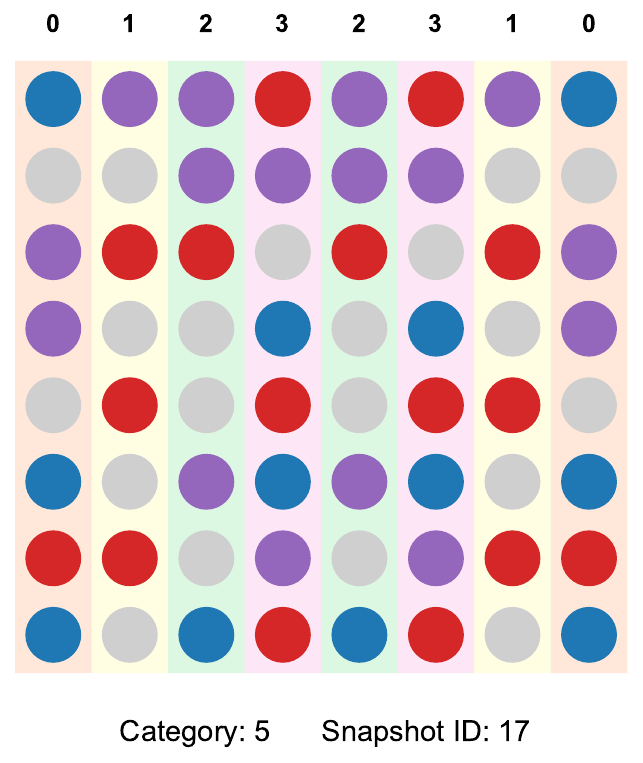}}\\
\subfloat{\includegraphics[scale=0.46]{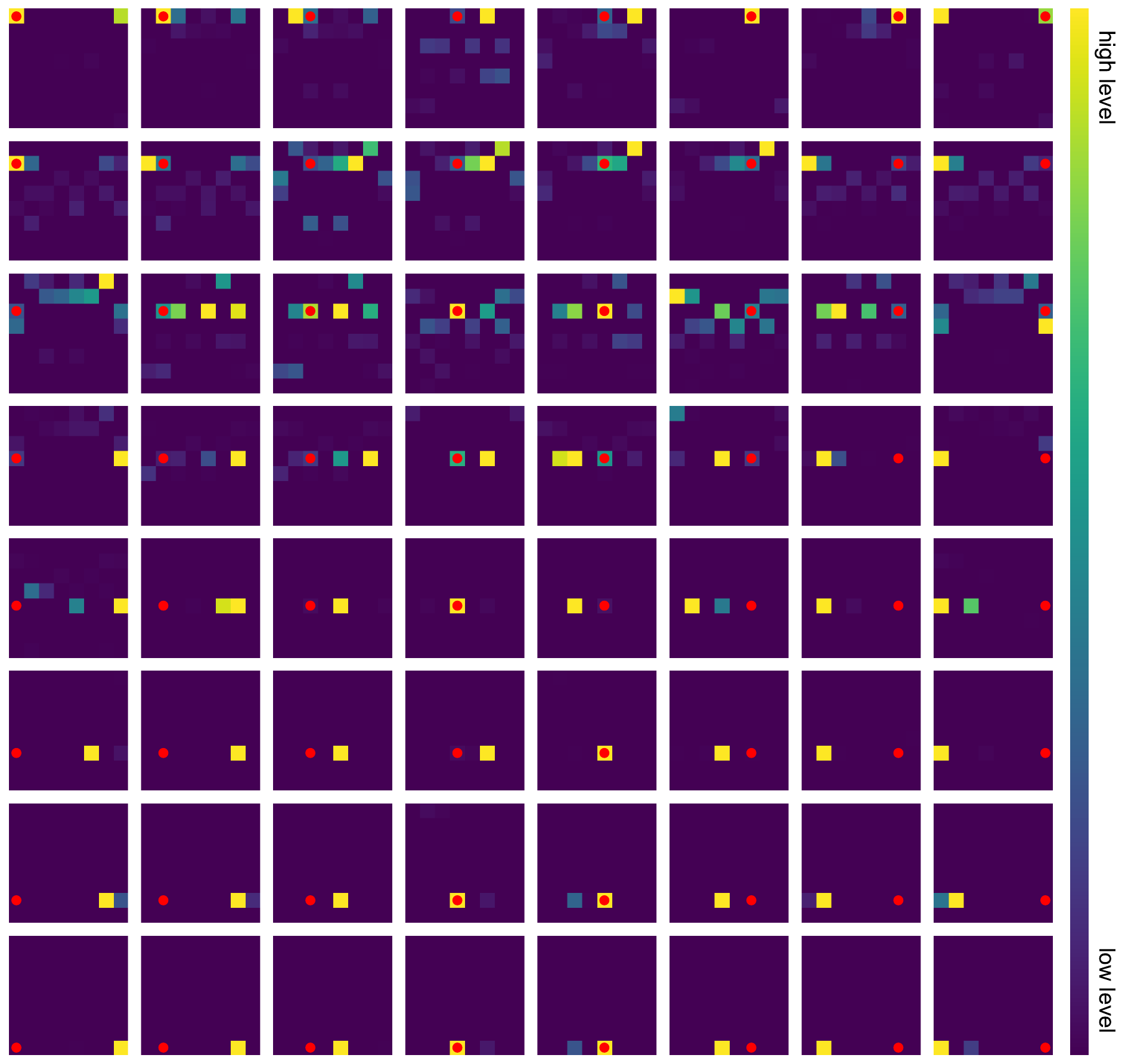}}\\[0.7em]
\begin{minipage}{0.95\textwidth}
    \BatchCaption{5}{Same as Fig.~\ref{fig:derange_cat1}-1, now for a snapshot from category~5 (\texttt{[2,3,1,0]}). Same as Fig.~\ref{fig:derange_cat1}-1, the visualization consists of an 8×8 array of subplots, each displaying an 8×8 grid of cells. In each subplot, attention scores $\mathcal{A}_{ij}$ are encoded by a color scale; the query position $i$ coincides with the subplot's position and is marked by a red circle. The prominent highlighting of the attention map indicates that the model has successfully captured the underlying correlations between corresponding columns in the left and right halves.}
    \label{fig:derange_cat5}
\end{minipage}
\end{figure*}

\begin{figure*}[hp]
\centering
\vspace{1.2em}
\subfloat{\includegraphics[scale=0.54]{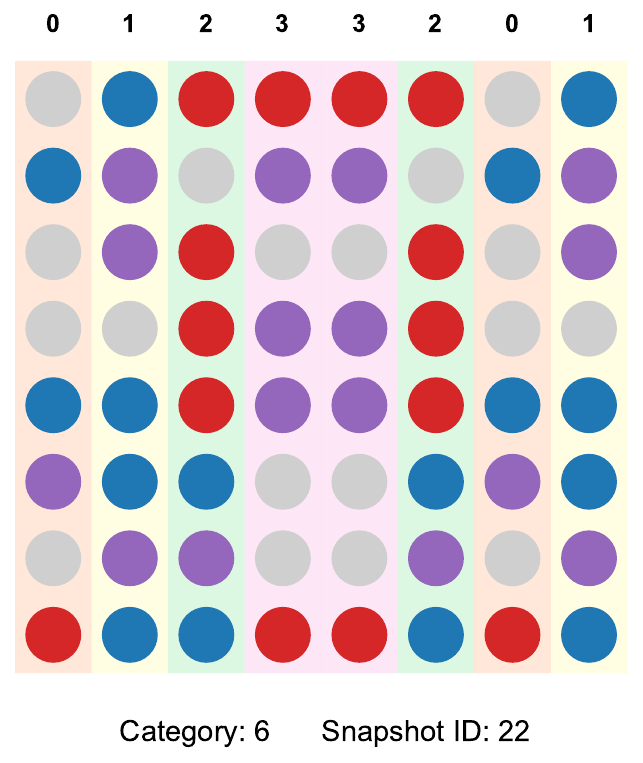}}\\
\subfloat{\includegraphics[scale=0.46]{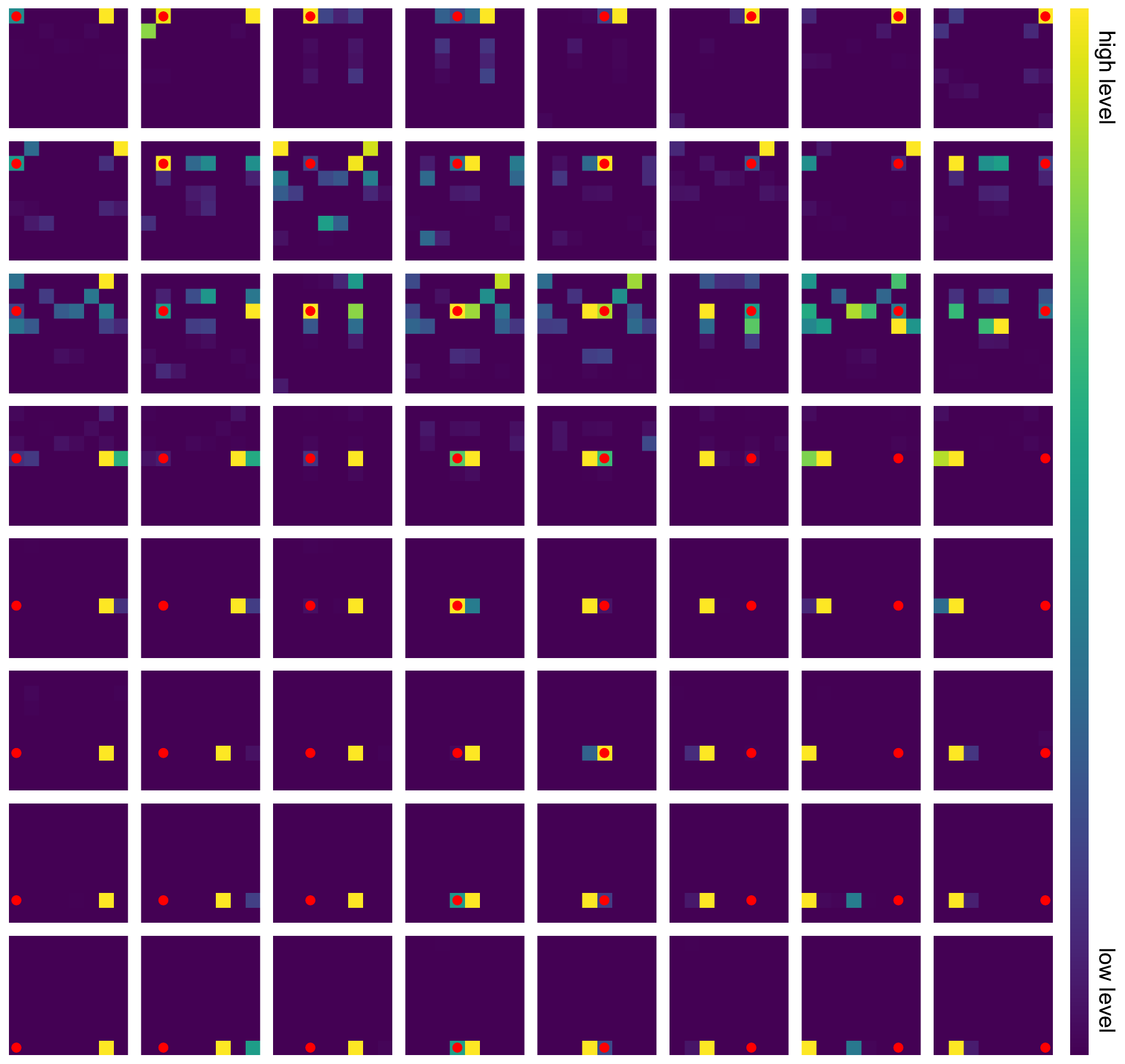}}\\[0.7em]
\begin{minipage}{0.95\textwidth}
    \BatchCaption{6}{Same as Fig.~\ref{fig:derange_cat1}-1, now for a snapshot from category~6 (\texttt{[3,2,0,1]}). Same as Fig.~\ref{fig:derange_cat1}-1, the visualization consists of an 8×8 array of subplots, each displaying an 8×8 grid of cells. In each subplot, attention scores $\mathcal{A}_{ij}$ are encoded by a color scale; the query position $i$ coincides with the subplot's position and is marked by a red circle. The prominent highlighting of the attention map indicates that the model has successfully captured the underlying correlations between corresponding columns in the left and right halves.}
    \label{fig:derange_cat6}
\end{minipage}
\end{figure*}

\BatchEnd

\BatchStart

\begin{figure*}[hp]
\centering
\vspace{1.2em}
\subfloat{\includegraphics[scale=0.54]{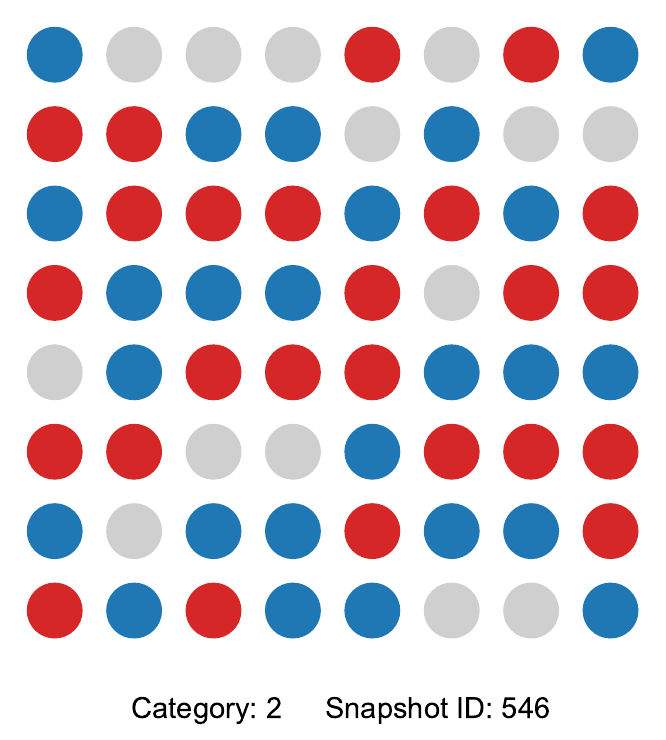}}\\
\subfloat{\includegraphics[scale=0.46]{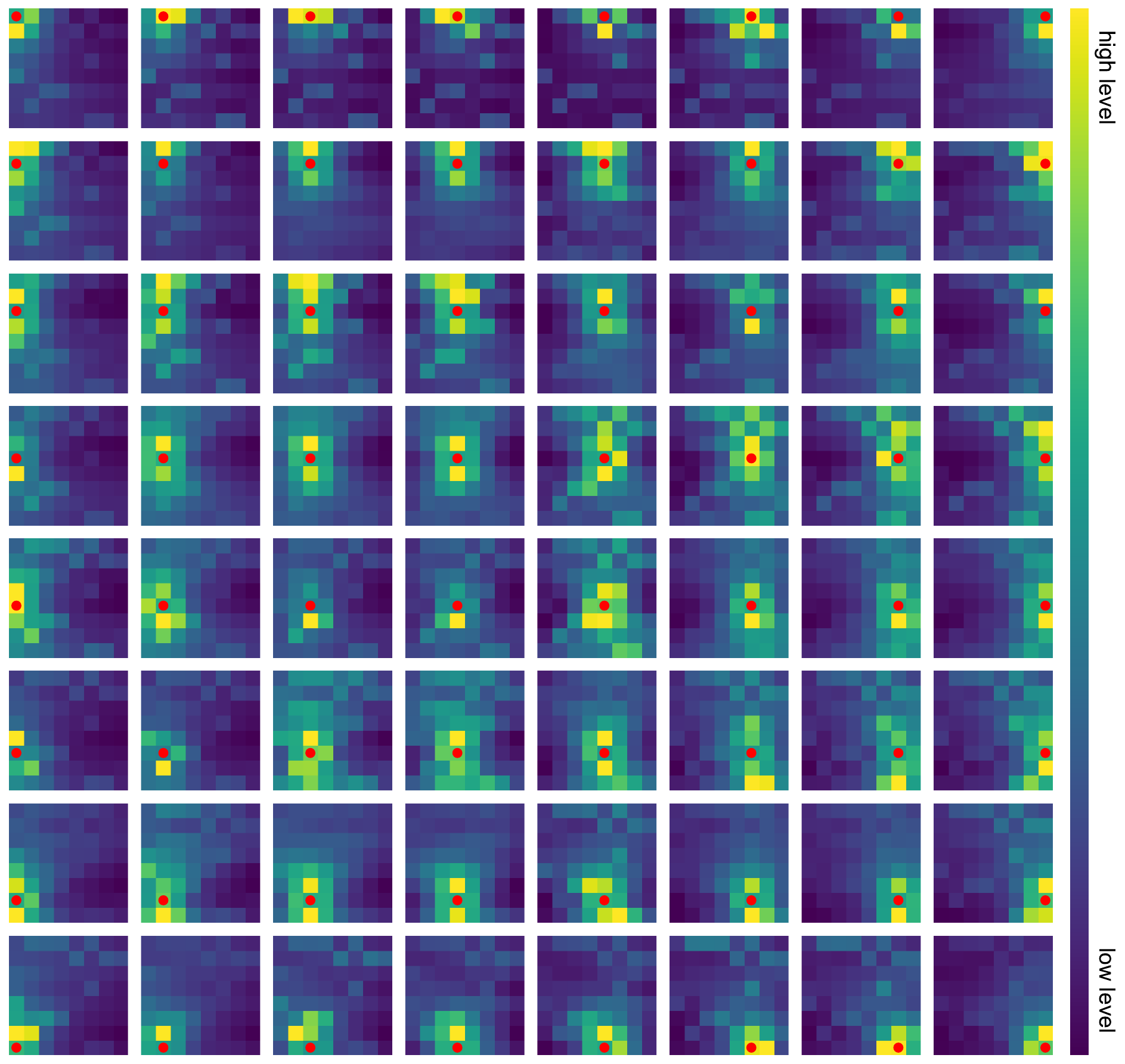}}\\[0.7em]
\begin{minipage}{0.95\textwidth}
    \BatchCaption{1}{Attention map (bottom) of the core model evaluated on a snapshot (top) at low temperature and \emph{over-doped} region. The visualization consists of an 8×8 array of subplots, each displaying an 8×8 grid of cells. In each subplot, attention scores $\mathcal{A}_{ij}$ are encoded by a color scale; the query position $i$ coincides with the subplot's position and is marked by a red circle.}
    \label{fig:map_2546}
\end{minipage}
\end{figure*}

\begin{figure*}[hp]
\centering
\vspace{1.2em}
\subfloat{\includegraphics[scale=0.54]{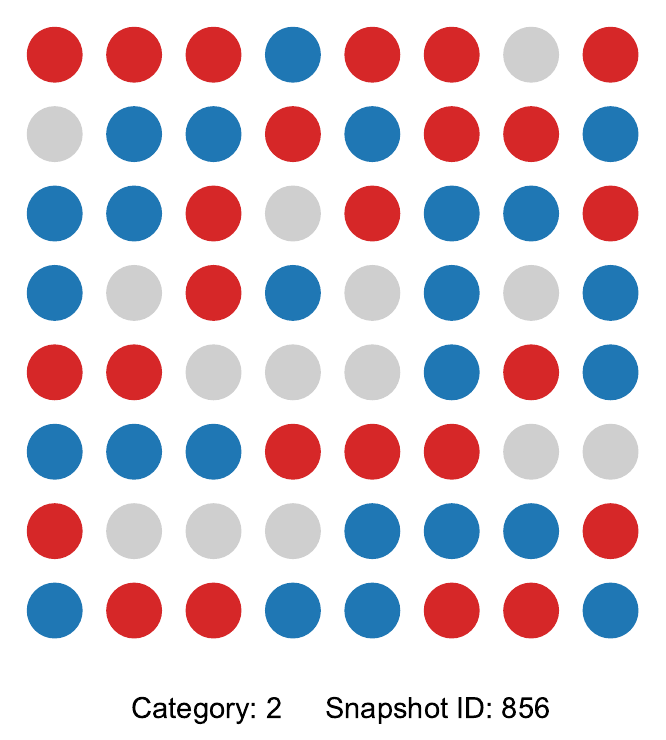}}\\
\subfloat{\includegraphics[scale=0.46]{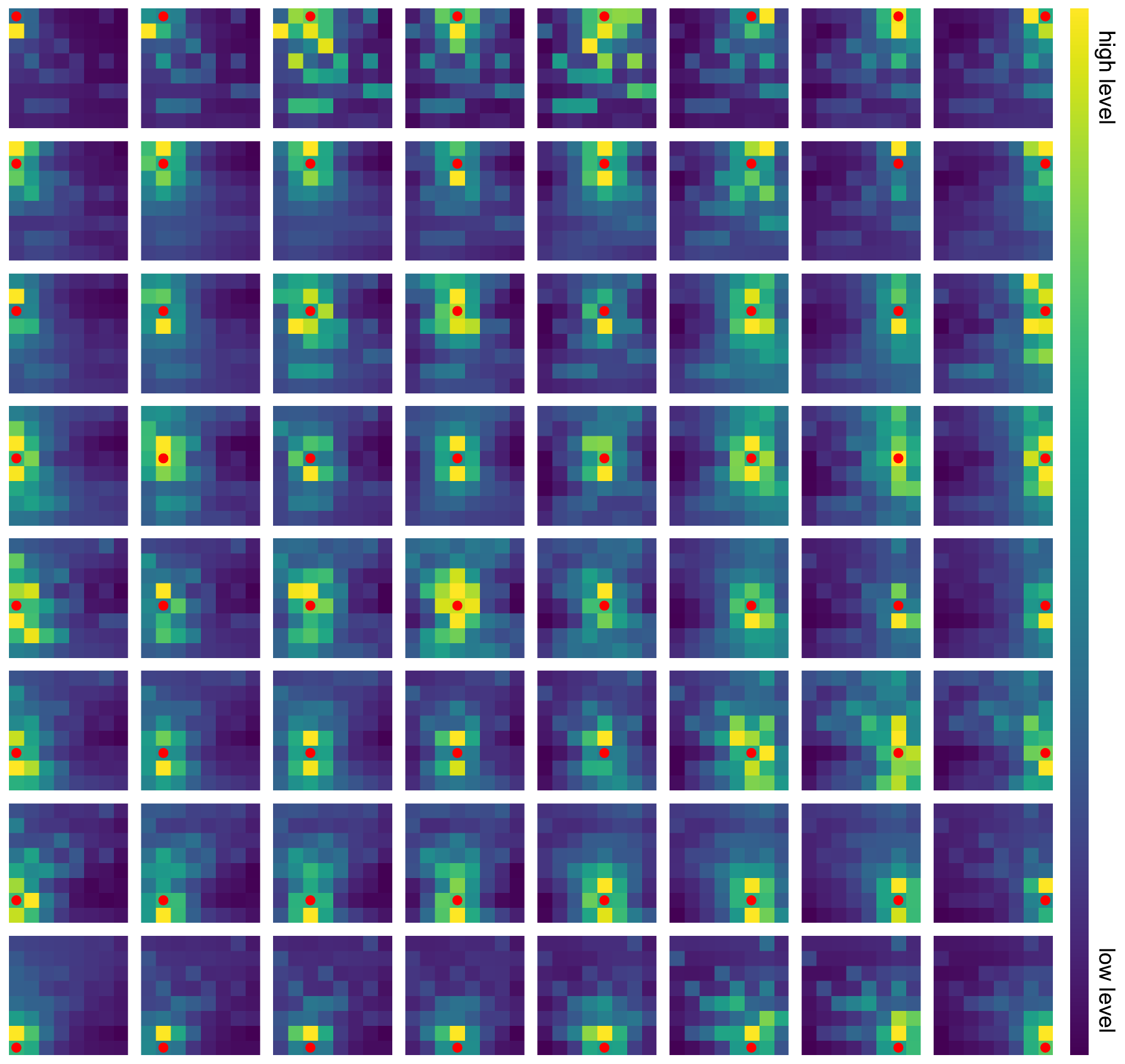}}\\[0.7em]
\begin{minipage}{0.95\textwidth}
    \BatchCaption{2}{Attention map (bottom) of the core model evaluated on a snapshot (top) at low temperature and \emph{over-doped} region. Same as Fig.~\ref{fig:map_2546}-1, the visualization consists of an 8×8 array of subplots, each displaying an 8×8 grid of cells. In each subplot, attention scores $\mathcal{A}_{ij}$ are encoded by a color scale; the query position $i$ coincides with the subplot's position and is marked by a red circle.}
    \label{fig:map_2856}
\end{minipage}
\end{figure*}

\begin{figure*}[hp]
\centering
\vspace{1.2em}
\subfloat{\includegraphics[scale=0.54]{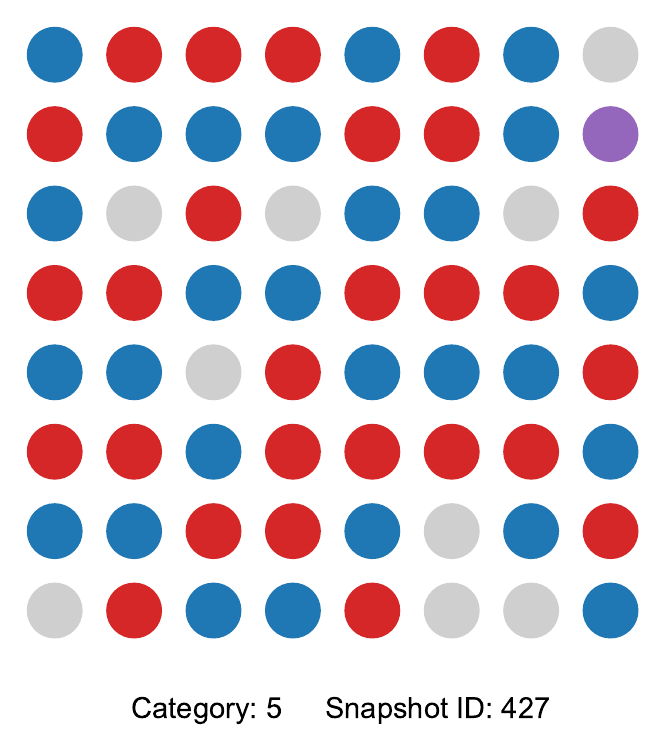}}\\
\subfloat{\includegraphics[scale=0.46]{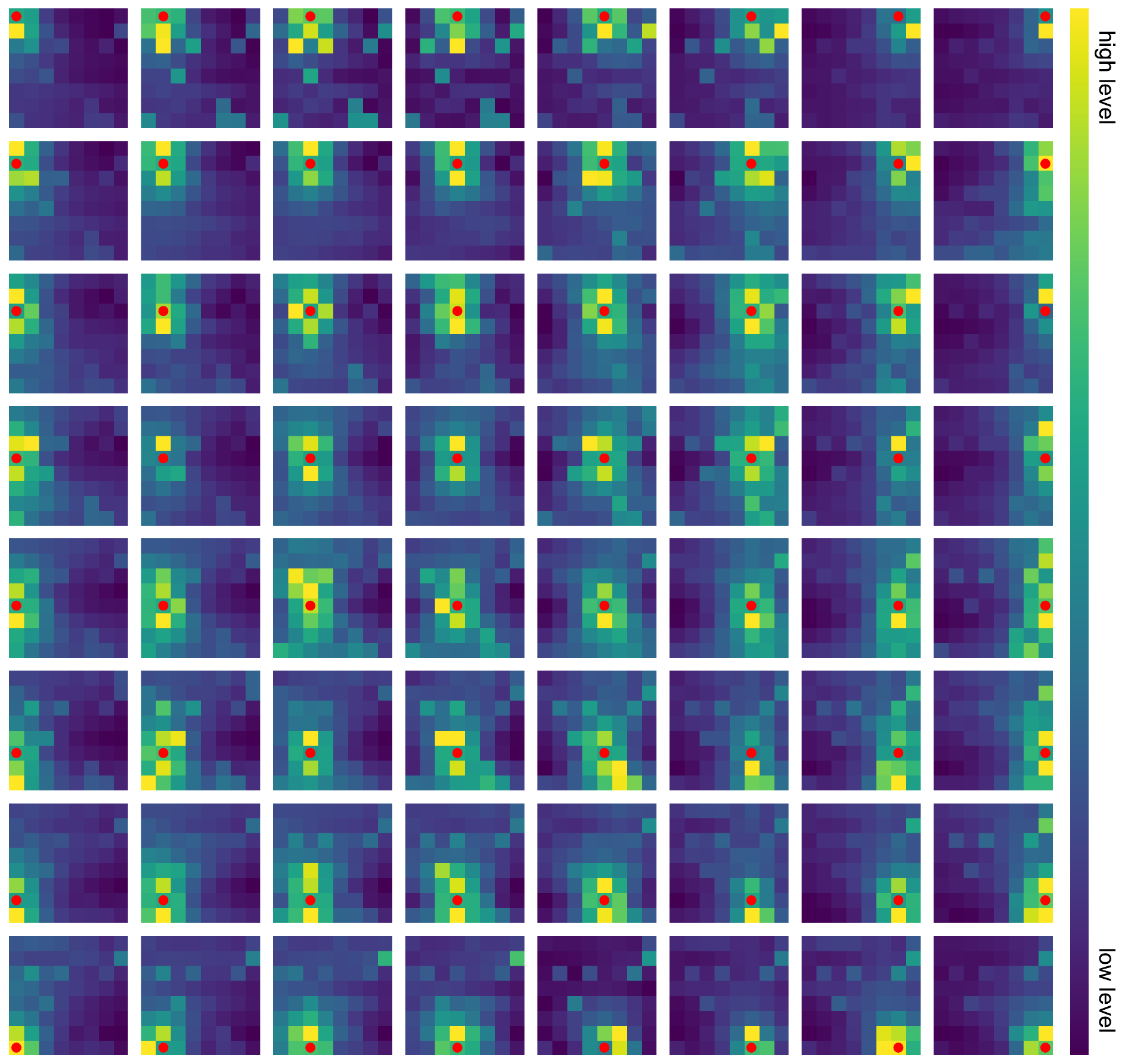}}\\[0.7em]
\begin{minipage}{0.95\textwidth}
    \BatchCaption{3}{Attention map (bottom) of the core model evaluated on a snapshot (top) at low temperature and \emph{medium-doped} region. Same as Fig.~\ref{fig:map_2546}-1, the visualization consists of an 8×8 array of subplots, each displaying an 8×8 grid of cells. In each subplot, attention scores $\mathcal{A}_{ij}$ are encoded by a color scale; the query position $i$ coincides with the subplot's position and is marked by a red circle.}
    \label{fig:map_5427}
\end{minipage}
\end{figure*}

\begin{figure*}
\centering
\vspace{1.2em}
\subfloat{\includegraphics[scale=0.54]{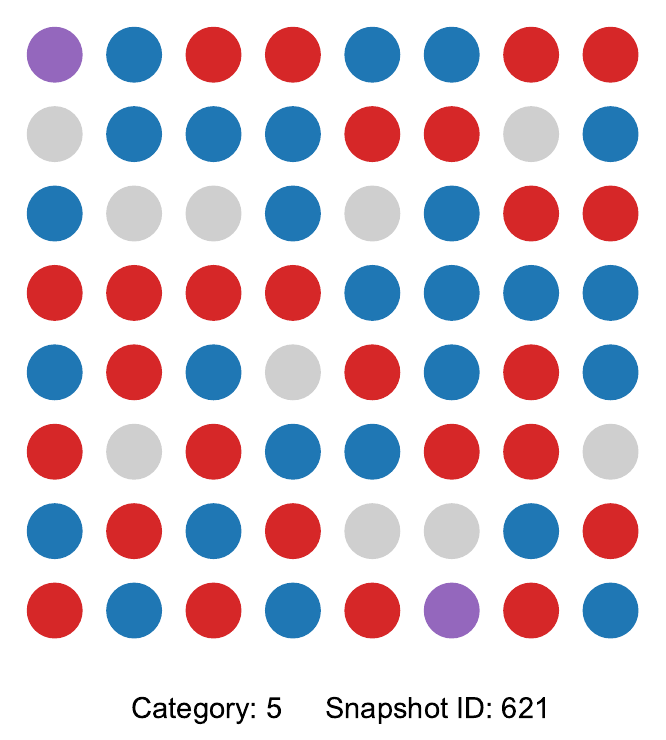}}
\hspace{2em}
\subfloat{\includegraphics[scale=0.46]{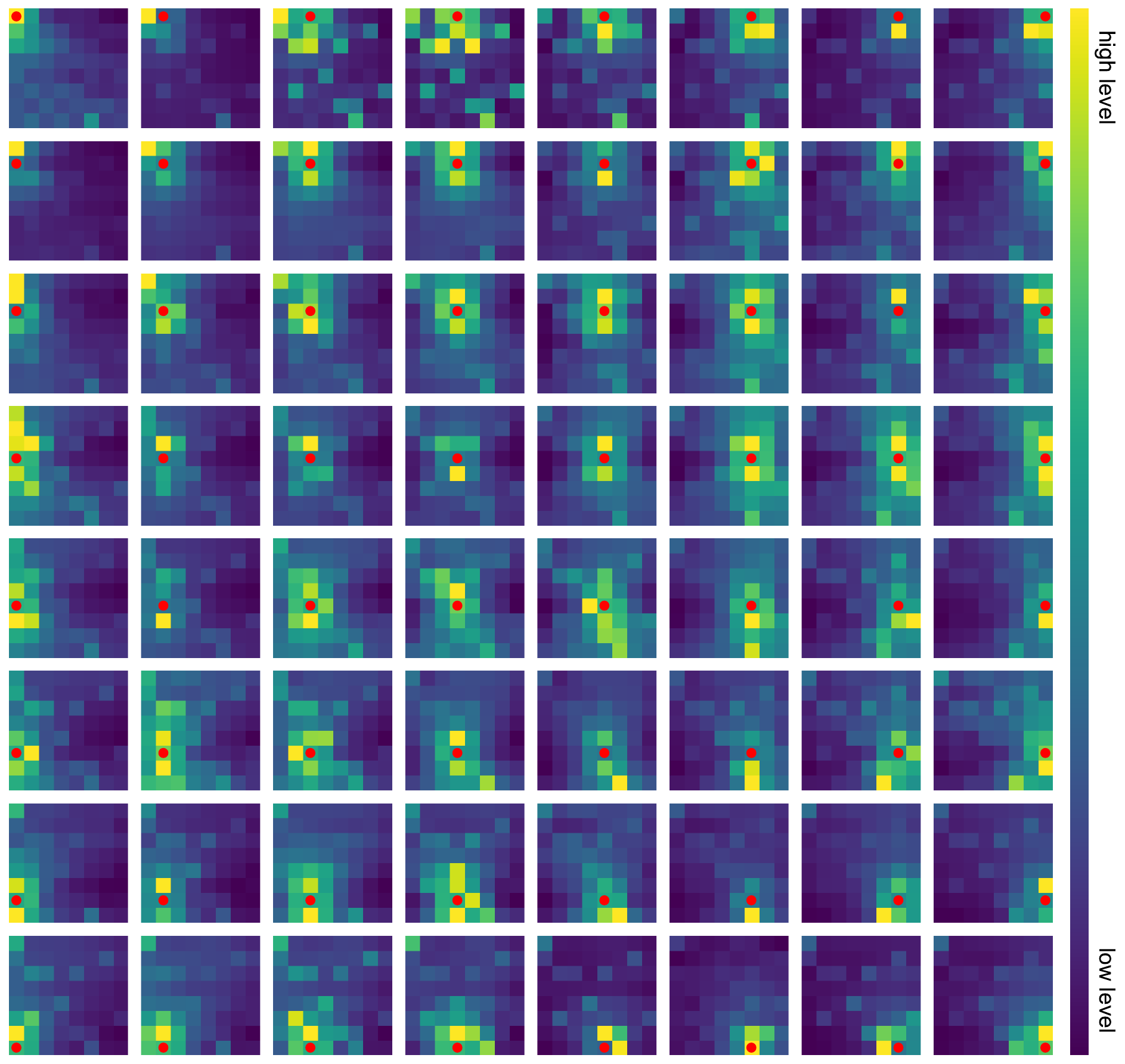}}\\[0.7em]
\begin{minipage}{0.95\textwidth}
    \BatchCaption{4}{Attention map (bottom) of the core model evaluated on a snapshot (top) at low temperature and \emph{medium-doped} region. Same as Fig.~\ref{fig:map_2546}-1, the visualization consists of an 8×8 array of subplots, each displaying an 8×8 grid of cells. In each subplot, attention scores $\mathcal{A}_{ij}$ are encoded by a color scale; the query position $i$ coincides with the subplot's position and is marked by a red circle.}
    \label{fig:map_5621}
\end{minipage}
\end{figure*}

\begin{figure*}[hp]
\centering
\vspace{1.2em}
\subfloat{\includegraphics[scale=0.54]{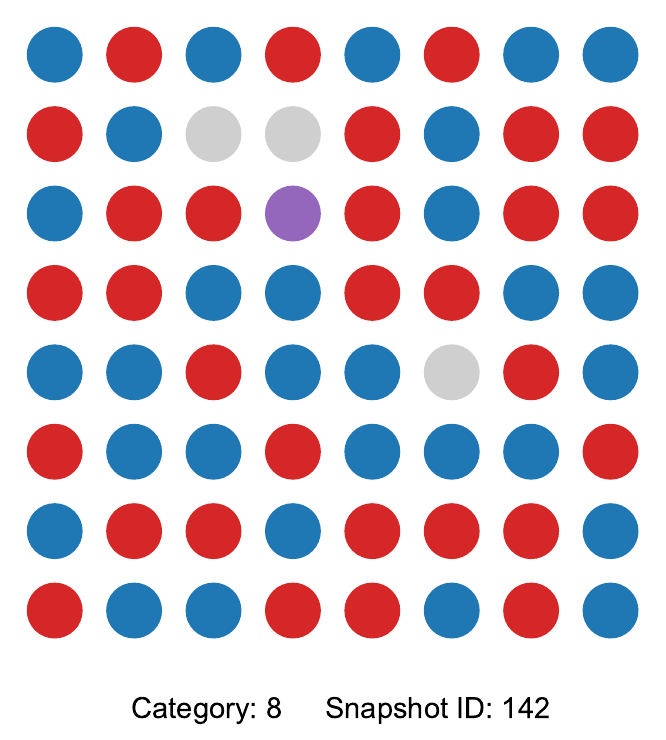}}\\
\subfloat{\includegraphics[scale=0.46]{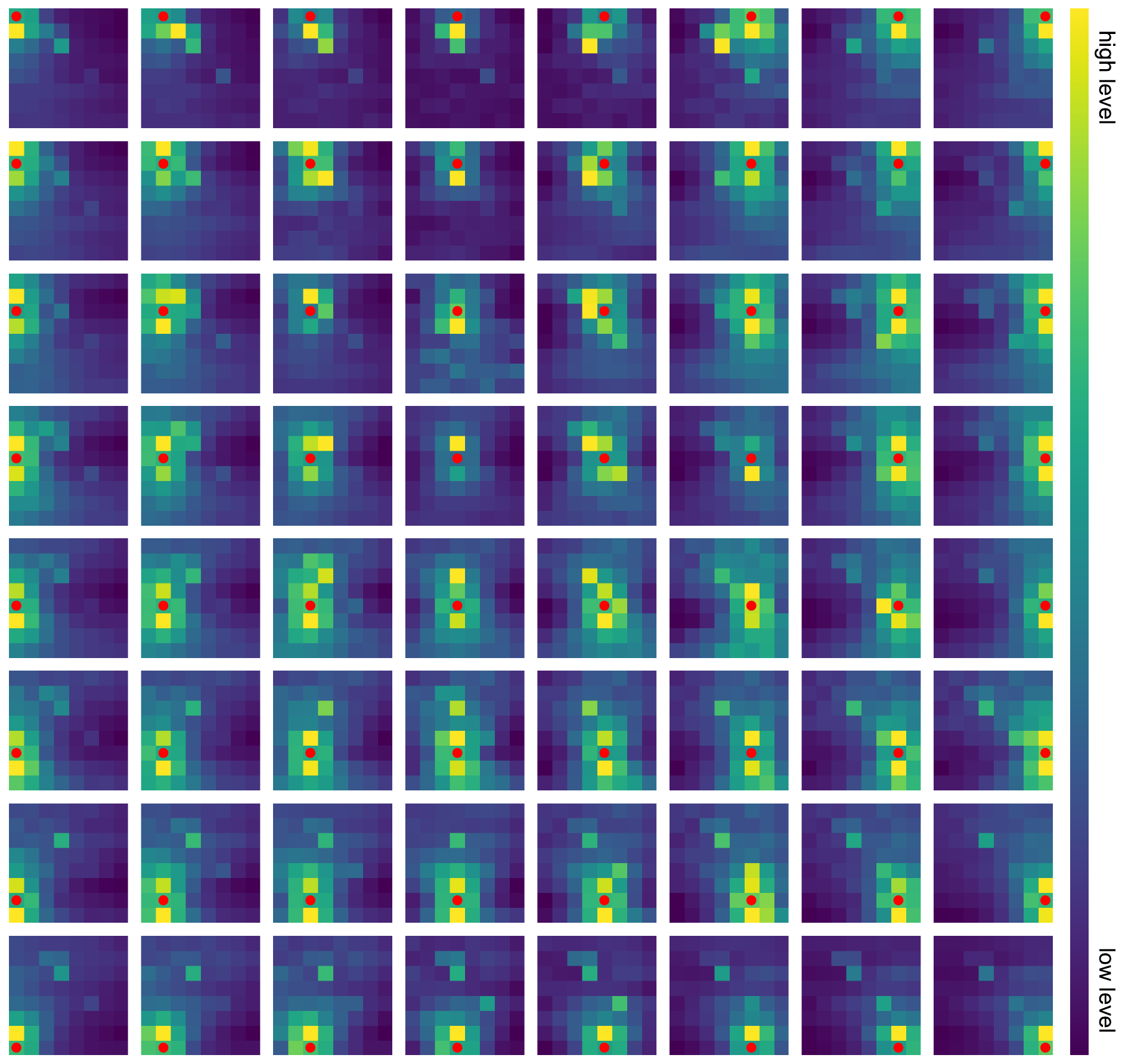}}\\[0.7em]
\begin{minipage}{0.95\textwidth}
    \BatchCaption{5}{Attention map (bottom) of the core model evaluated on a snapshot (top) at low temperature and \emph{under-doped} region. Same as Fig.~\ref{fig:map_2546}-1, the visualization consists of an 8×8 array of subplots, each displaying an 8×8 grid of cells. In each subplot, attention scores $\mathcal{A}_{ij}$ are encoded by a color scale; the query position $i$ coincides with the subplot's position and is marked by a red circle.}
    \label{fig:map_8142}
\end{minipage}
\end{figure*}

\begin{figure*}[hp]
\centering
\vspace{1.2em}
\subfloat{\includegraphics[scale=0.54]{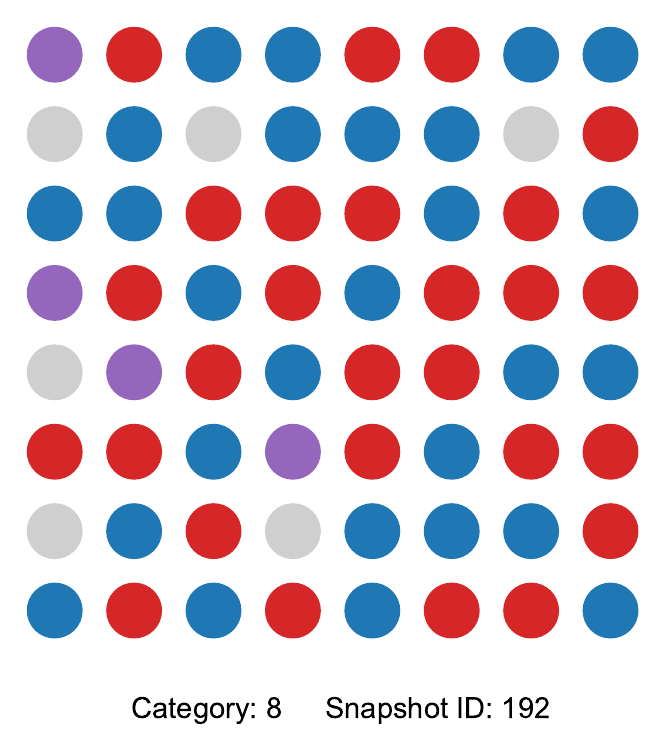}}\\
\subfloat{\includegraphics[scale=0.46]{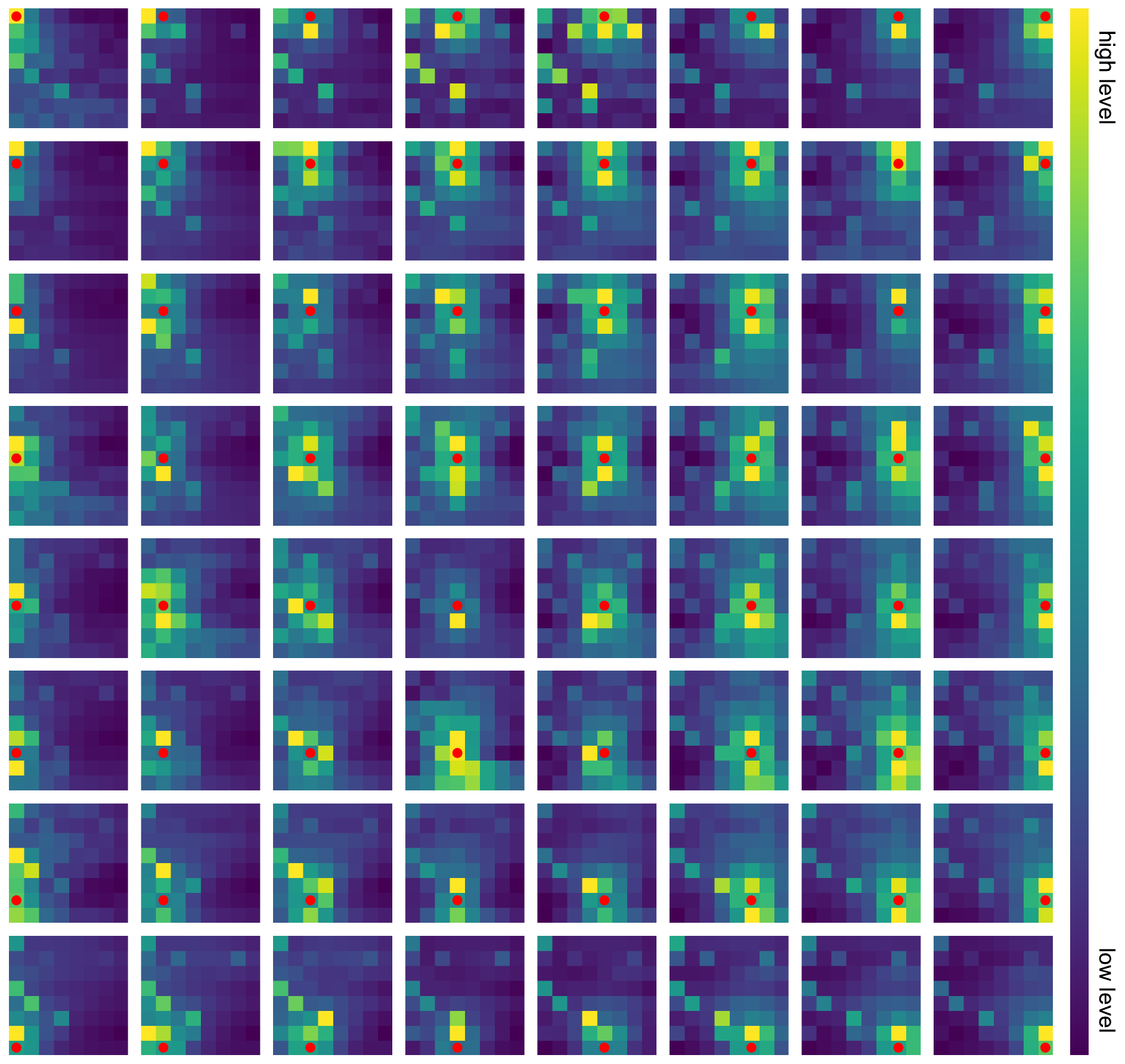}}\\[0.7em]
\begin{minipage}{0.95\textwidth}
    \BatchCaption{6}{Attention map (bottom) of the core model evaluated on a snapshot (top) at low temperature and \emph{under-doped} region. Same as Fig.~\ref{fig:map_2546}-1, the visualization consists of an 8×8 array of subplots, each displaying an 8×8 grid of cells. In each subplot, attention scores $\mathcal{A}_{ij}$ are encoded by a color scale; the query position $i$ coincides with the subplot's position and is marked by a red circle.}
    \label{fig:map_8192}
\end{minipage}
\end{figure*}

\BatchEnd

%

\endgroup
\makeatother

\end{document}